\documentclass[review]{elsarticle}

\usepackage{lineno,hyperref}
\modulolinenumbers[5]

\journal{Advances in Water Resources}

\usepackage[dvipsnames]{xcolor}
\usepackage{graphicx}
\usepackage{subcaption}
\usepackage{hyphenat}
\usepackage{amsmath}
\usepackage{adjustbox}

\graphicspath{ {./figures/} }

\begin{document}

\begin{frontmatter}

\title{Bed form-induced hyporheic exchange and geochemical hotspots}





\author{Faranak Behzadi\corref{cor1}}
\ead{faranak.behzadi@uc.edu}
\cortext[cor1]{Corresponding author}

\author{Corey D. Wallace}
\author{Dylan Ward}
\author{Mohamad Reza Soltanian}
\ead{soltanma@uc.edu}

\address{Department of Geology, University of Cincinnati, Cincinnati, OH, USA}

\begin{abstract}

Small-scale bed form topographies control hyporheic exchange and biogeochemical processes within aquatic sediments, which ultimately affect water quality and nutrient cycling at the watershed scale. The impact of three-dimensional (3D) and small-scale bed form topographies on hyporheic exchange and solute mixing is investigated in the present work. The effect of bed form morphologies on the development of zones of enhanced reaction rates (i.e., hot spots) is also studied. A computational fluid dynamics model to simulate river flow over bed forms is combined with a subsurface flow and multicomponent reactive solute transport model. A wide variety of bed form topographies are generated using geometric models by varying parameters controlling curvature as well as bed form wavelength and amplitude. Our results show that the pressure distribution at the sediment-water interface is strongly affected by the bed form geometry. Higher phase shifts in bed form shapes result in overall higher average velocity, larger zones of enhanced pressure and reaction rates, and higher amounts of solute exchange.
Moreover, the bed form shapes control the reaction process for a wide range of sediment conductivities. This study advances the understanding of the effects of complex and small scale morphological features on hyporheic exchange processes. 

\end{abstract}


\end{frontmatter}


\section{Introduction}

Surface water in rivers and streams mixes with groundwater as it is diverted along subsurface (hyporheic) flow paths, bringing with it a myriad of chemical solutes that are transported throughout the shallow streambed by advective and dispersive processes and exposed to microbially-active sediments. As a fully-coupled system, the interactions between surface water and groundwater control the delivery of limiting nutrients and the distribution and abundance of microorganisms in streams and the hyporheic zone. Transport processes constantly re-adjust spatially and temporally in response to geomorphic roughness features along the channel (e.g., bed form which is a single geometric element such as a ripple or a dune \cite{rl1982sedimentary}) and variations in streamflow. In particular, the size and shape of small bed forms strongly influence hyporheic exchange (\cite{ElliottBrooks1997,Gomez-Velez2015,BOANO20101367}), thus impacting the distribution of solutes throughout the hyporheic zone \cite{Buffington_Tonina2009,Harvey2012,Chen_Cardenas2018,BAYANICARDENAS20081382}. As a result, regions of enhanced biogeochemical reaction rates (i.e., hotspots) often develop as river water and groundwater coalesce in close contact with geochemically and microbially active sediment surfaces \cite{McClain2003,Lautz2008,Li2017}. Variable or turbulent flow over the non-uniform streambed further drives hyporheic flow and solute transport as high- and low-pressure zones develop around the bed form geometries, often referred to as bed form-induced hyporheic flow \cite{Thibodeaux1987,Cardenas_Wilson2007a,CARDENAS2007301,Marion2002,BOANO2007148}.

Across scales, the three-dimensional (3D) geometric configuration of bed forms (i.e., assemblage of bed forms
of a given type occurring in a specific bed area such as a dune bed configuration) introduces complexities into flow and solute exchange between surface water and groundwater \cite{Sedimentary1982,FLECKENSTEIN20101291}. Often such complexities are analyzed using numerical approaches, which provide a detailed understanding of the mechanics of bed form-induced hyporheic exchange (e.g., \cite{ElliottBrooks1997,Packman2000,Worman2002,KAZEZYILMAZALHAN200626,Trauth2014,Trauth2015,Li2020,Yabusaki2017}). The majority of these have focused on idealized, two-dimensional (2D) bed forms \cite{ElliottBrooks1997, Cardenas_Wilson2007a, Boano2015, Janssen2012, bardini2013small, sawyer2009hyporheic}. A few recent exceptions incorporate the third dimension into their investigations \cite{Chen_Cardenas2018,Marion2002,Tonina_Buffington2007,Chen2015,Zhou2014,Zheng2019}, but were exclusive to one simplified bed form type or shape.

W\"{o}rman et al. \cite{worman2006} proposed a semi-empirical method to study 3D hyporheic exchange by assigning pressure over a flat sediment-water interface following a Fourier-series spectrum translated from topographic data. 
They extended the simple 2D case in \cite{ElliottBrooks1997} using a Fourier-series to represent pressure variations along the sediment-water interface. 
For dynamic pressure head, they assigned a spatially periodic (e.g., sine or cosine) function derived from 2D pressure measurements over a triangular dune.
Recent studies \cite{Stonedahl2010,Stonedahl2012,Stonedahl2013} have applied the same approach based on 2D measurements presented in \cite{ElliottBrooks1997}.
Although their approach has not been tested for 3D cases, its application would neglect the complexity of 3D turbulent flow and its influence on hyporheic flow.

To provide a systematic analysis of 3D bed form induced hyporheic flow over ripples and dunes, Chen et al. \cite{Chen2015} simulated flow through a single 3D barchanoid dune with a lateral cross-section following a full cosine wave and a planform-straight crest (i.e., orthogonal to the mean flow at all positions).
They showed that despite the relatively simple bed form geometry, 3D bed forms generated more complex patterns of hyporheic flow and significantly greater hyporheic fluxes than the equivalent 2D setting in \cite{McLean1994}.

Other studies have expanded the analysis to include solute exchange through 3D domains. For example, Trauth et al.\cite{Trauth2014} simulated reactive solute transport through a 3D pool-riffle to determine that the flux and residence times of solutes, and the associated size of the reaction zones, are influenced by both the magnitude and direction of ambient groundwater flow.
They also showed that the complex 3D pressure distribution over the streambed can induce significant lateral hyporheic flow which results in a hyporheic exchange flow independent of riffle height, when higher than a specific threshold height.
In a similar study, Li et al. \cite{Li2020} simulated flow and reactive transport through a dune bed form to demonstrate the capability of a coupled surface water and groundwater model.
They compared the fully coupled simulation of surface-subsurface exchanges with the sequential approach results, and found out that the difference between sequential and coupled model results primarily depends on permeability, a key parameter in the magnitude of hyporheic exchange. As permeability increases and drives hyporheic exchange, coupled modeling becomes more imperative. 
By incorporating the migration of ripples into their numerical framework, Zheng et al. \cite{Zheng2019} found that the turnover exchange due to ripple migration has a large impact on reactant supply and reaction rates. 

The most relevant study to this work is by Chen et al. \cite{Chen_Cardenas2018}, which investigated the impacts of bed form three-dimensionality on hyporheic
flow. 
They calculated both the surface water flow over and hyporheic
flux through various 3D bed forms using a computational fluid dynamics (CFD) model implemented across a range of Reynolds Numbers. 
They found that  hyporheic exchange is sensitive to the geometric parameters which define the bed form three-dimensionality, particularly the bed form steepness and lateral elongation. However, the focus of their work was on hyporheic flow, and so did not consider the effects of bed forms on biogeochemical reaction processes. 

Given the importance of bed form-induced hyporheic exchange on reactive solute transport, a fundamental understanding of how variable bed form geometry influences the distribution of flow and biogeochemical reaction rates is needed, especially given its implications for remediation of both traditional (e.g., nitrate, phosphate) and emerging contaminants (e.g., Per- and polyfluoroalkyl substances (PFAS)).
However, none of the previous studies considered the effects of variable bed form geometry on the magnitude of exchange nor the rate of biogeochemical reaction processes in the hyporheic zone. 
Here a computational surface water-groundwater modeling framework is developed and used to investigate the effects of complex bed form geometries on hyporheic exchange flux, solute transport, and biogeochemical reaction processes. The details of the geometric models for generating bed form topographies and of the modeling framework are presented in Section \ref{sec:methodology}. The simulation results and implications are discussed in Section \ref{sec:results}.

\section{Methodology}
\label{sec:methodology}
\subsection{Streambed morphology}

\begin{figure}[!hb]
  \centering
  \begin{subfigure}{0.3\linewidth}
    \includegraphics[trim=1cm 0 1cm 0, clip, width=\linewidth]{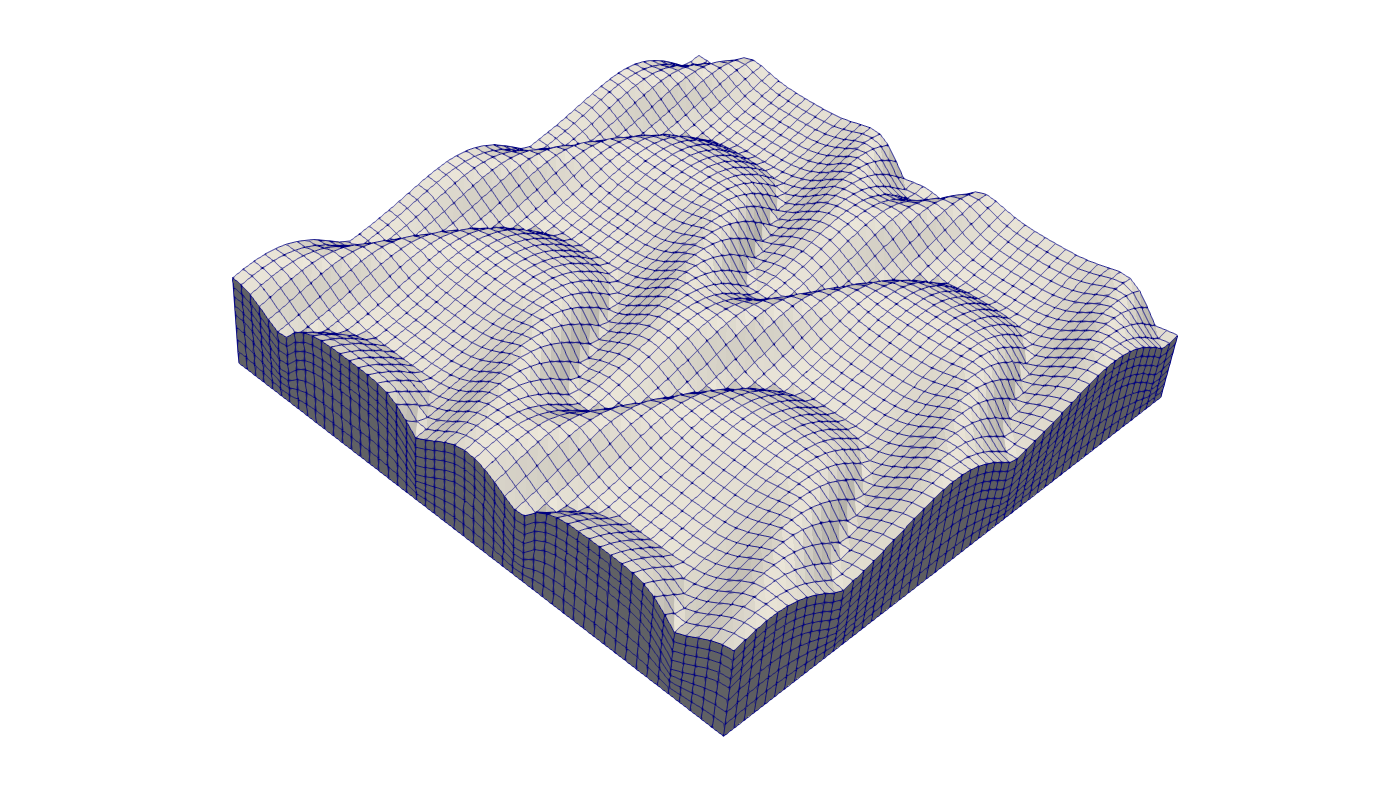}
    \caption{\(\Delta \varphi_1 = 90 ^{\circ}, \Delta \varphi_2 = 0 ^{\circ}\)}
  \end{subfigure}
  \begin{subfigure}{0.3\linewidth}
    \includegraphics[trim=1cm 0 1cm 0, clip, width=\linewidth]{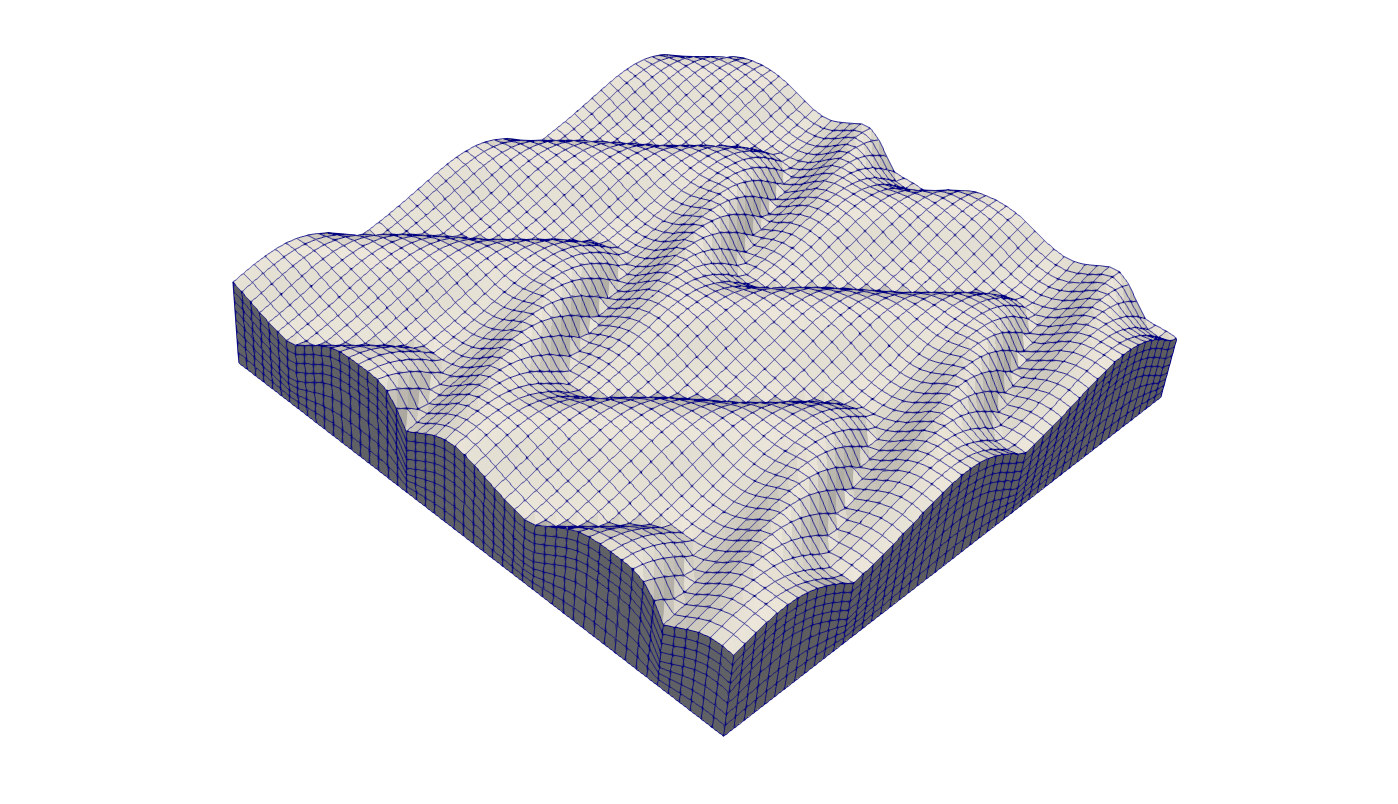}
    \caption{\(\Delta \varphi_1 = 180 ^{\circ}, \Delta \varphi_2 = 0 ^{\circ}\)}
  \end{subfigure}
  \begin{subfigure}{0.3\linewidth}
    \includegraphics[trim=1cm 0 1cm 0, clip, width=\linewidth]{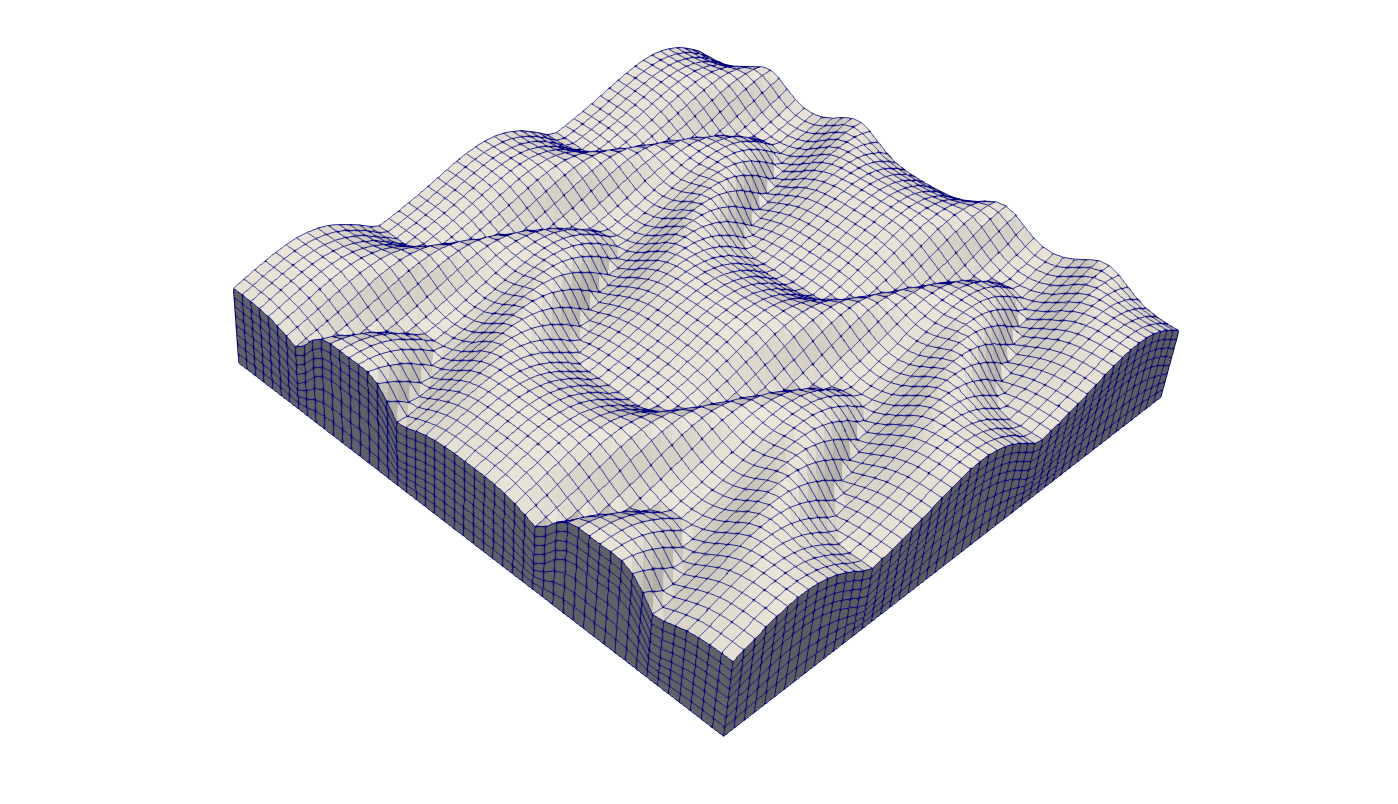}
    \caption{\(\Delta \varphi_1 = 270 ^{\circ}, \Delta \varphi_2 = 0 ^{\circ}\)}
  \end{subfigure}
  \begin{subfigure}{0.3\linewidth}
    \includegraphics[trim=1cm 0 1cm 0, clip, width=\linewidth]{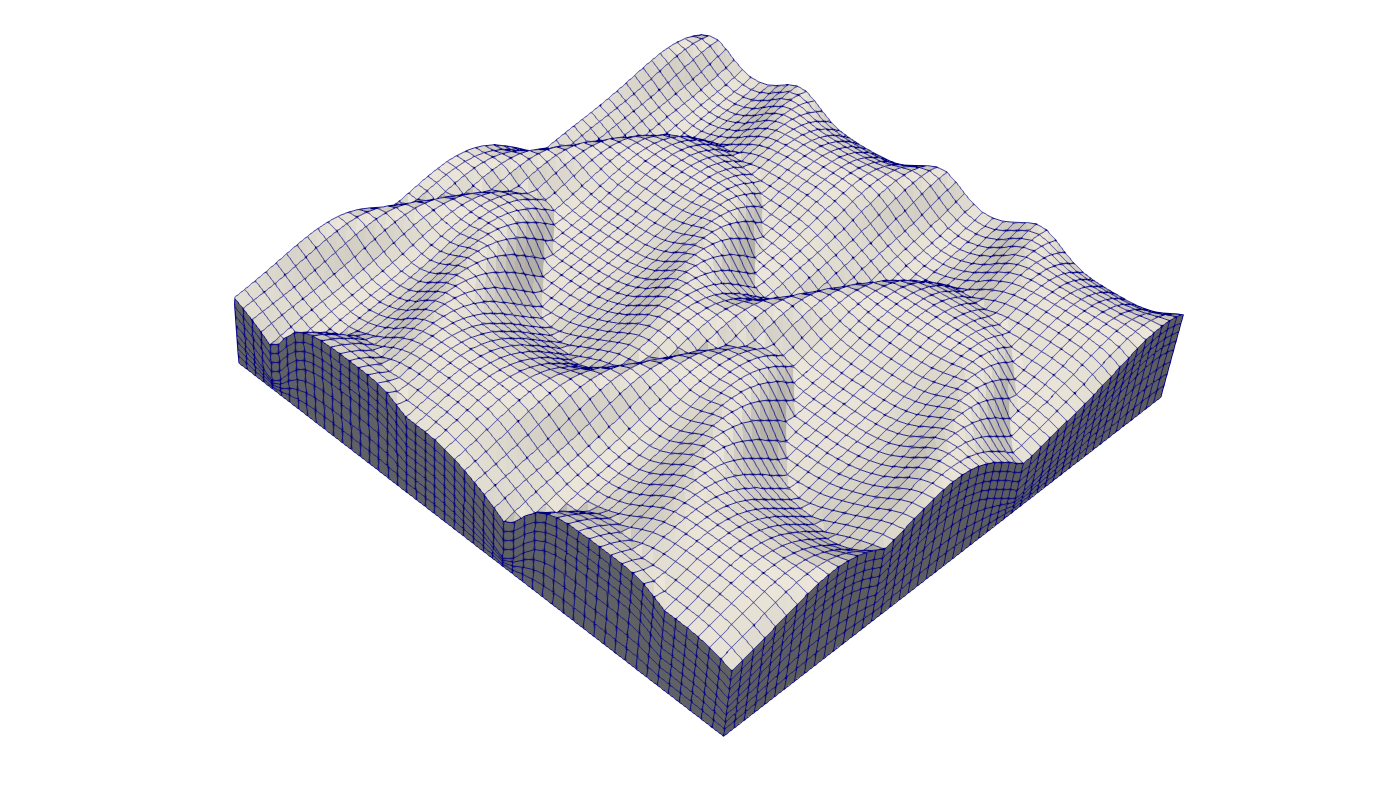}
    \caption{\(\Delta \varphi_1 = 90 ^{\circ}, \Delta \varphi_2 = 90 ^{\circ}\)}
  \end{subfigure}
  \begin{subfigure}{0.3\linewidth}
    \includegraphics[trim=1cm 0 1cm 0, clip, width=\linewidth]{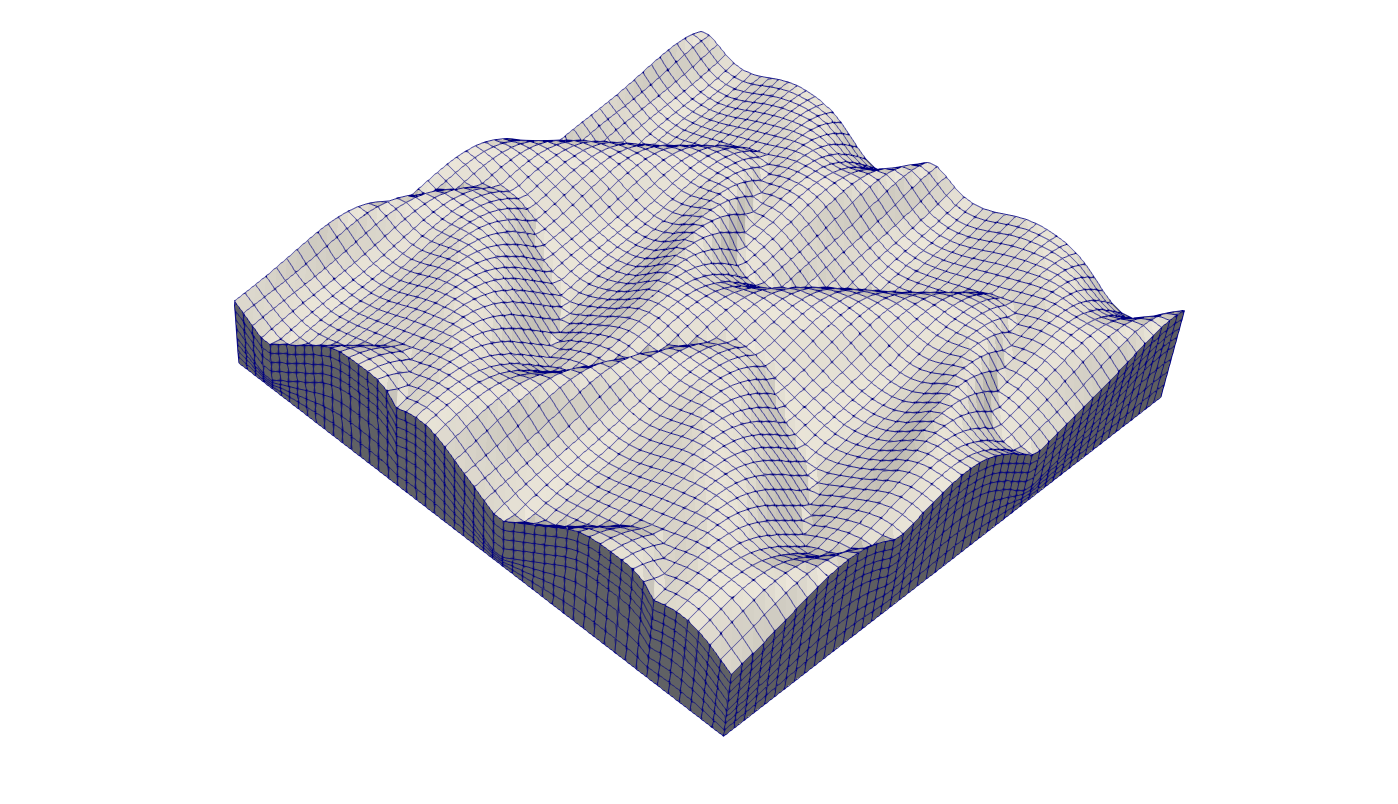}
    \caption{\(\Delta \varphi_1 = 180 ^{\circ}, \Delta \varphi_2 = 90 ^{\circ}\)}
  \end{subfigure}
  \begin{subfigure}{0.3\linewidth}
    \includegraphics[trim=1cm 0 1cm 0, clip, width=\linewidth]{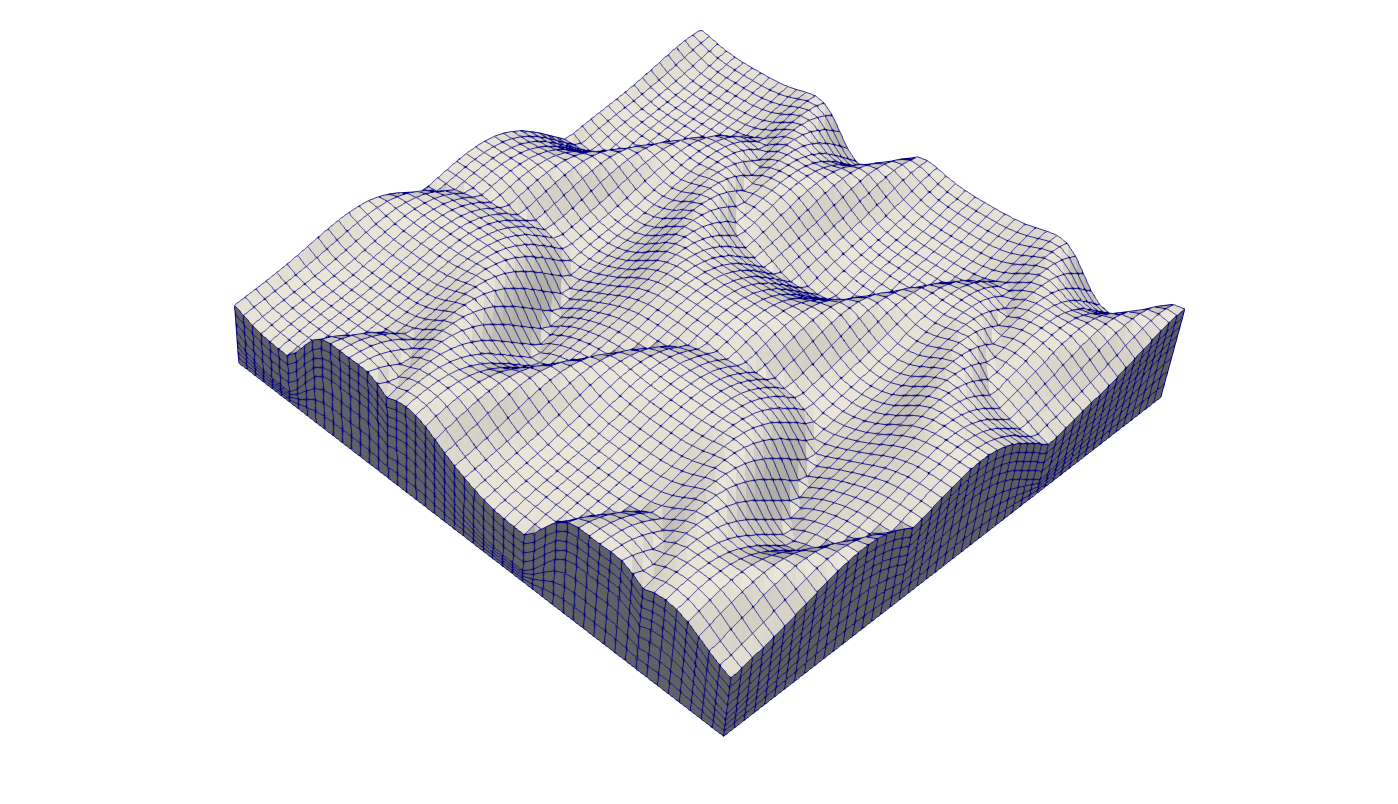}
    \caption{\(\Delta \varphi_1 = 270 ^{\circ}, \Delta \varphi_2 = 90 ^{\circ}\)}
  \end{subfigure}
  \begin{subfigure}{0.3\linewidth}
    \includegraphics[trim=1cm 0 1cm 0, clip, width=\linewidth]{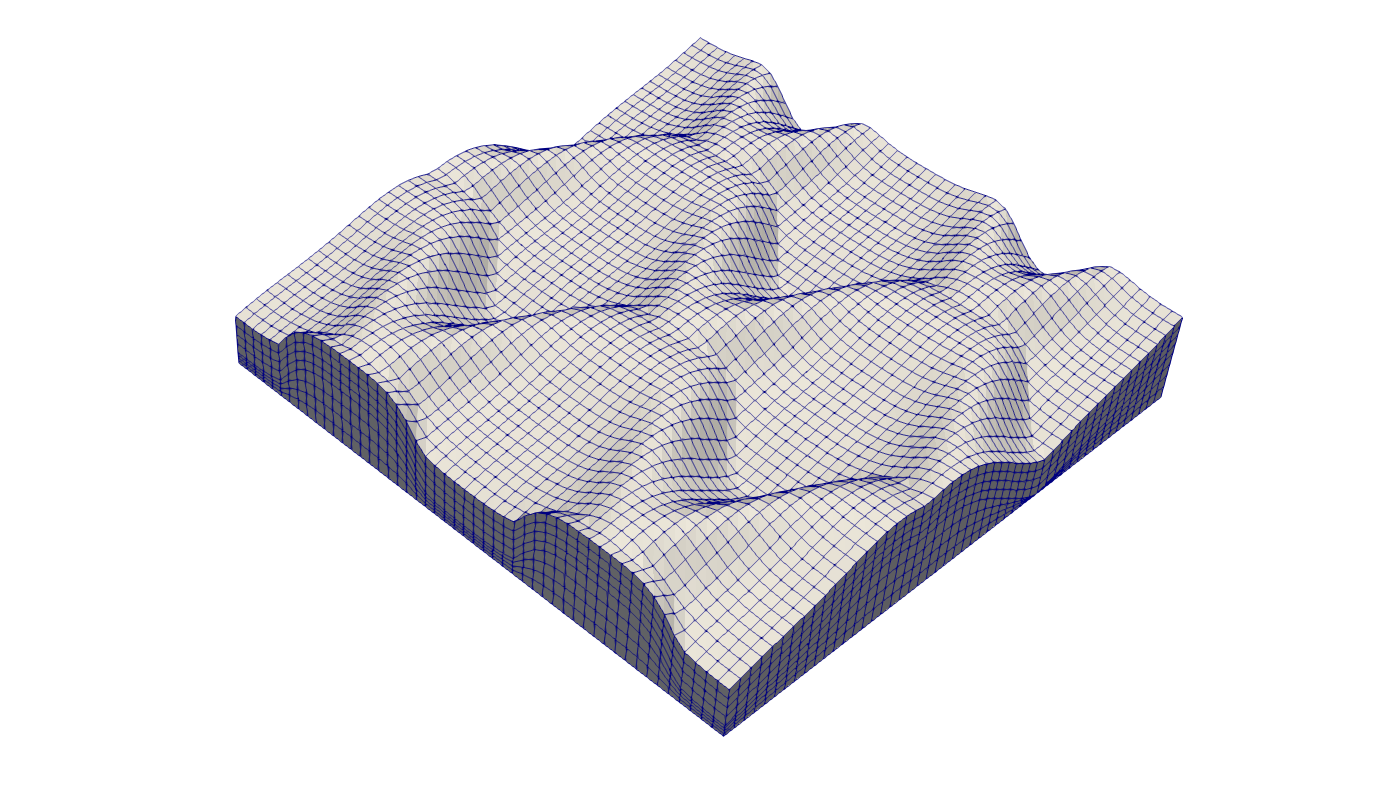}
    \caption{\(\Delta \varphi_1 = 90 ^{\circ}, \Delta \varphi_2 = 180 ^{\circ}\)}
  \end{subfigure}
  \begin{subfigure}{0.3\linewidth}
    \includegraphics[trim=1cm 0 1cm 0, clip, width=\linewidth]{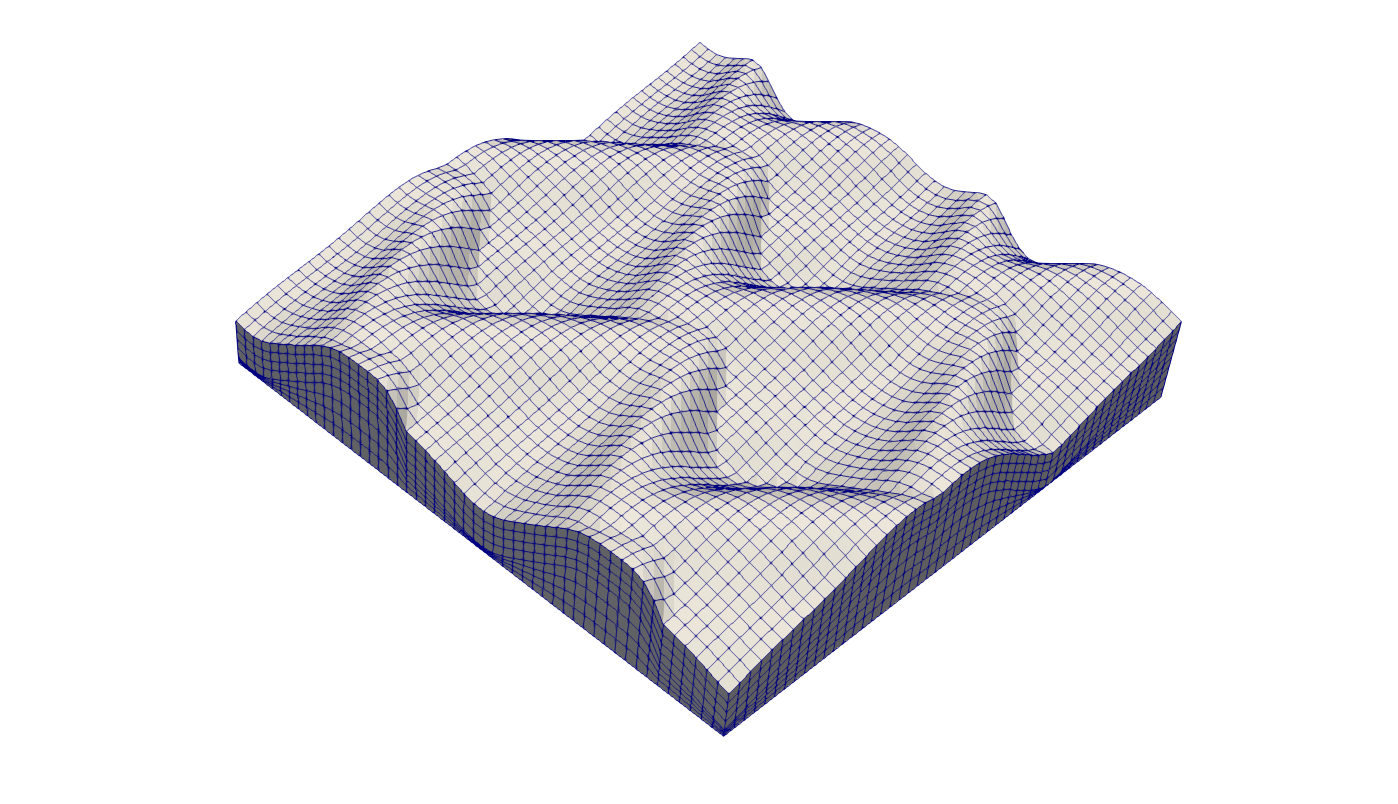}
    \caption{\(\Delta \varphi_1 = 180 ^{\circ}, \Delta \varphi_2 = 180 ^{\circ}\)}
  \end{subfigure}
  \begin{subfigure}{0.3\linewidth}
    \includegraphics[trim=1cm 0 1cm 0, clip, width=\linewidth]{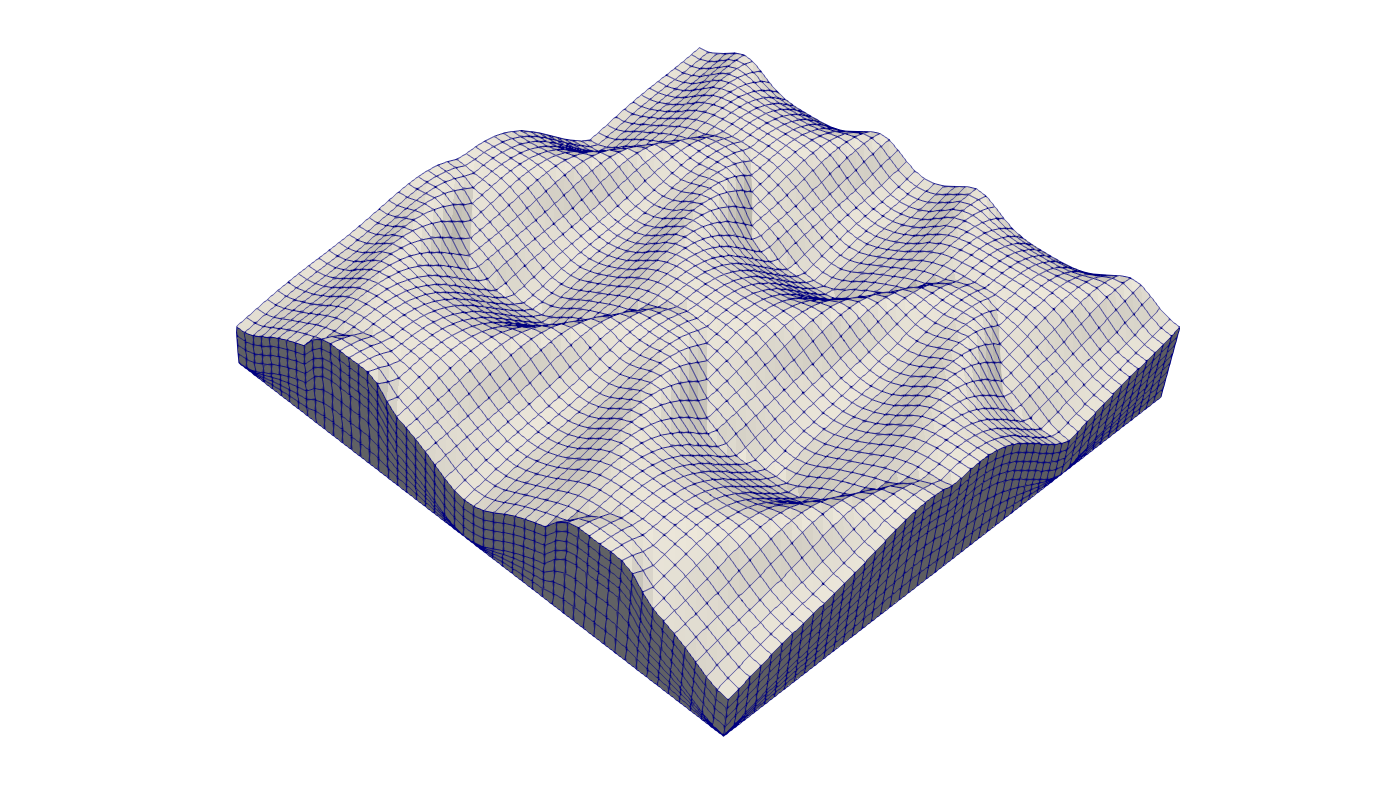}
    \caption{\(\Delta \varphi_1 = 270 ^{\circ}, \Delta \varphi_2 = 180 ^{\circ}\)}
  \end{subfigure}
    \caption{Three-dimensional bed form topographies generated for this study using the geometrical approach of Rubin and Carter \cite{Rubin_Carter} (details in Table \ref{table:alpha_beta}). The mean steepness (\(\eta\)) is 0.6 and the longitudinal (\(\lambda_L\)) and transverse wavelengths (\(\lambda_T\)) of the bed form are 0.75 (m) and 0.5 (m), respectively.}
  \label{fig:bedforms}
\end{figure}

This study follows the methodology of \cite{Chen_Cardenas2018} to create synthetic 3D bed forms for use in coupled numerical surface water-groundwater flow and reactive transport models.
Bed form morphology is generated using the geometric model from a widely used set of codes originally developed by Rubin \cite{Rubin1987} and later published by Rubin and Carter \cite{Rubin_Carter} as a set of MATLAB routines.  
More than eighty parameters are used in their model, which generates geomorphologically plausible, realistic, and  visually satisfying bed form morphologies that represent a wide range of dynamic environmental conditions (see examples in Figure \ref{fig:bedforms}). The focus is on relating the depositional and geomorphological processes and resulting characteristics to different scales of bed forms. Specifically, the models synthesize the bed form geometry by merging two sets of surfaces generated using sine functions. Here, nine representative bed forms are used to represent typical types of sand dunes under unidirectional river flow, such as barchanoid or linguoid forms with dune crest lines that are sinusoidal and with pronounced lobes and saddles.

The mathematical formulation for reproducing the bed form geometry may be written as
\begin{equation}
  z_i = \eta \frac{\lambda _L}{L} \left( 7.5 - 6 sin \left(\frac{\pi  L}{50 \lambda_L} m_i\right) - 1.5 sin \left(\frac{\pi  L}{25 \lambda_L} m_i\right) \right)
\end{equation}
where \(m_i\) is defined as:
\begin{equation}
  m_i = x - \lambda_L \frac{\varphi_i^{bf}}{360} - A_i^f sin \left( \frac{2 \pi}{\lambda_T} y +\frac{2 \pi}{360} \alpha_i  \right) - A_i^s sin \left( \frac{4 \pi}{\lambda_T} y +\frac{2 \pi}{360} \beta_i  \right)
\end{equation}
The bed form topographies (i.e., elevations) in the x-y plane are calculated as the maximum value of each set as:
\begin{equation}
z_{bed}(x,y) = max(z_i)
\end{equation}
where \(i = 1,2\) represent the first and second set of bed forms, respectively. The x and y are the Cartesian coordinates, 
and \(L\) is the length of the repeating block.
\(A_i^f\) and \(A_i^s\) are the planform amplitudes of the first and second sine curves of each set of bed forms, respectively. The \(\varphi_i^{bf}\) is the bed form phase (in degrees), \(\eta\) is the mean steepness (height-to-wavelength ratio), \(\lambda_L\) and \(\lambda_T\) are the longitudinal and transverse wavelengths of the bed form, respectively, and \(\alpha_i\) and \(\beta_i\) are the phases of the first and second planform sinuosity of each set of bed forms, respectively.

\begin{figure}[!hbt]
  \centering
  \begin{subfigure}{0.3\linewidth}
    \includegraphics[trim=1cm 0cm 3cm 7cm, clip, width=\linewidth]{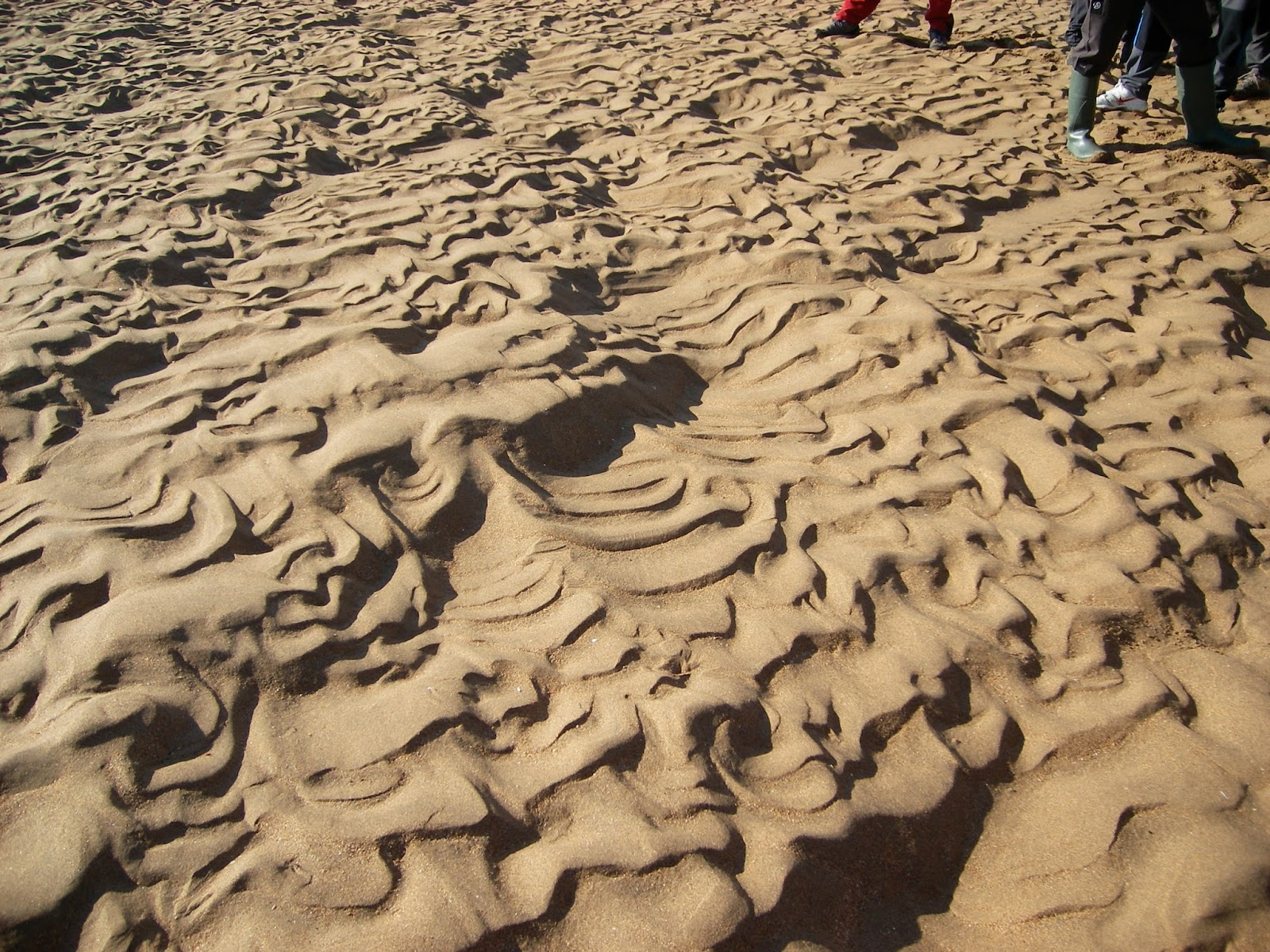}
    \caption{Linguoid ripples}
  \end{subfigure}
  \begin{subfigure}{0.19\linewidth}
    \includegraphics[trim=0cm 3cm 0cm 0cm, clip, width=\linewidth]{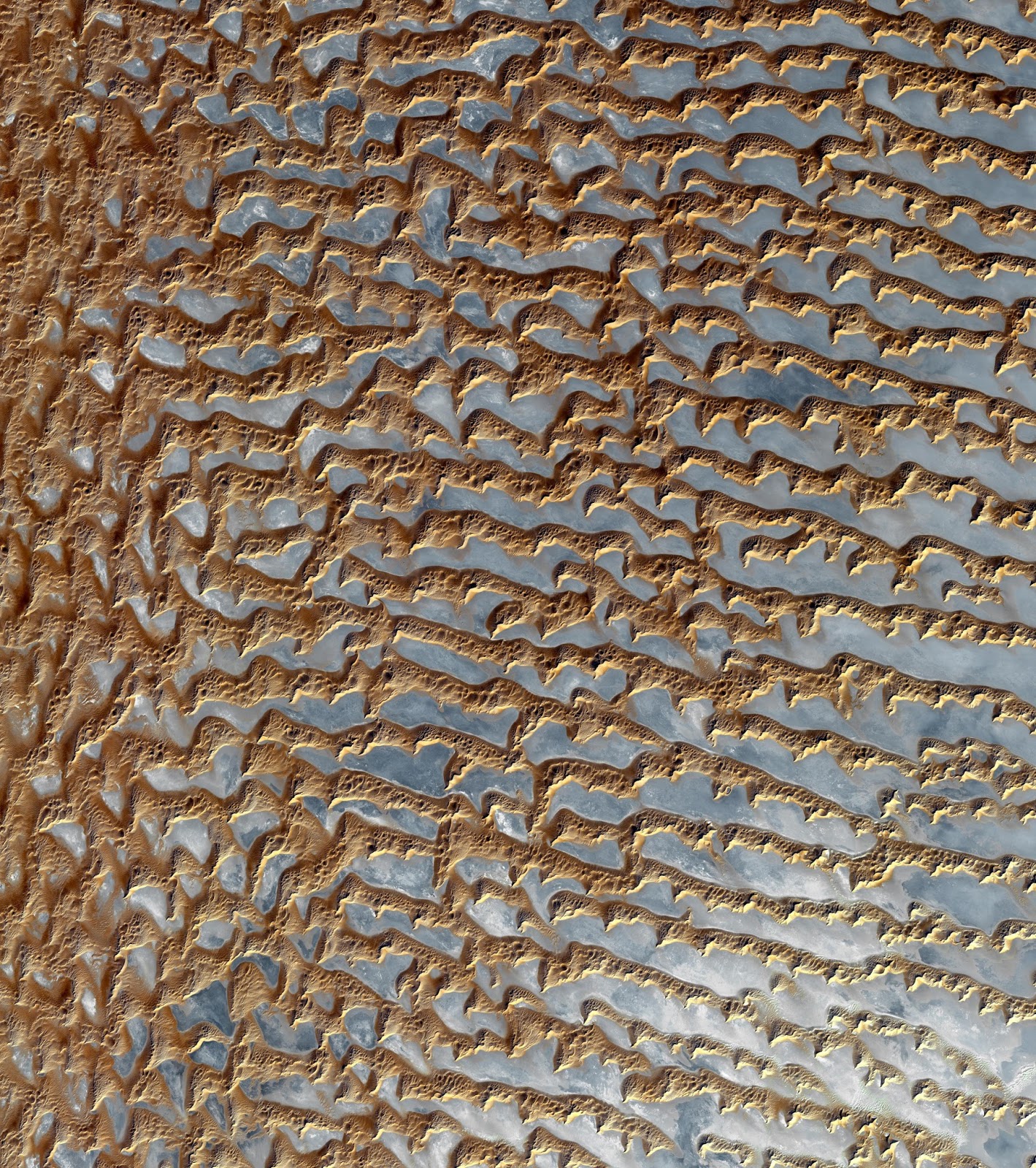}
    \caption{Sand dunes}
  \end{subfigure}
  \begin{subfigure}{0.41\linewidth}
    \includegraphics[trim=7cm 1cm 1cm 9cm, clip, width=\linewidth]{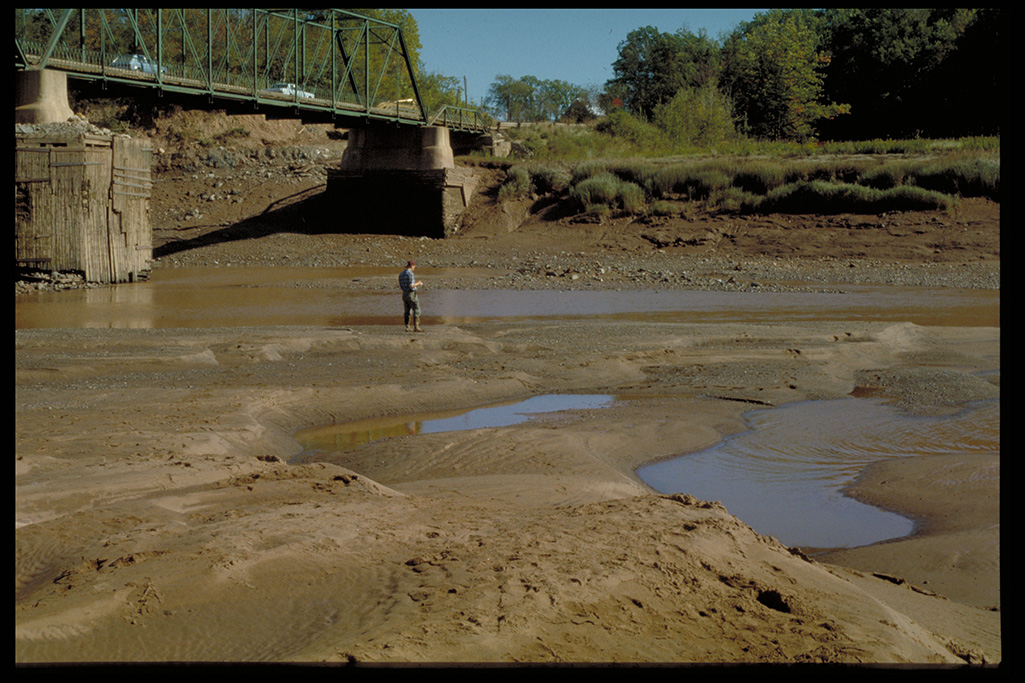}
    \caption{Lunate dunes}
  \end{subfigure}
    \caption{Photos of bed forms produced by currents; a) Linguoid ripples \cite{ripple_dune}, b) Sand dunes \cite{ripple_dune}, c) Lunate (crescent-shaped) dunes in sand and gravel, Kennetcook River, Nova Scotia, Canada. Modified from \cite{lunate}.}
  \label{fig:bedforms_in_nature}
\end{figure}

In this study, bed form shapes are specified by two phase shift parameters:
\begin{gather}
    \Delta \varphi_1 = \beta_1 - \alpha_1  \\
    \Delta \varphi_2 = \alpha_2 - \alpha_1 
\end{gather}  
with \(\beta_2 = \beta_1\).
Figure \ref{fig:bedforms} and Table \ref{table:alpha_beta} present details of the nine different bed forms.
Other parameters are held constant throughout the computations (Table \ref{table:bedform_constants}). The overall bed form geometry changes based on variations in the phase shift parameters (Figure \ref{fig:bedforms}).For instance, a $\Delta \varphi_1$ value of $90 ^{\circ}$ represents linguoid bed forms while a $\Delta \varphi_1$ value of $270 ^{\circ}$ represents lunate (crescent-shaped) bed forms. There are also more subtle differences between bed forms that are better observed in 2D cross-sections. For example, while linguoid ripples produce an angle to the flow and have a random shape rather than a W-type shape, the stoss sides of lunate ripples  are curved rather than having a lee slope. Photos of such bed forms exposed in aquatic environments are shown in Figure \ref{fig:bedforms_in_nature}.

\begin{table}[!htb]
\caption{Phase shifts (in degrees) used to change the bed form patterns in Figure \ref{fig:bedforms} ($\alpha_1 = 0 ^{\circ}$).}
\footnotesize
\centering
\begin{tabular}{|c|c c c c|} 
 \hline
 Bed form & $\alpha_2$ & $\beta_1 = \beta_2$ & $\Delta \varphi_1$ & $\Delta \varphi_2$ \\ 
 \hline\hline
 a &  0 & 90 & 90  & 0   \\
 \hline
 b &  0 & 180 & 180 & 0   \\
 \hline
 c &  0 & 270 & 270 & 0   \\
 \hline
 d &  90 & 90 & 90  & 90  \\
 \hline
 e &  90 & 180 & 180 & 90  \\
 \hline
 f &  90 & 270 & 270 & 90  \\
 \hline
 g &  180 & 90 & 90  & 180 \\
 \hline
 h &  180 & 180 & 180 & 180 \\
 \hline
 i &  180 & 270 & 270 & 180 \\ 
 \hline
\end{tabular}
\label{table:alpha_beta}
\end{table}

\begin{table}[!hbtp]
\caption{Constants of bed form geometry model \cite{Chen_Cardenas2018}.}
\footnotesize
\centering
\begin{tabular}{|c c c c c c c|} 
 \hline
  \(A_i^f\) & \(A_i^s\) & \(\lambda_L\) &
 \(\lambda_T\) & \(\eta\) &  \(\varphi_1^{bf}\)
 & \(\varphi_2^{bf}\) \\
 \hline \hline
  13 cm & 3 cm  & 75 cm & 50 cm & 0.6 & $0.0^{\circ}$ & $180.0^{\circ}$  \\
 \hline
\end{tabular}
\label{table:bedform_constants}
\end{table}

\subsection{Modeling framework for hyporheic exchange processes}

\begin{figure}[!hb]
  \centering  
  \includegraphics[trim=0cm 1cm 0cm 2cm, clip, width=\linewidth]{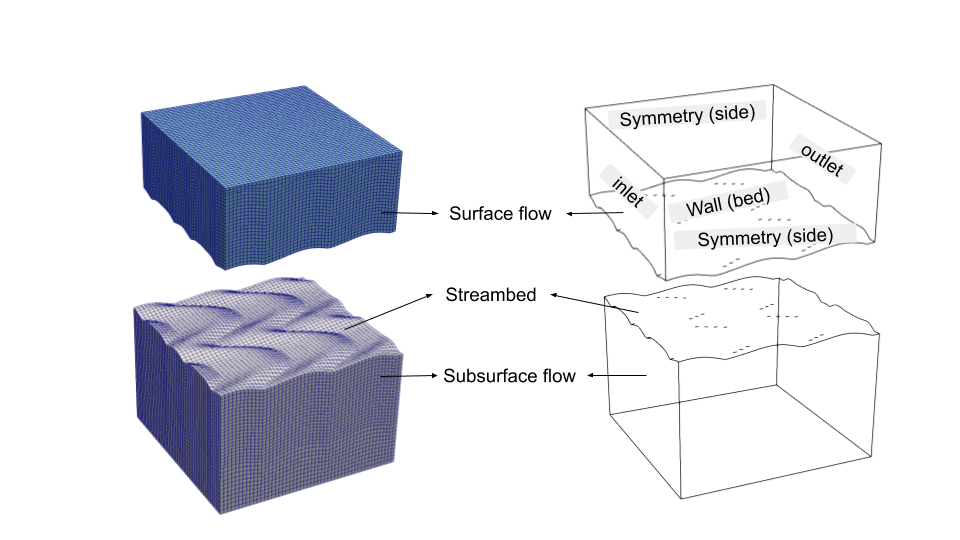}
  \caption{Illustration of the computational domain for surface and subsurface simulation. The bed form shown here coresponds to bed form (a) in Figure \ref{fig:bedforms}a. Flow direction is from left to right. Boundary conditions for surface water are presented here, while a zero flux condition is applied to the sides and bottom of the groundwater domain to characterize and quantify hyporheic exchange between the surface water and groundwater.}
  \label{fig:domain}
\end{figure}

A modeling framework named SimSGI (simulation of surface water and groundwater interactions) is developed to more realistically simulate bed form-driven hyporheic exchange processes. SimSGI integrates two advanced models of flow and mass transport: a surface water solver (OpenFOAM) \cite{OpenFoam} and a subsurface simulator (PFLOTRAN) \cite{pflotran-paper,pflotran-web-page,pflotran-user-ref}. SimSGI also uses LaGriT \cite{LaGriT}, an advanced and open-source mesh generation tool, to create numerical meshes for the different bed form geometries discussed above. 
After the streambed topographies are geometrically simulated, LaGrit generates computational grids based on the resulting surfaces and the SimSGI framework provides interface scripts for using them in both the surface water (OpenFOAM) and groundwater (PFLOTRAN) domains (Figure \ref{fig:domain}). 
After solving the turbulent surface flow in OpenFOAM, the resulting pressure distribution over the bed surface is assigned as the top boundary condition for the groundwater domain, which is then used to solve the groundwater flow and reactive solute transport in PFLOTRAN.

\subsubsection{Simulation of turbulent surface flow over 3D bed forms}

The SimSGI framework simulates surface water flow over 3D bed forms using the open-source CFD toolbox OpenFOAM \cite{OpenFoam}, wherein the three-dimensional Navier-Stokes equations are solved to fully capture the effects of turbulent surface water flow on hyporheic exchange.
The finite volume approach for the numerical discretization, the PISO algorithm for pressure-velocity coupling, and the shear-stress transport (SST) \(k-\omega\) model for the turbulence closure are applied.

\paragraph{Turbulent model equations} 
The Reynolds-averaged Navier-Stokes (RANS) equations for an incompressible, homogeneous fluid is used to simulate surface water flow in this study \cite{LAUNDER1974269,Janssen2012}:
\begin{gather}
    \frac{\partial U_i}{\partial x_i}=0 \\
    \rho U_j \frac{\partial U_i}{\partial x_j} = -\frac{\partial P}{\partial x_i} + \frac{\partial}{\partial x_j}\left ( 2 \mu S_{ij} - \rho  \overline{{u}'_j {u}'_i}\right )
\end{gather}
where \(U_i\) and \(u'_i\) are the time-averaged velocity and the fluctuations in instantaneous velocity components in \(x_i\) directions, respectively, with \(i\) = 1,2, 
\(\rho\) and \(\mu\) are fluid density and dynamic viscosity, \(P\) is the time-averaged pressure, and \(S_{ij}\) is the strain rate tensor specified as:
\begin{equation}
    S_{ij}=\frac{1}{2}\left ( \frac{\partial U_i}{\partial x_j} + \frac{\partial U_j}{\partial x_i} \right )
\end{equation}
The Reynolds stress is defined as
\begin{equation}
    \tau_{ij}= - \overline{{u}'_j {u}'_i} = \nu_t (2 S_{ij}) - \frac{2}{3} \delta_{ij} k
\end{equation}
where \(\nu_t\) is the kinematic eddy viscosity, \(\delta_{ij}\) represents Kronecher delta, and \(k\) is the turbulent kinetic energy.
The equations governing the turbulence kinetic energy and specific dissipation rate for the SST model are \cite{Menter2003}:

\begin{equation}
    \frac{\partial (\rho k)}{\partial t} + \frac{\partial (\rho U_i k)}{\partial x_i} = \tilde{P}_k - \beta^* \rho k \omega + \frac{\partial}{\partial x_i} \left [ (\mu + \sigma_k \mu_t) \frac{\partial k}{\partial x_i} \right ]
\end{equation}
\begin{equation}
    \begin{split}
    \frac{\partial (\rho \omega)}{\partial t} + \frac{\partial (\rho U_i \omega)}{\partial x_i} = & \alpha \rho S^2 - \beta \rho \omega^2 + \frac{\partial}{\partial x_i} \left [ (\mu + \sigma_{\omega} \mu_t) \frac{\partial \omega}{\partial x_i} \right ] + \\
    & 2 (1-F_1) \rho \sigma_{\omega2} \frac{1}{\omega} \frac{\partial k}{\partial x_i} \frac{\partial \omega}{\partial x_i}
    \end{split}
\end{equation}
where the first blending function \(F_1\) is defined by
\begin{equation}
    F_1 = \tanh \left \{ \left \{ min\left [ max(\frac{\sqrt{k}}{\beta^* \omega y},\frac{500 \nu}{y^2 \omega}), \frac{4 \rho \sigma_{\omega2} k}{CD_{k \omega} y^2} \right ] \right \}^4 \right \}
\end{equation}
with \( CD_{k\omega} = max\left ( 2 \rho \sigma_{\omega2} \frac{1}{\omega} \frac{\partial k}{\partial x_i} \frac{\partial \omega}{\partial x_i} \right ) \) and \(y\) is the distance from the wall. 
The turbulent eddy viscosity is defined as
\begin{equation}
   \nu_t = \frac{a_1 k}{max(a_1 \omega, S F_2)}
\end{equation}
where \(S\) is the invariant measure of the strain rate and \(F_2\) is a second blending function defined by
\begin{equation}
    F_2 = \tanh \left [\left [ max\frac{2 \sqrt{k}}{\beta^* \omega y}, \frac{500 \nu}{y^2 \omega}) \right ]^2 \right ]
\end{equation}
In the SST model, a production limiter \(\tilde{P}_k\) is used to prevent the build-up of turbulence stagnation regions
\begin{equation}
    \tilde{P}_k = min(P_k, 10 \beta^* \rho k \omega)
\end{equation}
where \( P_k = \mu_t \frac{\partial U_i}{\partial x_j} \left ( \frac{\partial U_i}{\partial x_j} + \frac{\partial U_j}{\partial x_i} \right )\).
Each constant in the above formulations is calculated by  \(\phi = \tilde{\phi}_1 F_1 + \tilde{\phi}_2 (1-F_1) \). Required constants and coefficients for turbulent closure model are presented in Table \ref{table:turbulence_constants}.

\begin{table}[!hb]
\caption{Constants for the shear-stress transport (SST) \(k-\omega\) model \cite{Menter2003}}
\footnotesize
\centering
\begin{tabular}{|c c c c c c c c c|} 
 \hline
  \(\beta^*\) & \(\tilde{\alpha}_1\) & \(\tilde{\alpha}_2\) &
 \(\tilde{\beta}_1\) & \(\tilde{\beta}_2\) &  \(\tilde{\sigma}_{k1}\)
 & \(\tilde{\sigma}_{k2}\) & \(\tilde{\sigma}_{\omega 1}\) &  \(\tilde{\sigma}_{\omega 2}\) \\
 \hline
 0.09  & $\frac{5}{9}$  & 0.44 & $\frac{3}{40}$ & 0.0828 & 0.85 &  1.0 & 0.5 & 0.856 \\
 \hline
\end{tabular}
\label{table:turbulence_constants}
\end{table}
\paragraph{Initialization}
Following the approach of Chen et al. \cite{Chen_Cardenas2018}, the bed form height (H) is considered here as the characteristic length, and thus the Reynolds Number for  turbulent flow is defined as:
\begin{equation}
Re = U_{ave} H / \nu
\end{equation}
where \(U_{ave}\) is the average horizontal velocity of the surface flow (m/s) 
and \(\nu\) is the kinematic viscosity of the fluid (m\(^2\)/s). 
Other required parameters are $L = 1$ (m) and the width of the domain which is from $y = 0.37$ (m) to $y = 1.37$ (m).
A fixed velocity inlet, fixed pressure outlet, no-slip bed, and symmetrical sides are implemented as boundary conditions for the surface flow (Figure \ref{fig:domain}). The grid spacing of the surface water computational domain is 0.02 (m) in all directions, resulting in 87,500 computational grid cells.

\subsubsection{Simulation of subsurface flow and reaction processes within hyporheic zone}

After the surface water dynamics have been simulated using OpenFOAM, SimSGI feeds the output to PFLOTRAN, a massively parallel subsurface flow and multicomponent reactive transport code which has seen broad acceptance within the scientific community \cite{pflotran-paper,pflotran-web-page,pflotran-user-ref, dwivedi2018hot, dwivedi2018geochemical}. 
Groundwater flow and reactive solute transport are simulated over a 3D, representative volume of riverbed (Figure \ref{fig:domain}). The resulting velocity field is used to solve the advection\hyp{}dispersion\hyp{}reaction equation for solute transport:
\begin{equation}
\frac{\partial C}{\partial t} = \nabla \cdot (\textbf{\textit{D}} \nabla C)-\nabla \cdot \textbf{\textit{v}}C + \sum R_i
\end{equation}
where \textit{C} is solute concentration (mol/L), \textbf{\textit{v}} is fluid velocity (m/s), and \textit{R\textsubscript{i}} is the sum of all reactions that influence the solute. The hydrodynamic dispersion coefficient tensor, \textbf{\textit{D}} = {\textit{D\textsubscript{ij}}}, is:
\begin{equation}
    D_{ij} = (\alpha_L-\alpha_T)\frac{v_i v_j}{\textbf{\textit{v}}} + \alpha_T \textbf{\textit{v}} \delta_{ij} + \tau D_m
\end{equation}
where \textit{\ensuremath{\tau}} is tortuosity, \textit{D\textsubscript{m}} is the molecular diffusion coefficient in porous media (m\textsuperscript{2}/s), \textit{\ensuremath{\alpha\textsubscript{L}}} is longitudinal dispersivity (m), \textit{\ensuremath{\alpha\textsubscript{T}}} is transverse dispersivity (m), and \textit{\ensuremath{\delta}\textsubscript{ij}} is the Kronecker delta function \cite{Bear1972}.

 Simulating an elementary, representative reaction addresses the primary goal of understanding the response of reaction hotspots (i.e., zones of enhanced reaction rates) to bed form topography without the complexities associated with real\hyp{}world reaction systems. Therefore, in this work, a simple, surrogate microbial reaction (\textit{A}+\textit{B}$\rightarrow$\textit{C}) is represented using multiple\hyp{}Monod kinetics:
\begin{equation}
    R_i = k_{max}X\frac{C_A}{K_A + C_A}\frac{C_B}{K_B + C_B}
\end{equation}
where \textit{R\textsubscript{i}} is the rate of the \textit{i\textsuperscript{th}} reaction, \textit{C\textsubscript{A}} and \textit{C\textsubscript{B}} are concentrations of solutes A and B, respectively (mg/L), \textit{K\textsubscript{A}} and \textit{K\textsubscript{B}} are the half\hyp{}saturation constants (mg/L), \textit{k\textsubscript{max}} is the maximum specific uptake rate (h\textsuperscript{-1}), and \textit{X} is microbial biomass (mg/L).
\begin{table}[!htb]
\caption{Model parameters for groundwater flow and transport simulation.}
\footnotesize
\centering
\begin{tabular}{|c c c|} 
 \hline
 Parameter & Value & Reference\\ 
 \hline\hline
 \textit{k\textsubscript{x,y}} (horizontal permeability; m\textsuperscript{2}) & 9.1$\times$10\textsuperscript{-12} & \cite{Freeze1979} \\
 \hline
 \textit{k\textsubscript{z}} (vertical permeability; m\textsuperscript{2}) & 9.1$\times$10\textsuperscript{-13} & \cite{Freeze1979} \\
 \hline
 \textit{$\phi$} (porosity) & 0.4 & \cite{Freeze1979} \\ 
 \hline
 \textit{$\alpha$\textsubscript{L}} (longitudinal dispersivity; m) & 0.005 & \cite{Freeze1979} \\
 \hline
 \textit{$\alpha$\textsubscript{T}} (transverse dispersivity; m) & 0.005 & \cite{Freeze1979}  \\
 \hline
 \textit{D\textsubscript{m}} (molecular diffusion coefficient; m\textsuperscript{2}/s) & 5$\times$10\textsuperscript{-11} & \cite{Freeze1979} \\
 \hline
 \textit{X} (microbial biomass; mg/L) & 0.14 & \cite{Gu2007} \\
 \hline
 \textit{K\textsubscript{A}} (half saturation constant for A; mol/kg) & 6$\times$10\textsuperscript{-6} & \\
 \hline
 \textit{K\textsubscript{B}} (half saturation constant for B; mol/kg) & 6$\times$10\textsuperscript{-6} & \\
 \hline
 \textit{k\textsubscript{AB}} (maximum specific reaction rate, hr\textsuperscript{-1}) & 1 & \\
 \hline
 Initial solute concentration  & & \\
 \hline
 A (mol/kg) & 1$\times$10\textsuperscript{-10} &  \\
 \hline
 B (mol/kg) & 1$\times$10\textsuperscript{-10} &  \\
 \hline 
 C (mol/kg) & 1$\times$10\textsuperscript{-10} &  \\
 \hline
 Surface water solute concentration & & \\
 \hline
 A (mol/kg) & 1$\times$10\textsuperscript{-04} &  \\
 \hline
 B (mol/kg) & 1$\times$10\textsuperscript{-04} &  \\
 \hline
 C (mol/kg) & 1$\times$10\textsuperscript{-10} &  \\
 \hline
\end{tabular}
\label{table:subsurface_parameters}
\end{table}

The groundwater flow field and advection-dispersion-reaction equations are solved using the finite-volume approach in PFLOTRAN \cite{pflotran-user-ref}.
The model domain is uniformly discretized with 0.02 (m) horizontal and vertical resolution, resulting in 82,500 computational grid cells for the subsurface domain. 
All groundwater simulations are run for simulation time of 480 hours to allow solute concentrations to reach the steady state. 
The top of the groundwater domain is reciprocal to the bottom of the surface water domain, wherein the top boundary cells are assigned steady pressures based on output from surface water simulations. 
No-flow boundaries with zero solute flux are applied for the bottom and sides of the subsurface domain to characterize exchange solely between the river and groundwater.
The initial hydraulic head is set to 0.5 (m) of standing water above the domain.
Surface water concentrations and initial concentrations in groundwater are specified so that solutes are only supplied by surface water infiltration and in-situ microbial production (Table \ref{table:subsurface_parameters}).

\section{Results and discussions}
\label{sec:results}

Surface water attributes, bed form shape, and sediment characteristics change the pressure patterns on the sediment-water interface and consequently affect the distribution and reaction of solutes moving through the sediment. 
In this section, the effects of the aforementioned factors are numerically evaluated.
Following the work in \cite{Chen_Cardenas2018}, all computations carried out with \(H=0.04\) (m), \(Re=3000\), and a surface water depth of \(0.5\) (m), which results in an average surface flow velocity of \(U_{ave} = 0.075\) (m/s), a turbulent energy \(k = 2.9 \times 10^{-5}\) (m\(^2\)/s\(^2\)), and a specific turbulent dissipation rate of \(\omega = 3.52\) (s\(^{-1}\)). 
To capture the maximum solute penetration depth in the groundwater domain, the sediment thickness is set to \(0.6\) (m).
The average total CPU time for all surface-subsurface simulations is 1385 (sec).

\begin{figure}[!htbp]
  \centering
    \includegraphics[trim=0cm 0cm 0cm 0cm, clip, width=\linewidth]{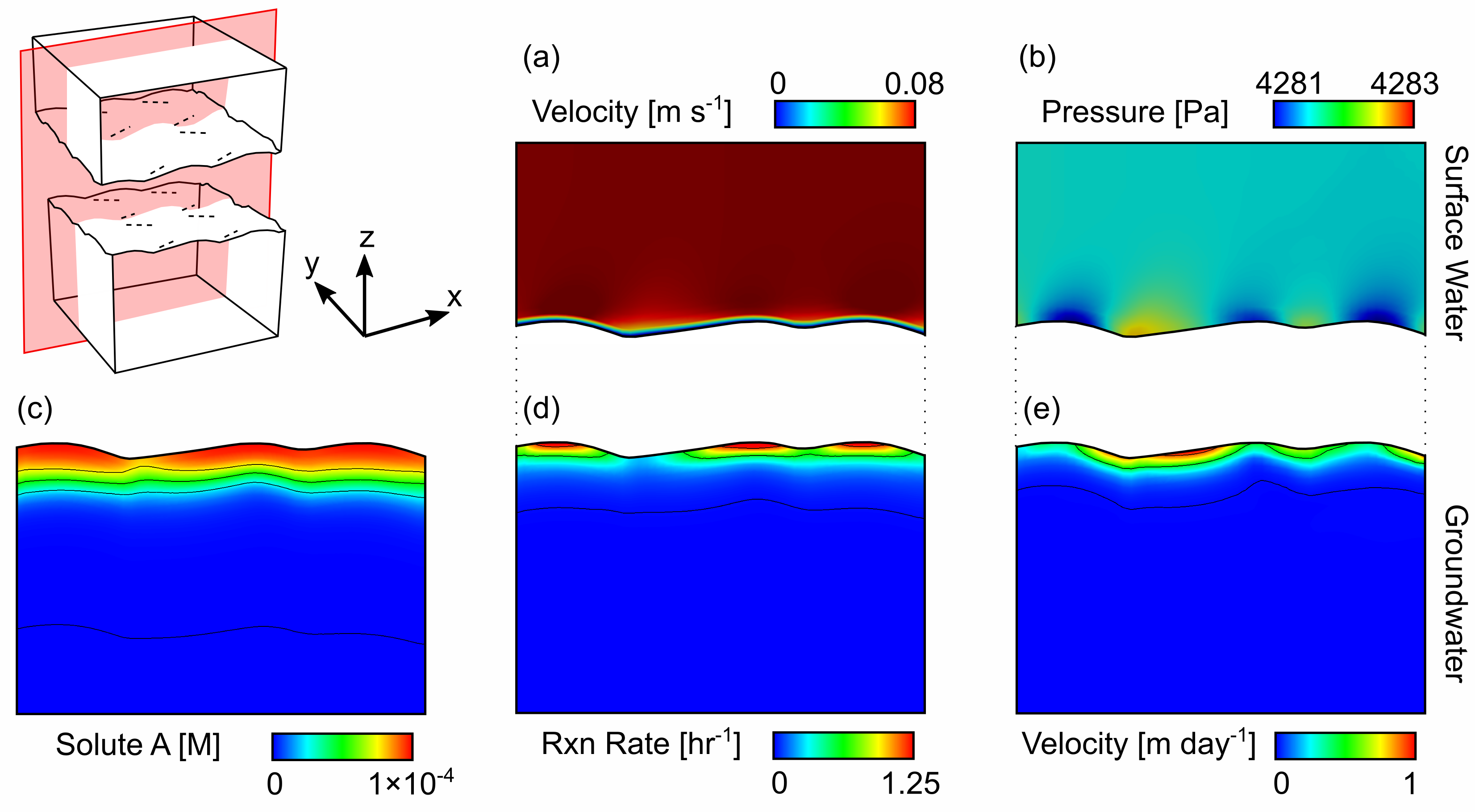}
    \caption{Contour plots for a) surface flow velocity, b) surface flow pressure, c) distribution of solute A in groundwater, d) reaction rate in sediment, and e) subsurface flow velocity.}
  \label{fig:contours}
\end{figure}

For illustration purposes, a cross section of bed form (e) at y = 0.98 (m) with simulated distributions of pressure, velocity, and solute transport characteristics are presented in Figure \ref{fig:contours}. 
Bed form shape and size determine the flow velocity and pressure gradients over the bed form. Higher pressure zones are correlated with higher pore water velocity at the sediment-water interface (Figures \ref{fig:contours}b and \ref{fig:contours}e).
Pressure gradients drive water and solute to penetrate into the sediment (Figure \ref{fig:contours}c), and lower pressure zones are related to higher reaction rates (Figure \ref{fig:contours}d). 

Table \ref{table:max_min} summarizes the simulation results for each bed form, which will be discussed in details in the following sections. 
The geometry of bed form (i) generates the maximum pressure gradient, which not only resulted in the highest average groundwater velocity and greatest solute flux, but also drove the highest reaction rates and largest reaction zone volume.

\begin{table}[!htb]
\caption{Simulated results at the sediment-water interface (SWI) for each bed form after the system reaches the equilibrium state. $p_{max}$ and $p_{min}$ are the maximum and minimum pressure (Pa) applied on the streambed from the surface water. $V_{avg}$ is the average groundwater velocity (m/day) at SWI. $R_{avg}$ is the average reaction rate (mol L$^{-1} s^{-1}$). $q_A$ and $q_C$ are the net exchange flux of solute A and solute C (mol/day) between water and sediment after the system reaches the steady state. $V_{rz}$ is the volume of reaction zone (m$^3$).}
\footnotesize
\centering
\resizebox{\columnwidth}{!}{%
\begin{tabular}{|c c c c c c c c c|} 
\hline
 Bed form & p$_{max}$ & p$_{min}$ & $\frac{\Delta p}{p_{max}} (\%)$ & $V_{avg}$ & $R_{avg}$  & $q_A$ & $q_C$ & $V_{rz}$\\ 
 \hline\hline
a       & 4283.59 & 4280.99 & 0.0606 & 0.6361   & 1.39E-10    & 0.00290 & -0.00282  & 0.137 \\ \hline
b       & 4283.31 & 4281.21 & 0.0491 & 0.6486   & 1.39E-10    & 0.00289 & -0.00282  & 0.136 \\ \hline
c       & 4283.22 & 4281.34 & 0.0439  & 0.6345  & 1.40E-10   & 0.00291 & -0.00283  & 0.137 \\ \hline
d       & 4283.31 & 4280.95 & 0.0550  & 0.7137  & 1.67E-10   & 0.00347 & -0.00336  & 0.170 \\ \hline
e       & 4283.23 & 4281.05 & 0.0509  & 0.7407  & 1.66E-10   & 0.00346 & -0.00336  & 0.168 \\ \hline
f       & 4283.06 & 4281.19 & 0.0438 & 0.7200   & 1.65E-10    & 0.00343 & -0.00333  & 0.167 \\ \hline
g       & 4283.35 & 4280.91 & 0.0572 & 0.7766   & 1.75E-10    & 0.00363 & -0.00355  & 0.170 \\ \hline
h       & 4283.35 & 4280.88 & 0.0578 & 0.8057   & 1.76E-10    & 0.00366 & -0.00357  & 0.172 \\ \hline
i       & 4283.46 & 4280.85 & 0.0610 & 0.8097   & 1.77E-10    & 0.00369 & -0.00360  & 0.173 \\ \hline
\end{tabular}
}
\label{table:max_min}
\end{table}

\subsection{Pressure distribution at the surface-subsurface interface}

Turbulent surface water flow generates non-uniform pressure distributions across the sediment-water interface in response to bed form shape, with more complex patterns associated with increasing \(\Delta \varphi_2 \). For example, as \(\Delta \varphi_2 \) increases for both linguoid (\(\Delta \varphi_1 \) = $90 ^{\circ}$) and lunate (\(\Delta \varphi_1 \) = $270 ^{\circ}$) bed forms, the flow regime shifts as adjacent bed forms become staggered, resulting in larger high-pressure zones on the lee faces of the bed forms. In-phase (\(\Delta \varphi_2 \) = $0 ^{\circ}$) and out-of-phase (\(\Delta \varphi_2 \) = $180 ^{\circ}$) bed forms display symmetrical pressure distributions with aligned low- and high-pressure zones, while the pressure distribution across intermediate bed forms (\(\Delta \varphi_2 \) = $90 ^{\circ}$) is more erratic. In contrast, as \(\Delta \varphi_1 \) increases the high and low pressure zones becomes more uniformly distributed across the bed form geometry, trending away from focused high pressure zones along bed form troughs that drive downwelling conditions. Spatial changes in the pressure distribution alter the pressure gradient, and thus control hyporheic exchange and the flux of solutes between the river and groundwater. Further, larger pressure zones ultimately affect hotspot formation by increasing both the volume of the reaction zone and the overall reaction rates (Table \ref{table:max_min}).

\begin{figure}[!ht]
  \centering
  \begin{subfigure}{0.3\linewidth}
    \includegraphics[trim=7cm 3cm 6cm 0cm, clip, width=\linewidth]{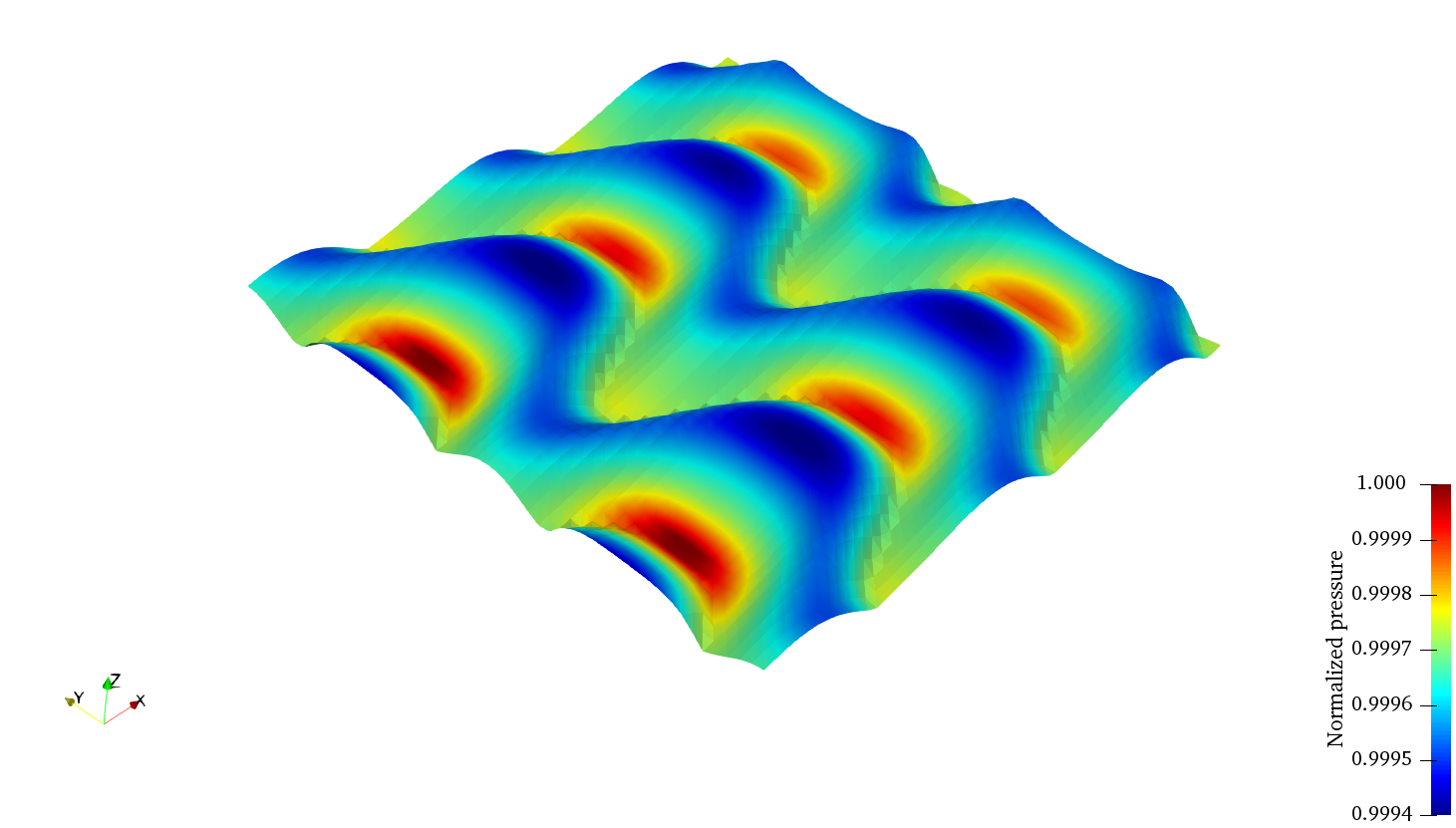}
    \caption{\(\Delta \varphi_1 = 90 ^{\circ}, \Delta \varphi_2 = 0 ^{\circ}\)}
  \end{subfigure}
  \begin{subfigure}{0.3\linewidth}
    \includegraphics[trim=7cm 3cm 6cm 0cm, clip, width=\linewidth]{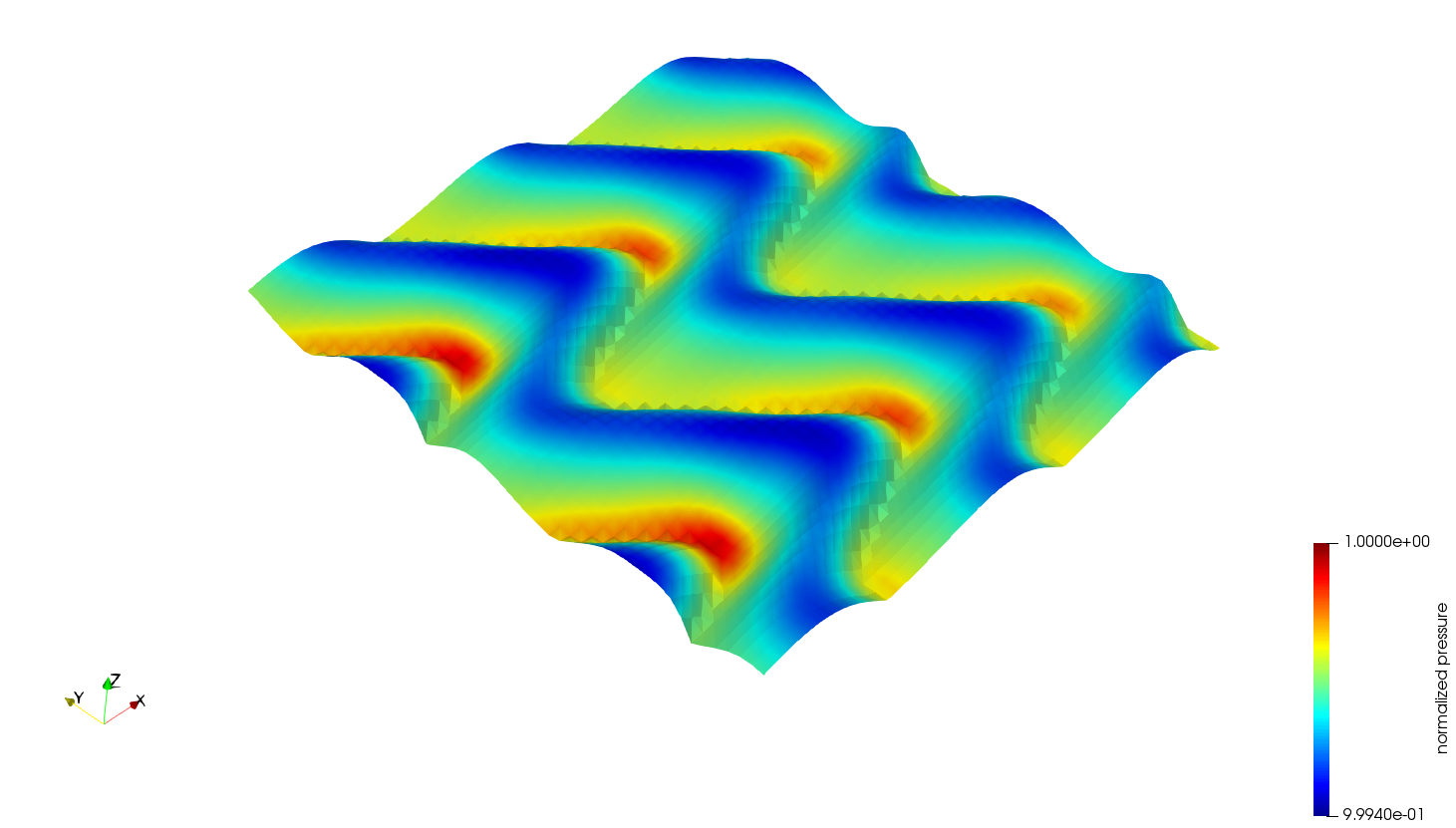}
    \caption{\(\Delta \varphi_1 = 180 ^{\circ}, \Delta \varphi_2 = 0 ^{\circ}\)}
  \end{subfigure}
  \begin{subfigure}{0.3\linewidth}
    \includegraphics[trim=7cm 3cm 6cm 0cm, clip, width=\linewidth]{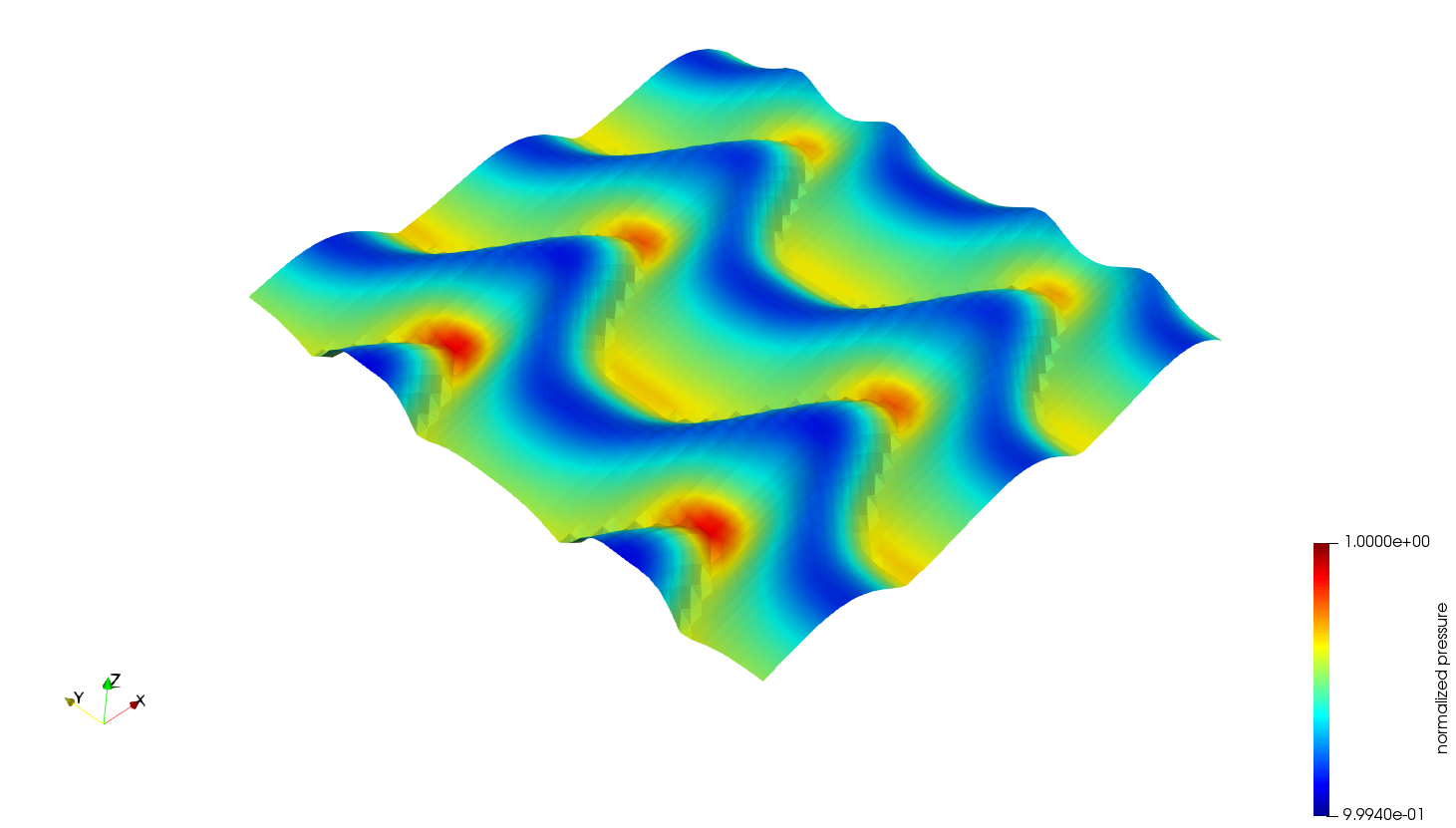}
    \caption{\(\Delta \varphi_1 = 270 ^{\circ}, \Delta \varphi_2 = 0 ^{\circ}\)}
  \end{subfigure}
  \begin{subfigure}{0.3\linewidth}
    \includegraphics[trim=7cm 3cm 6cm 0cm, clip, width=\linewidth]{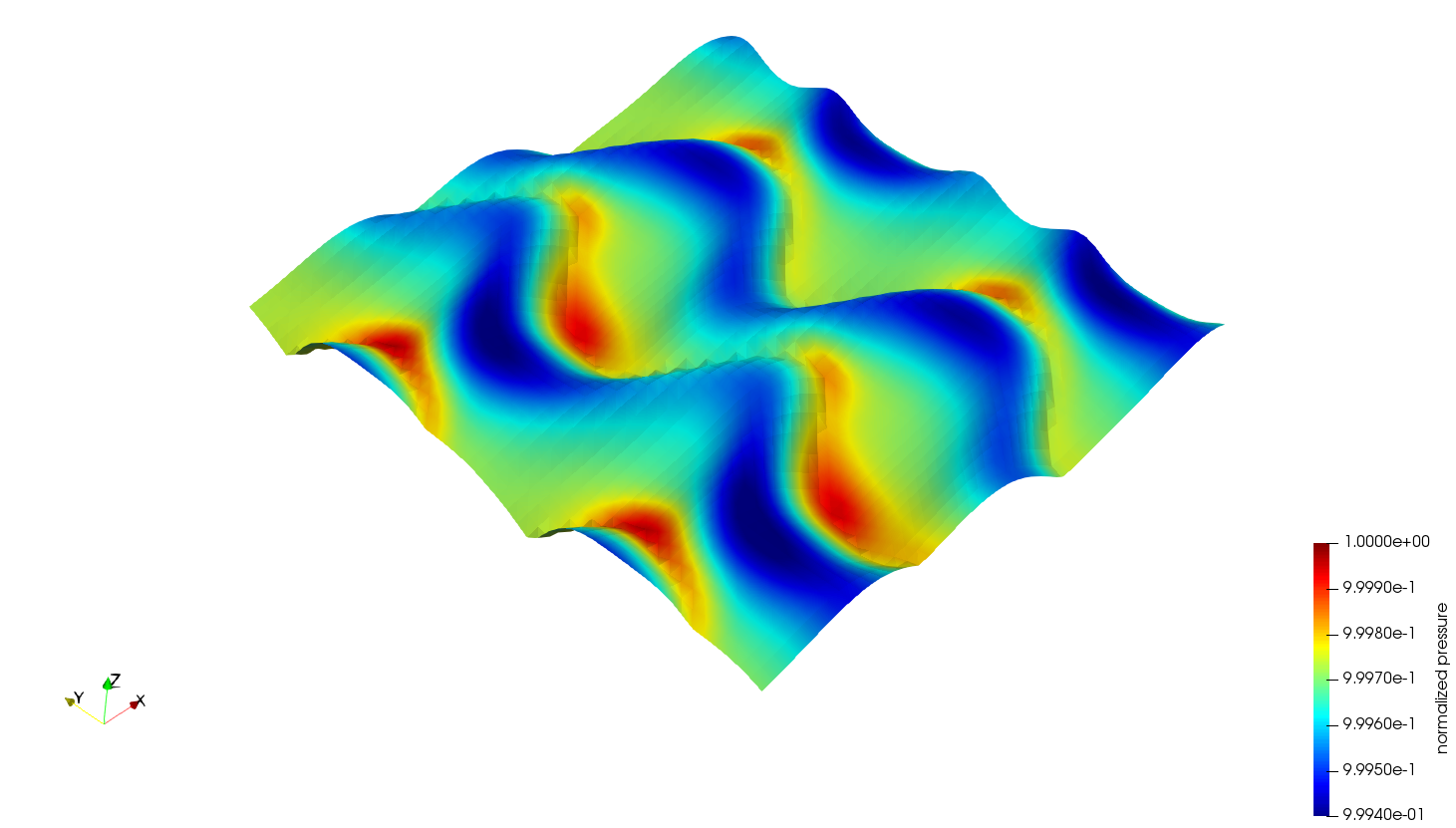}
    \caption{\(\Delta \varphi_1 = 90 ^{\circ}, \Delta \varphi_2 = 90 ^{\circ}\)}
  \end{subfigure}
  \begin{subfigure}{0.3\linewidth}
    \includegraphics[trim=7cm 3cm 6cm 0cm, clip, width=\linewidth]{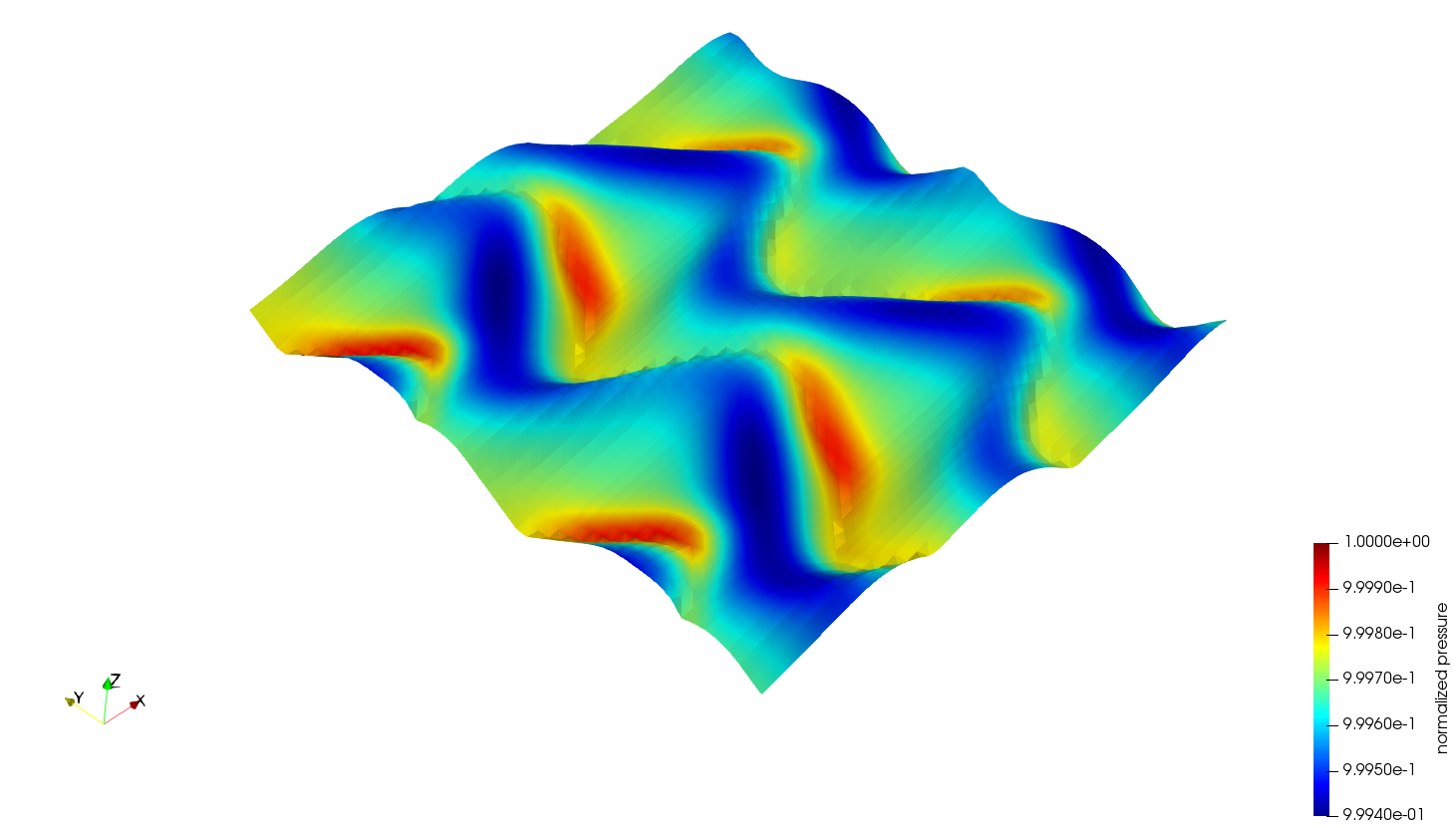}
    \caption{\(\Delta \varphi_1 = 180 ^{\circ}, \Delta \varphi_2 = 90 ^{\circ}\)}
  \end{subfigure}
  \begin{subfigure}{0.3\linewidth}
    \includegraphics[trim=7cm 3cm 6cm 0cm, clip, width=\linewidth]{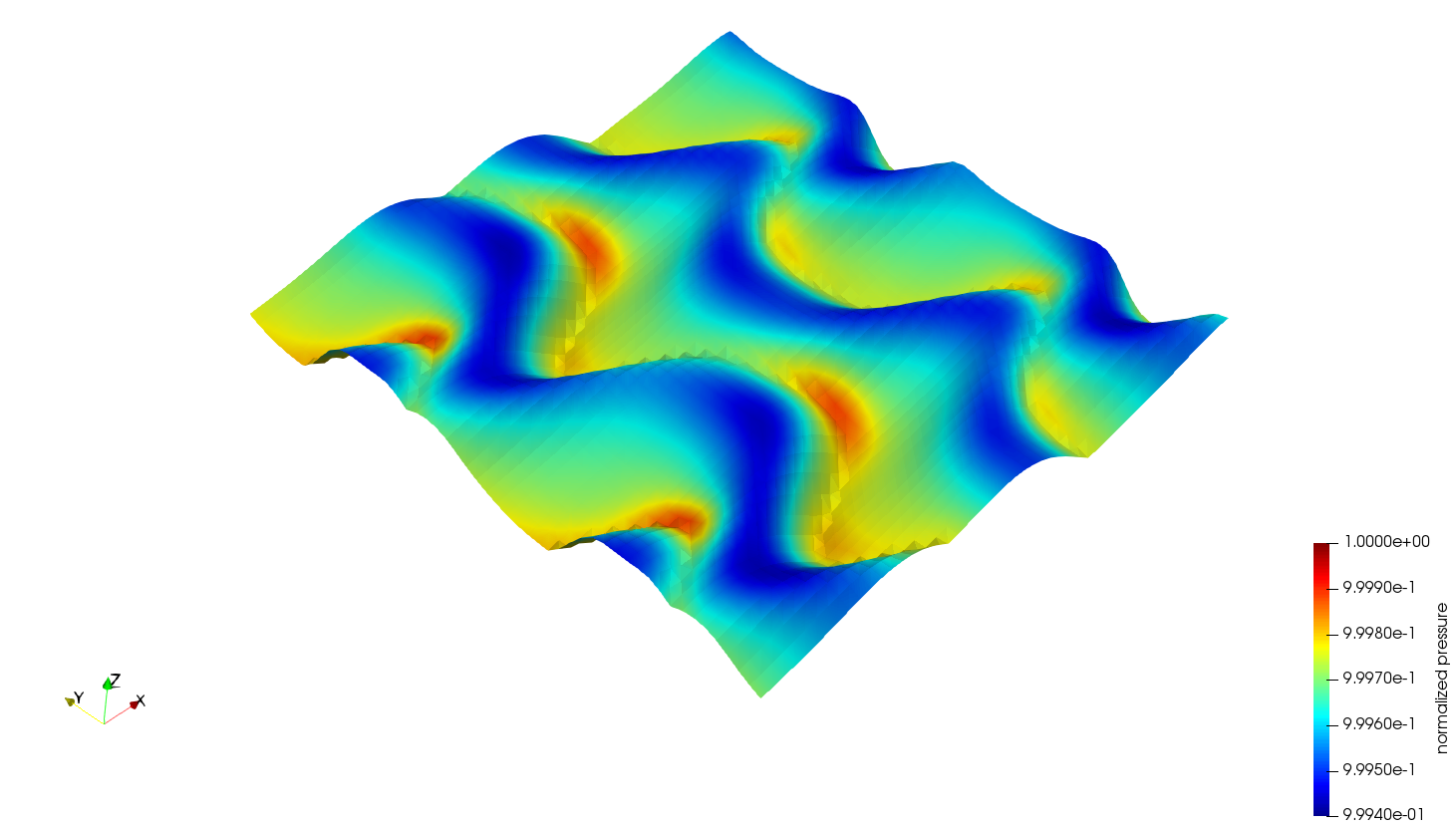}
    \caption{\(\Delta \varphi_1 = 270 ^{\circ}, \Delta \varphi_2 = 90 ^{\circ}\)}
  \end{subfigure}
  \begin{subfigure}{0.3\linewidth}
    \includegraphics[trim=7cm 3cm 6cm 0cm, clip, width=\linewidth]{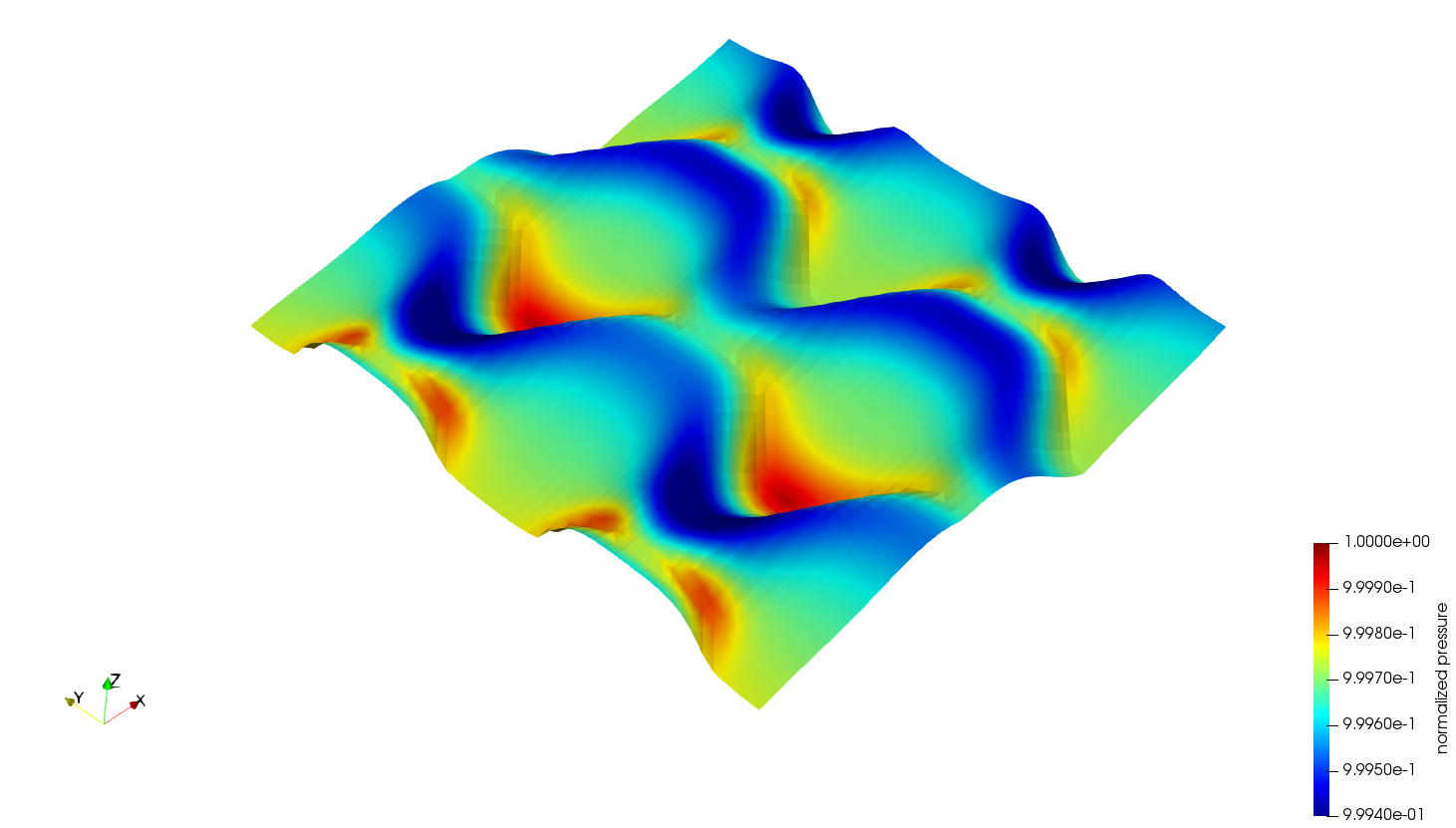}
    \caption{\(\Delta \varphi_1 = 90 ^{\circ}, \Delta \varphi_2 = 180 ^{\circ}\)}
  \end{subfigure}
  \begin{subfigure}{0.3\linewidth}
    \includegraphics[trim=7cm 3cm 6cm 0cm, clip, width=\linewidth]{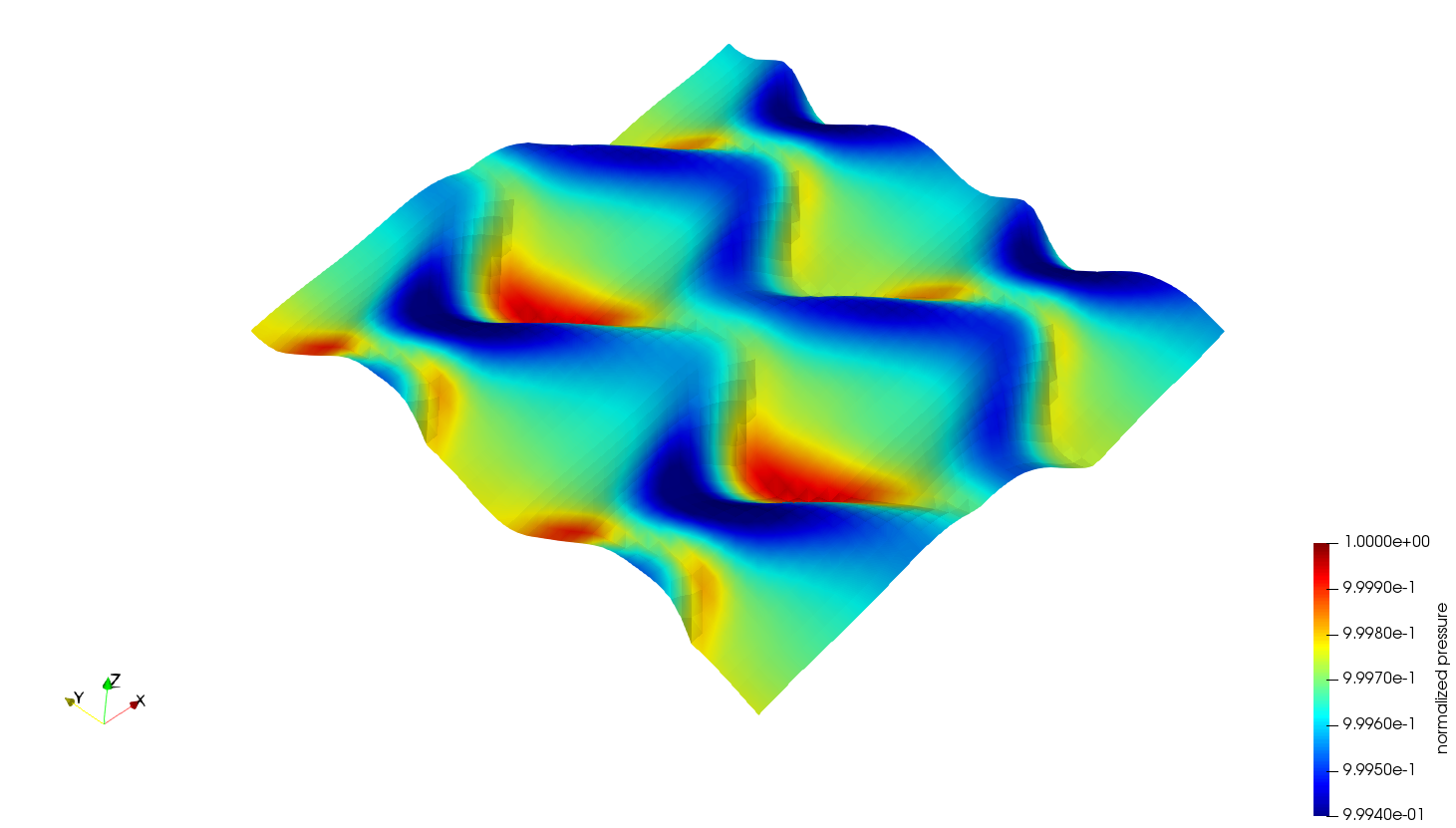}
    \caption{\(\Delta \varphi_1 = 180 ^{\circ}, \Delta \varphi_2 = 180 ^{\circ}\)}
  \end{subfigure}
  \begin{subfigure}{0.3\linewidth}
    \includegraphics[trim=7cm 3cm 6cm 0cm, clip, width=\linewidth]{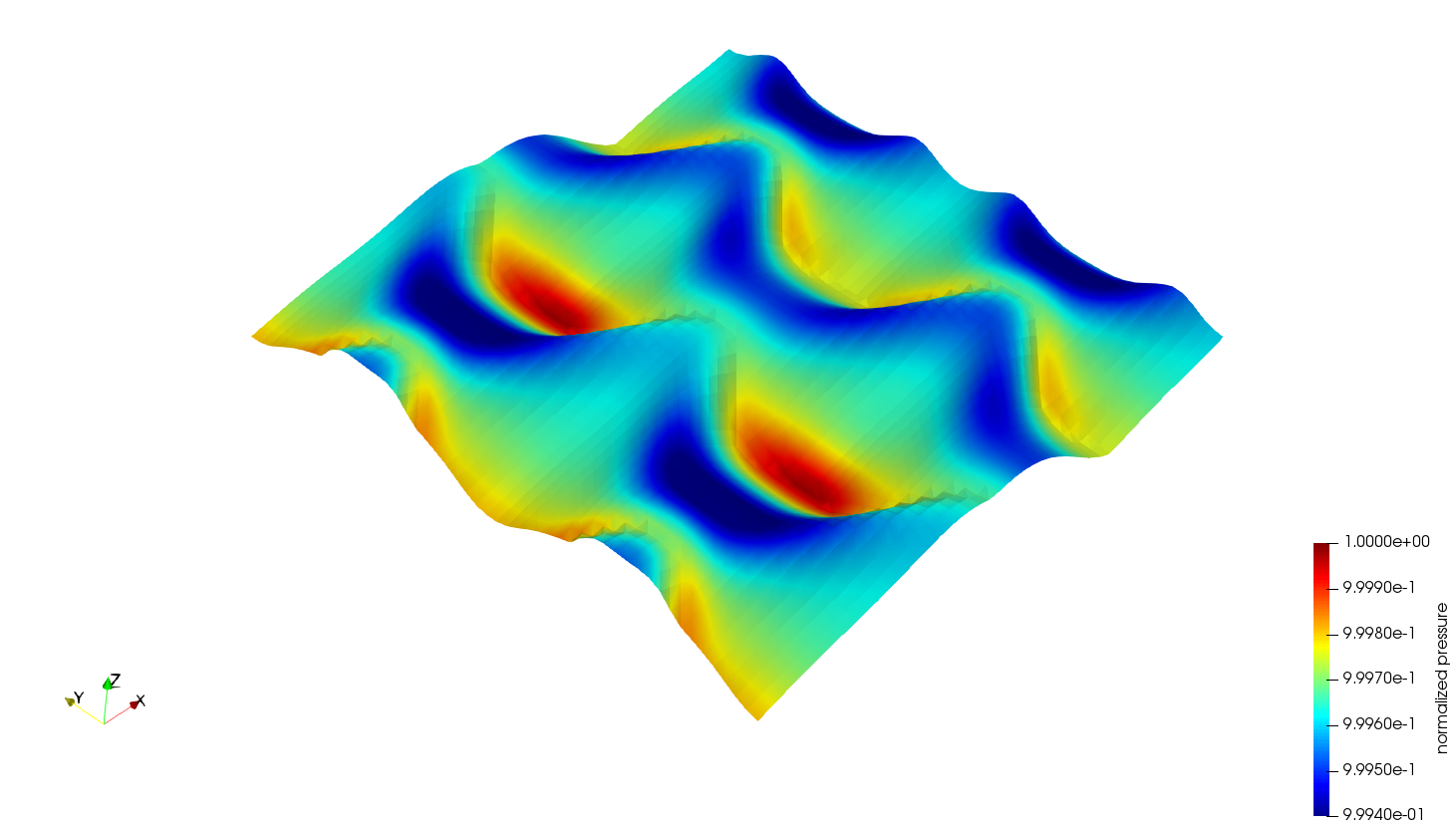}
    \caption{\(\Delta \varphi_1 = 270 ^{\circ}, \Delta \varphi_2 = 180 ^{\circ}\)}
  \end{subfigure}
  \begin{subfigure}{0.6\linewidth}
    \includegraphics[width=\linewidth]{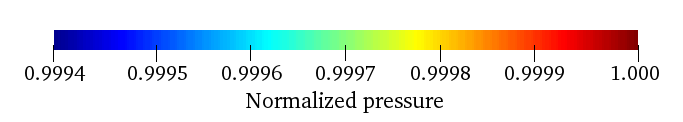}
  \end{subfigure}
    \caption{Distribution of normalized pressure on the surface-subsurface interface. The normalized pressure for each bed form is defined as \(\frac{p}{\{p_{max}\}_a}\), in which \(\{p_{max}\}_a\) is the maximum pressure in bed form (a).}
  \label{fig:bed_pressure}
\end{figure}

\subsection{Solute exchange at the sediment-water interface}
\label{Sol_Ex_SWI}

\begin{figure}[!ht]
  \centering
  \begin{subfigure}{0.48\linewidth}
    \includegraphics[trim=0cm 0cm 0cm 0cm, clip, width=\linewidth]{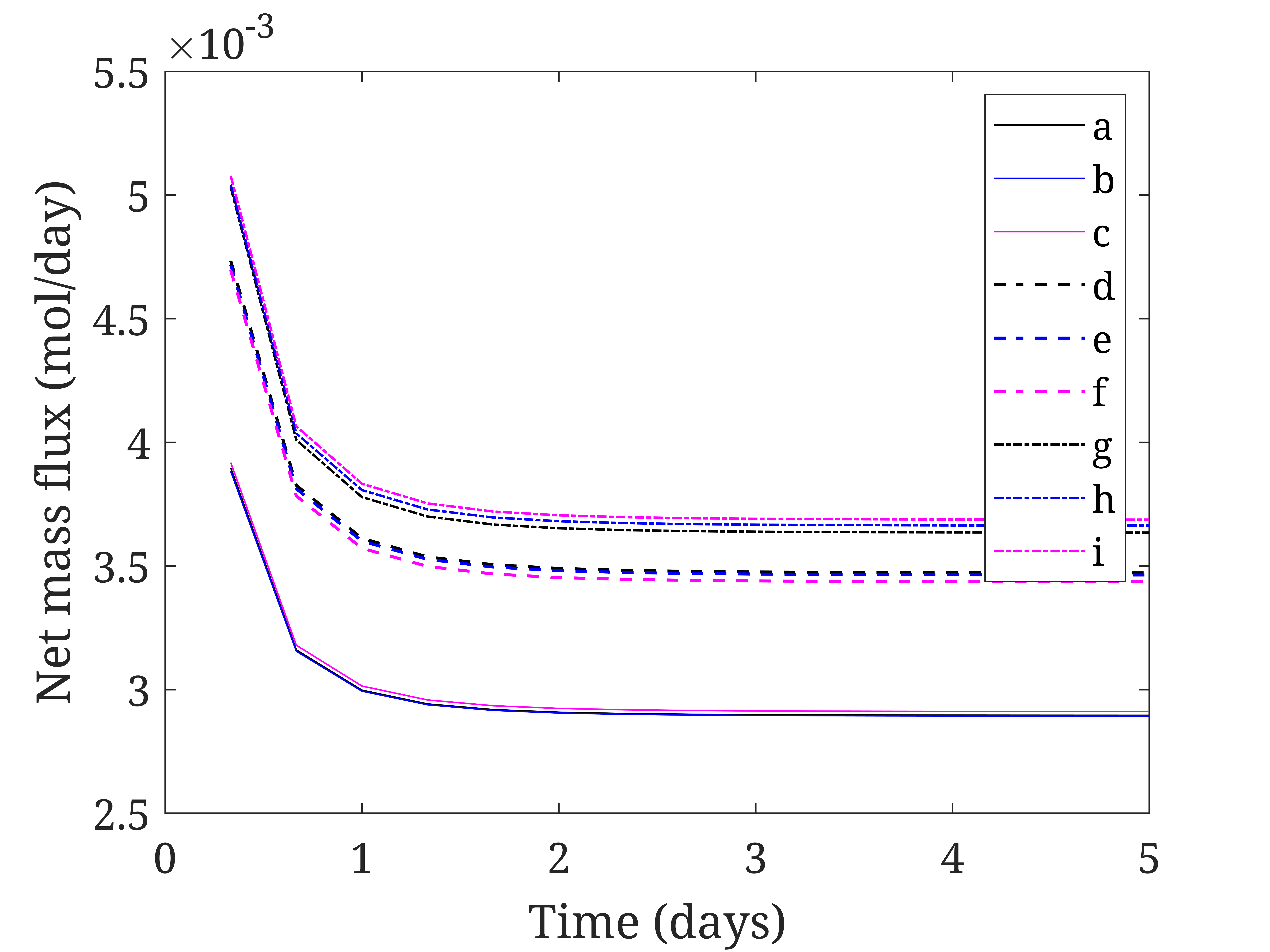}
    \caption{Solute A}
  \end{subfigure}
  \begin{subfigure}{0.48\linewidth}
    \includegraphics[trim=0cm 0cm 0cm 0cm, clip, width=\linewidth]{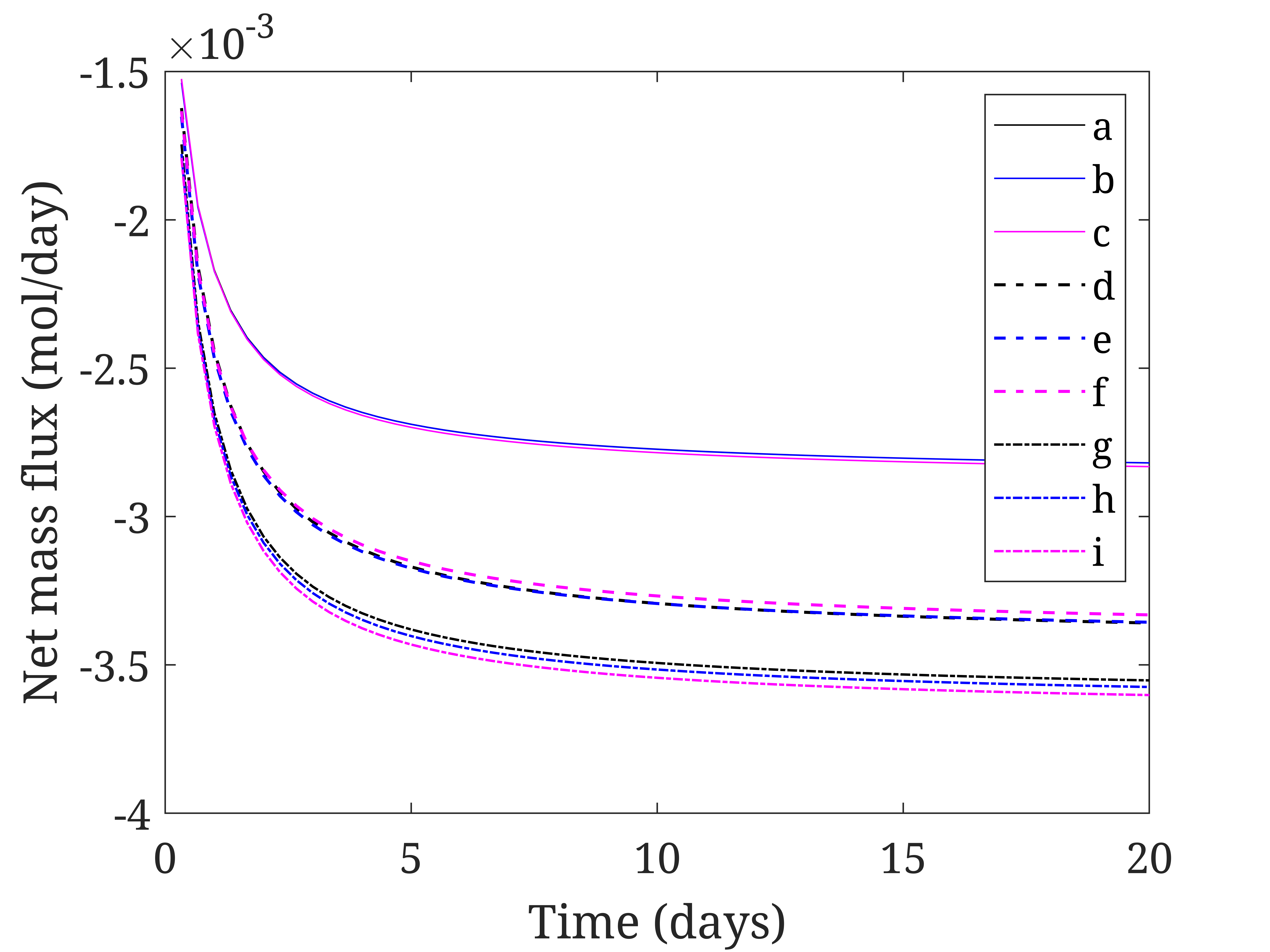}
    \caption{Solute C}
  \end{subfigure}
    \caption{Solute exchange at sediment-water interface over time. Note that solute B is identical to solute A and has similar behavior.}
  \label{fig:solute_at_SWI}
\end{figure}

During the first few days of simulation solutes are only infiltrating until the amounts of infiltration and exfiltration are stabilized and the system reaches an equilibrium.
As shown in Figures \ref{fig:solute_at_SWI}a and \ref{fig:solute_at_SWI}b, the net exchange flux of solute A is positive meaning that infiltration of solute A at the streambed is more than its exfiltration. It should be noted that characteristics of solute A and solute B are identical and thus they show the same behavior.
Net flux of solute C through the streambed is negative, which indicates that solute C is mostly leaving the subsurface and going back to the surface water.

Moreover, the comparison of exchange fluxes for all bed forms in Figures \ref{fig:solute_at_SWI}a and \ref{fig:solute_at_SWI}b shows that as \(\Delta \varphi_2\) grows, the net flux of solute A transferring from surface into subsurface, as well as the net flux of solute C leaving the sediment increase. 
In other words, bed forms with larger \(\Delta \varphi_2\) transfer higher amounts of solute.
In particular, for lunate bed forms (c, f, i) the absolute value of solute transferred (both solute A and solute C) is growing as \(\Delta \varphi_2\) is increasing from bed form (c) to bed form (i). 
Linguoid bed forms (a, d, g) shows similar trend.
This is also shown in Table \ref{table:max_min} that higher \(\Delta \varphi_2\) results in higher average velocity, larger reaction rate and higher amounts of solute exchange. Bed form (i) has a phase shift (\(\Delta \varphi_2\)) of 180$^\circ$ compared to bed form (a) which leads to 21\% change in average velocity and average reaction rate. The solute exchange flux is also increased by 21\% for bed form (i).

\begin{figure}[!hb]
  \centering
  \begin{subfigure}{0.48\linewidth}
    \includegraphics[trim=0cm 0cm 0cm 0cm, clip, width=\linewidth]{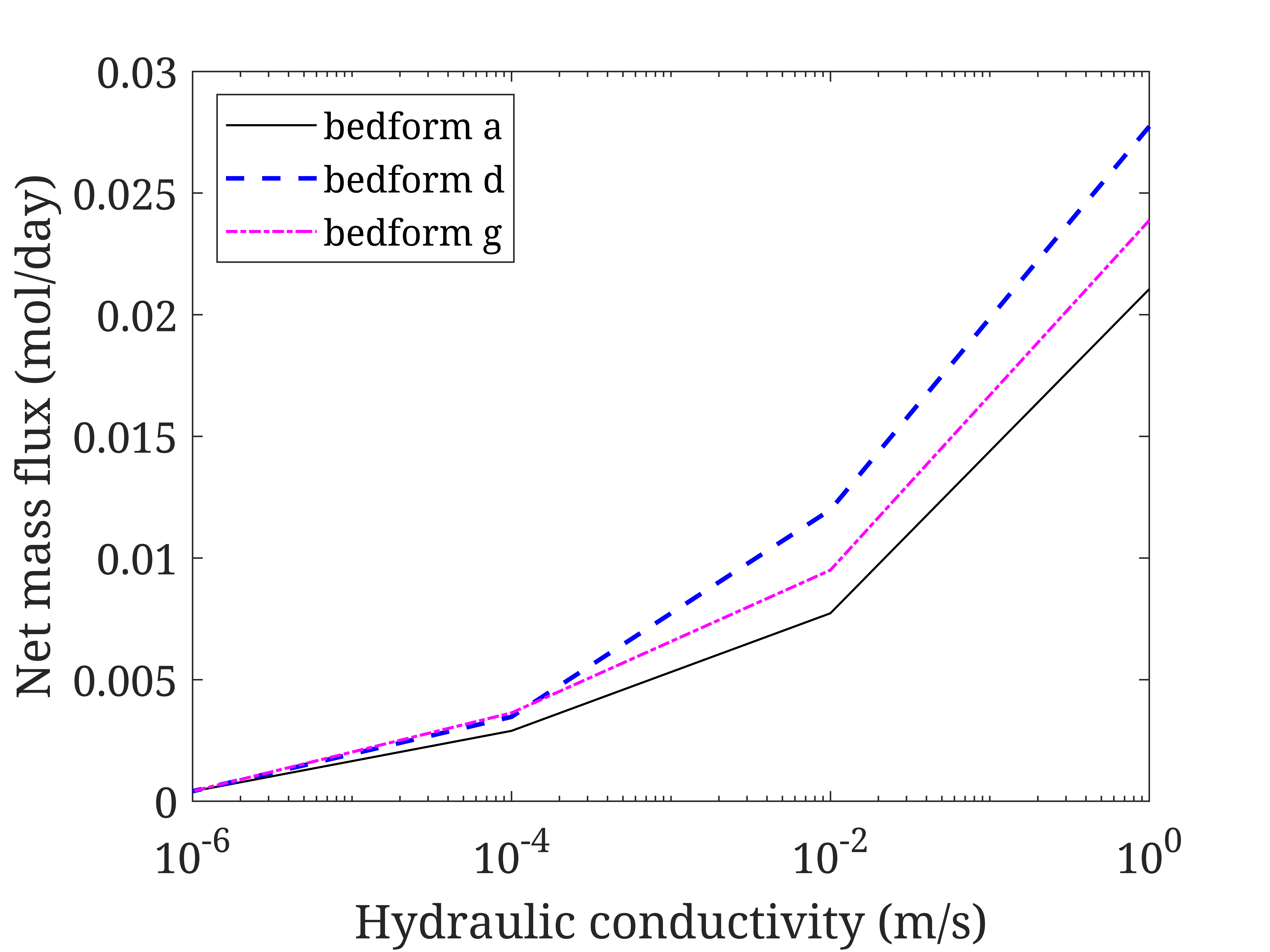}
    \caption{Solute A}
  \end{subfigure}
  \begin{subfigure}{0.48\linewidth}
    \includegraphics[trim=0cm 0cm 0cm 0cm, clip, width=\linewidth]{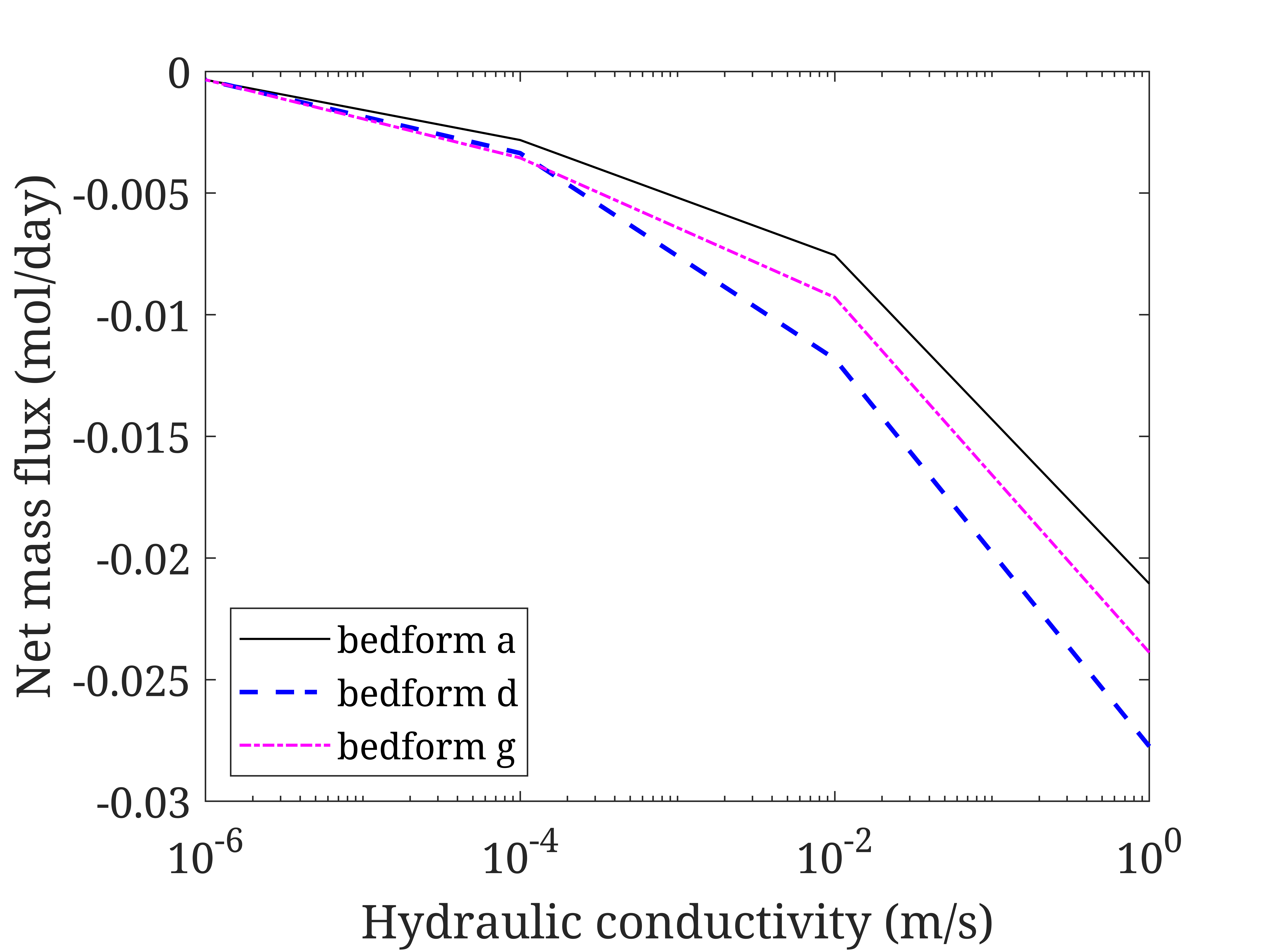}
    \caption{Solute C}
  \end{subfigure}
    \caption{Mass flux sensitivity to hydraulic characteristics of sediment. Solute B is identical to solute A and has similar behavior.}
  \label{fig:mas_perm_SWI}
\end{figure}

Furthermore, both streambed morphology and hydraulic conductivity ($K$) influence the solute exchange amount between surface water and groundwater. 
Figure \ref{fig:mas_perm_SWI} displays the equilibrated mass flux at time = 20 days for linguiod type bed forms and with different $K$. Other bed form types such as lunate with the same \(\Delta \varphi_1\) shows similar trends.
As shown, increasing $K$ will enhance the amount of entering mass of solute A from surface water to the hyporheic zone. However, the amount of solute C moving out of the hyporheic zone and into the stream decreases. It is also clear that if sediment is more conductive the differences between linguoid bed forms are amplified in terms of solute mass flux.

\subsection{Solute dynamics in the hyporheic zone}

The residence time of solutes infiltrating from river water to riverbed sediments can provide critical information about biogeochemical reaction times \cite{gomez2014hydrogeomorphic, gomez2015denitrification}. In fact, it is known that residence time distributions can determine the type and rate of many processes occurring in surface, near surface, and deep environments from solely physical to complex biogeochemical processes\cite{cardenas2008surface}. Therefore, there have been several recent studies directed at unraveling mechanisms controlling solute residence times for hyporheic exchange processes at multiple scales \cite{cardenas2007potential, kollet2008demonstrating, worman2006, worman2007fractal}. PFLOTRAN does not explicitly simulate residence time distributions using particle tracking, but it does use the concept of groundwater age \cite{gardner2015high}. Here, this capability is used to quantify residence times for infiltrated (i.e., A and B) and produced (i.e., C) solutes to further study the impacts of bed form topographies on hyporheic processes. 

\begin{figure}[!ht]
  \centering
  \begin{subfigure}{0.48\linewidth}
    \includegraphics[trim=0cm 0cm 0cm 0cm, clip, width=\linewidth]{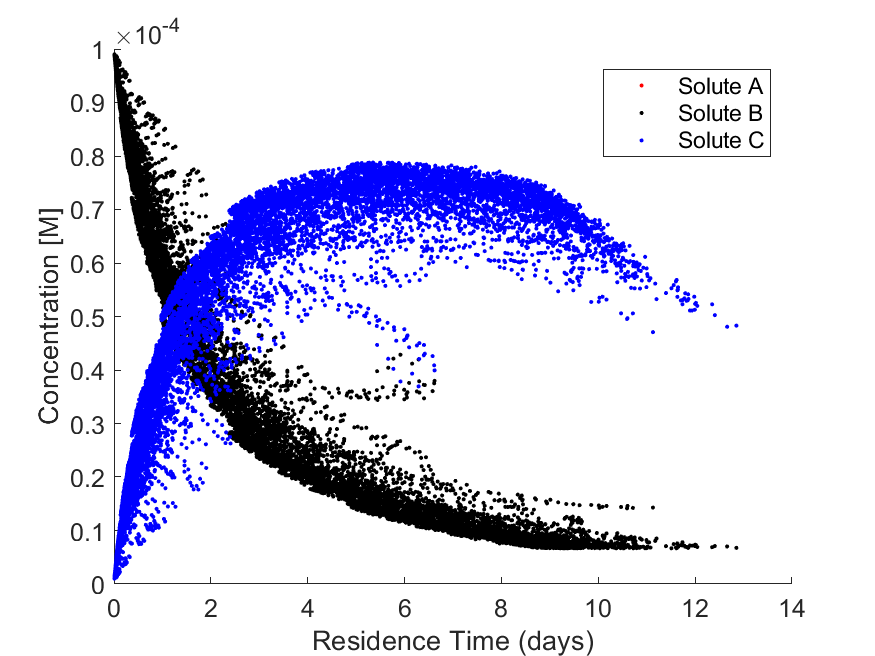}
    \caption{bed form (a) }
  \end{subfigure}
  \begin{subfigure}{0.48\linewidth}
    \includegraphics[trim=0cm 0cm 0cm 0cm, clip, width=\linewidth]{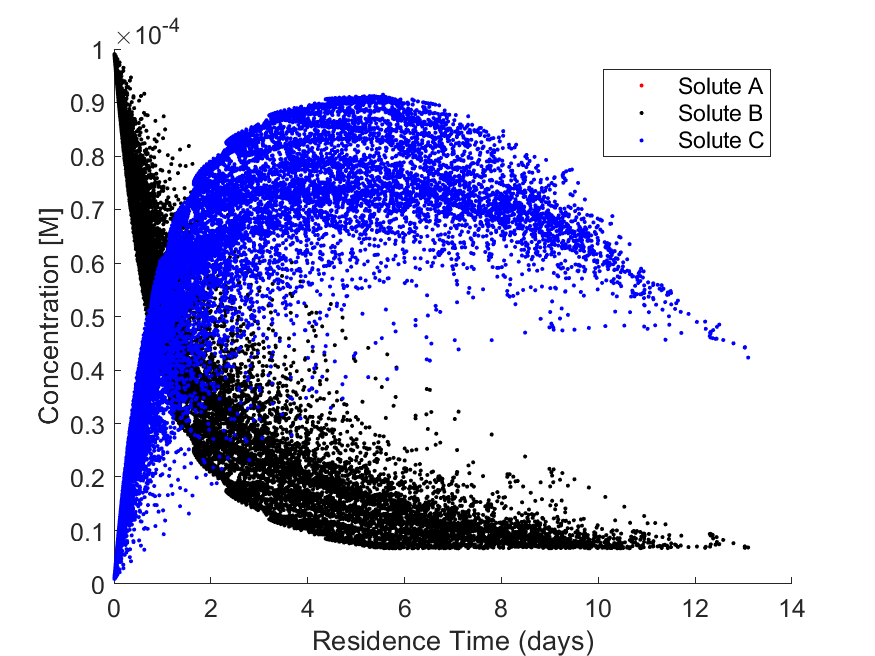}
    \caption{bed form (d)}
  \end{subfigure}
  \begin{subfigure}{0.48\linewidth}
    \includegraphics[trim=0cm 0cm 0cm 0cm, clip, width=\linewidth]{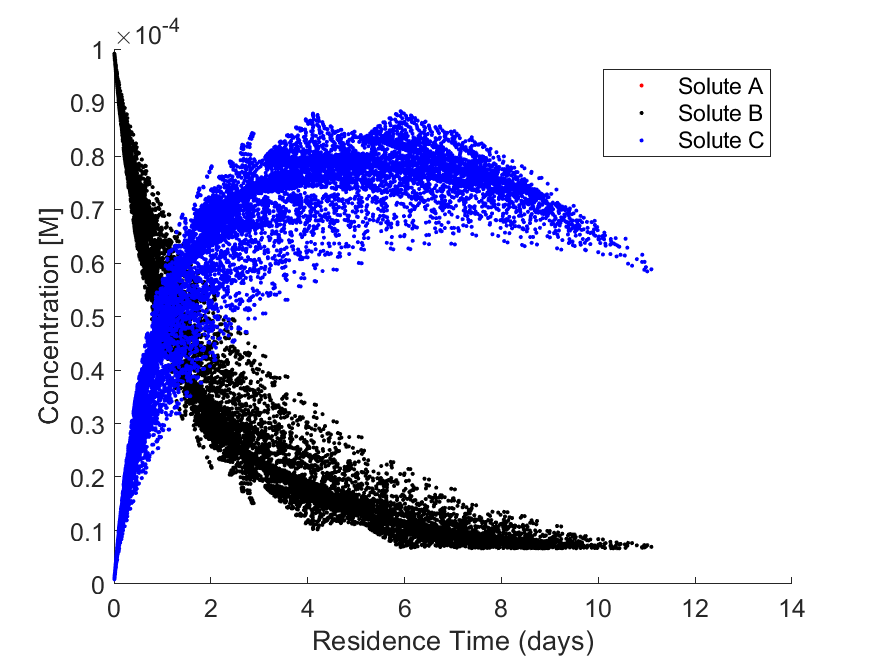}
    \caption{bed form (g)}
  \end{subfigure}
    \caption{Duration of concentration level at all computational cells for various bed forms. Since bed forms with equal \(\Delta \varphi_2\) show similar trend, one bed form is presented for each group.}
  \label{fig:solute_at_cells}
\end{figure}

Bed form geometry controlled the extent of solute transformations by regulating residence times within the streambed. At steady-state (\textit{t}=20 days), residence times are lowest in the shallow streambed, associated with higher groundwater velocities and shorter hyporheic flowpaths than deeper regions (Figure \ref{fig:solute_at_cells}). Longer residence times (up to 14 days) in the deeper streambed allowed for more complete solute transformation, however, resulting in lower concentrations of A and B and the highest concentrations of C. As a governing factor in the dynamics of groundwater flow patterns, the phase shift in the second planform sinuosity (\(\Delta \varphi_2\)) strongly controlled the distribution of residence times based on bed form geometry. Average residence times increased from $2.08\pm0.01$ days to $2.52\pm0.03$ days as phase shift increased from \(\Delta \varphi_2\)=0$^{\circ}$ to \(\Delta \varphi_2\)=180$^{\circ}$. Though solute concentration profiles followed a similar trend for all bed form geometries, with solutes A and B decreasing to roughly 0.1 M and solute C reaching up to 0.9 M as cell residence time increased, concentrations are much more variable throughout the domain when bed forms are out-of-phase (\(\Delta \varphi_2\)=90$^{\circ}$). For example, the average standard deviation of the concentration of solute C is 568\% and 351\% larger in bed forms where \(\Delta \varphi_2\)=90$^{\circ}$ than those where \(\Delta \varphi_2\)=0$^{\circ}$ and \(\Delta \varphi_2\)=180$^{\circ}$, respectively.

\begin{figure}[!ht]
  \centering
  \begin{subfigure}{0.48\linewidth}
    \includegraphics[trim=0cm 0cm 0cm 0cm, clip, width=\linewidth]{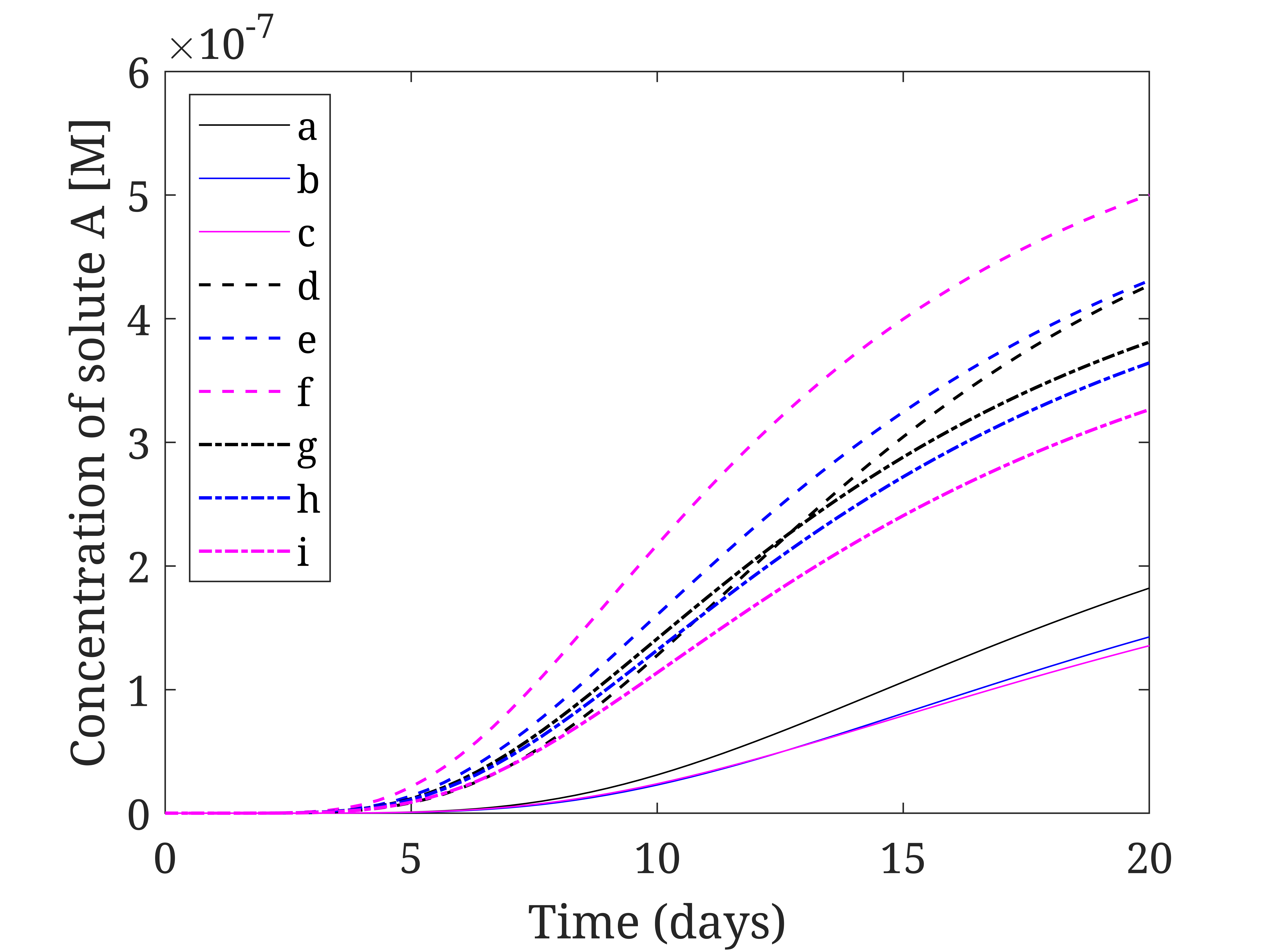}
    \caption{}
  \end{subfigure}
  \begin{subfigure}{0.48\linewidth}
    \includegraphics[trim=0cm 0cm 0cm 0cm, clip, width=\linewidth]{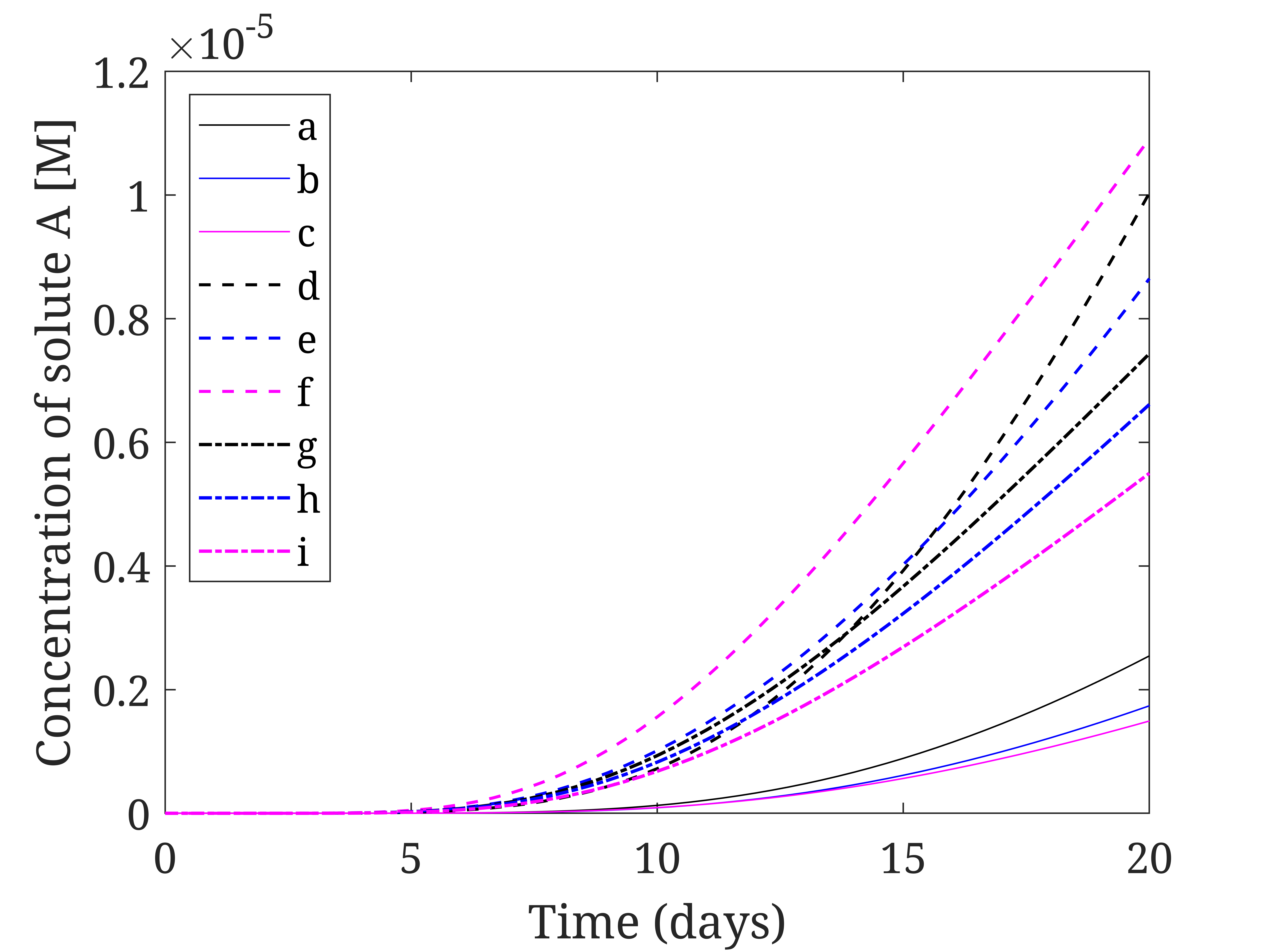}
    \caption{}
  \end{subfigure}
    \caption{Solute concentration time series at depth=0.2 (m) a) Solute A, b) Solute C, (Solute A and solute B have identical trends)}
  \label{fig:solute_timeseries}
\end{figure}

Temporal variation of concentration can also be described as time series at a particular depth. As shown in Figure \ref{fig:solute_timeseries}, at 0.2 m deep in the center of the subsurface domain, solute concentrations for out-of-phase bed forms (\(\Delta \varphi_2 \neq 0\)) experience sharp increases while they gradually increase for in-phase bed forms (\(\Delta \varphi_2 = 0\)).
This outcome agrees well with the exchange flux results in section (\ref{Sol_Ex_SWI}), which demonstrated that the out-of-phase bed forms allow for a greater amount of solute exchange.

\subsection{Sensitivity analysis on volume of reaction zone}

\begin{figure}[!ht]
  \centering
  \begin{subfigure}{0.3\linewidth}
    \includegraphics[trim=0cm 0cm 0cm 0cm, clip, width=\linewidth]{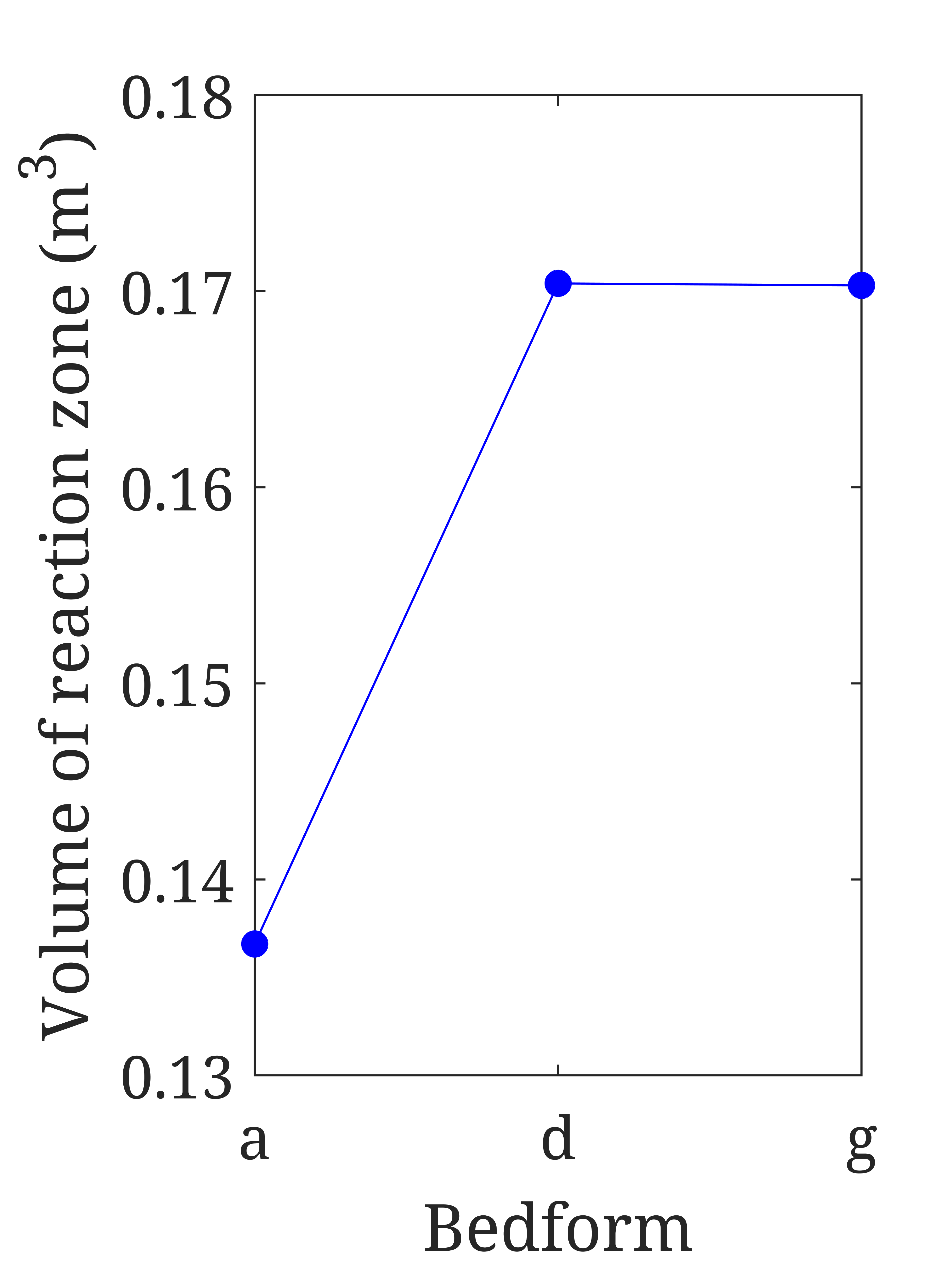}
    \caption{}
  \end{subfigure}
  \begin{subfigure}{0.3\linewidth}
    \includegraphics[trim=0cm 0cm 0cm 0cm,clip,width=\linewidth]{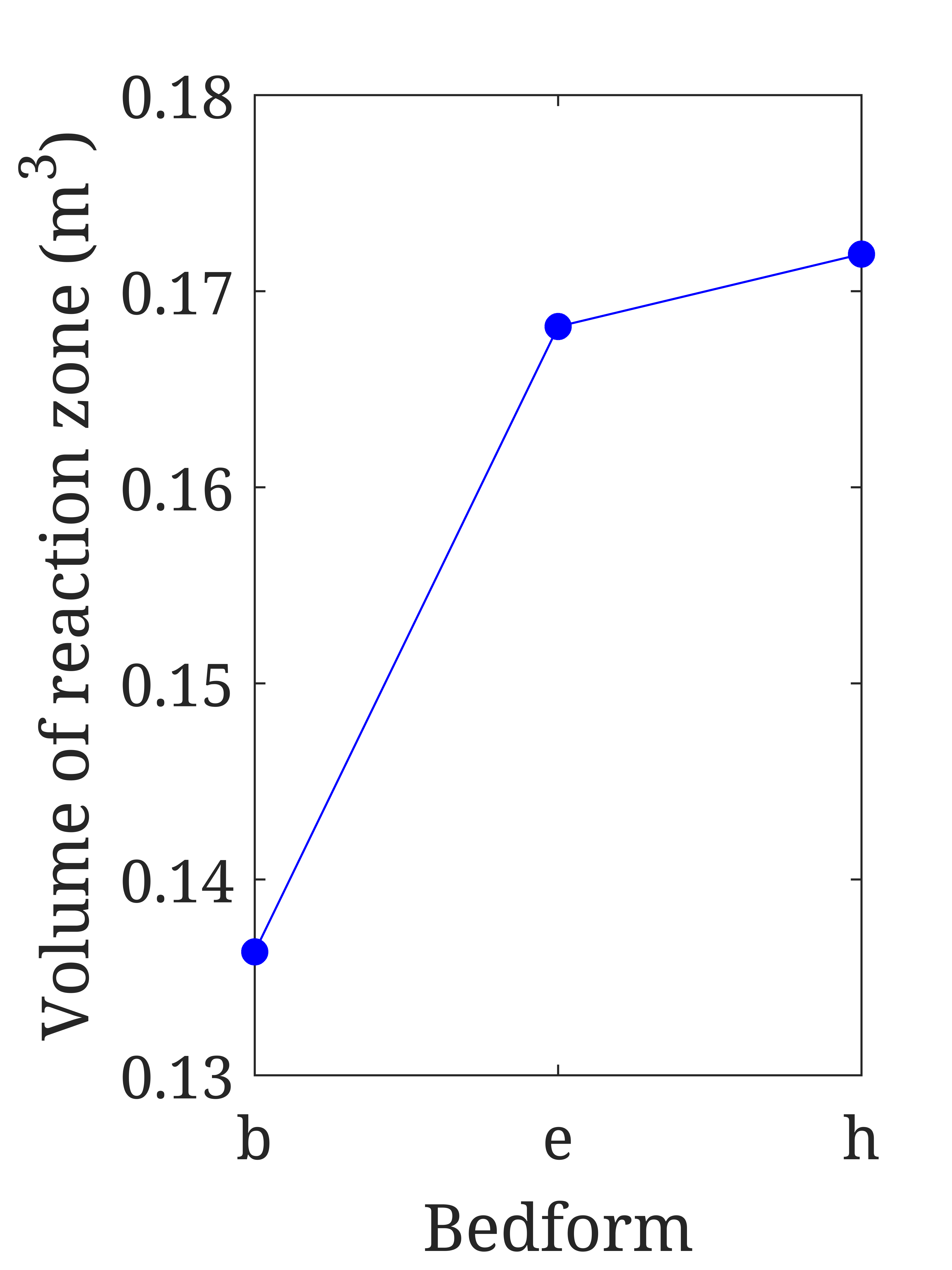}
    \caption{}
  \end{subfigure}
  \begin{subfigure}{0.3\linewidth}
    \includegraphics[trim=0cm 0cm 0cm 0cm,clip,width=\linewidth]{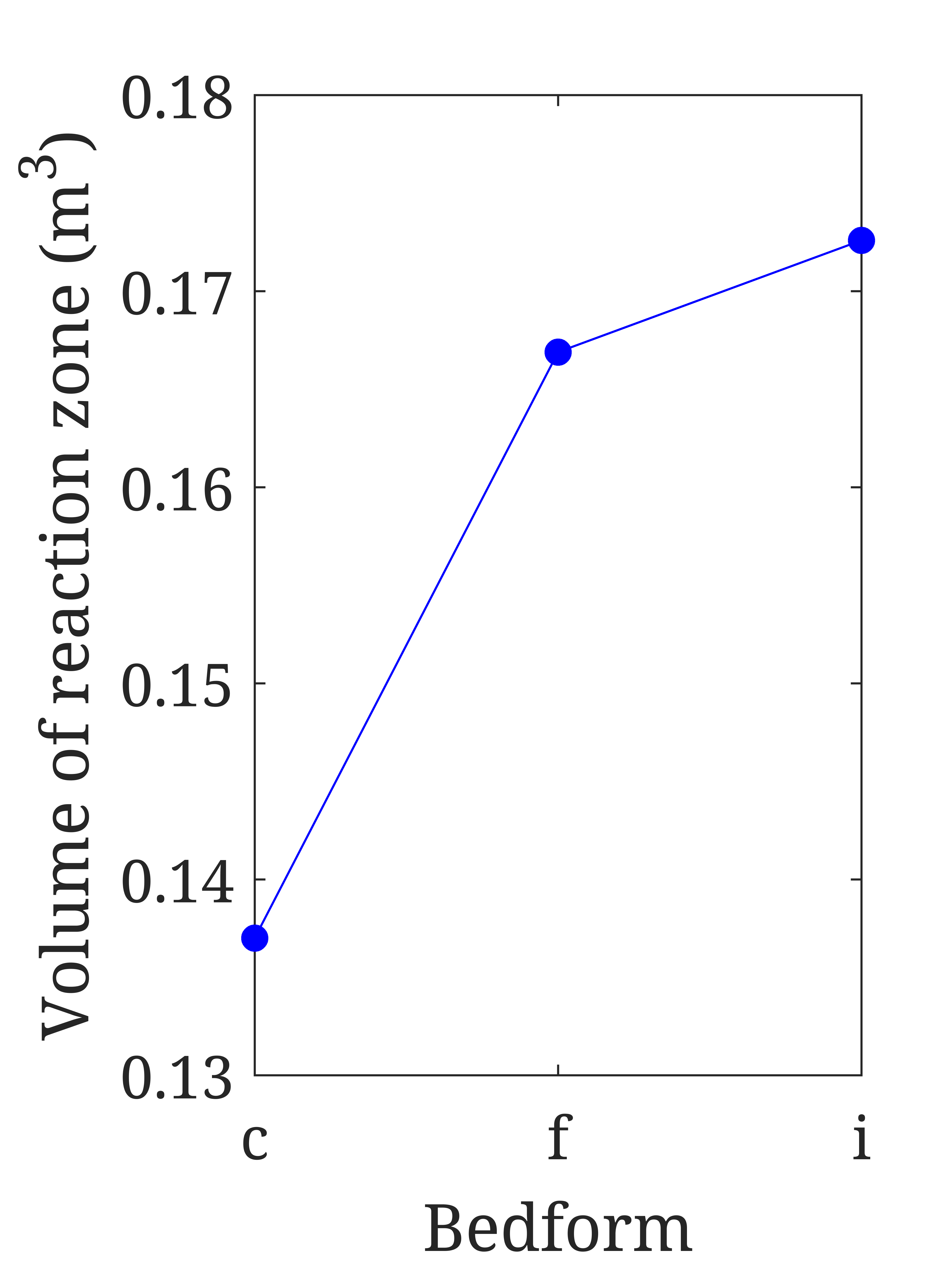}
    \caption{}
  \end{subfigure}
    \caption{Change in volume of reaction zone with bed form shapes; a) Linguoid type bed forms $\Delta \varphi_1 = 90^{\circ}$, b) Bed form with $\Delta \varphi_1 = 180^{\circ}$, c) Lunate type bed forms $\Delta \varphi_1 = 270^{\circ}$. As $\Delta \varphi_2$ increases in a given set of bed form types the volume of reaction zone is also enhanced. }
  \label{fig:reaction_zone_size_bedform}
\end{figure}

The volume of the reaction zone, \(V_{rz}\), is defined where reaction rates are greater than 1\% of the maximum rate (\textit{k\textsubscript{max}}).
For each bed form type (e.g., linguiod versus lunate), \(V_{rz}\) is most sensitive to \(\Delta \varphi_2\), and increases for out-of-phase bed forms (\(\Delta \varphi_2 \neq 0\)) (Figures \ref{fig:reaction_zone_size_bedform}).
Table \ref{table:max_min} also shows that with a 180$^\circ$ phase shift, \(V_{rz}\) grows by 20\%.

\begin{figure}[!ht]
  \centering
  \begin{subfigure}{0.48\linewidth}
    \includegraphics[trim=0cm 0cm 0cm 0cm, clip, width=\linewidth]{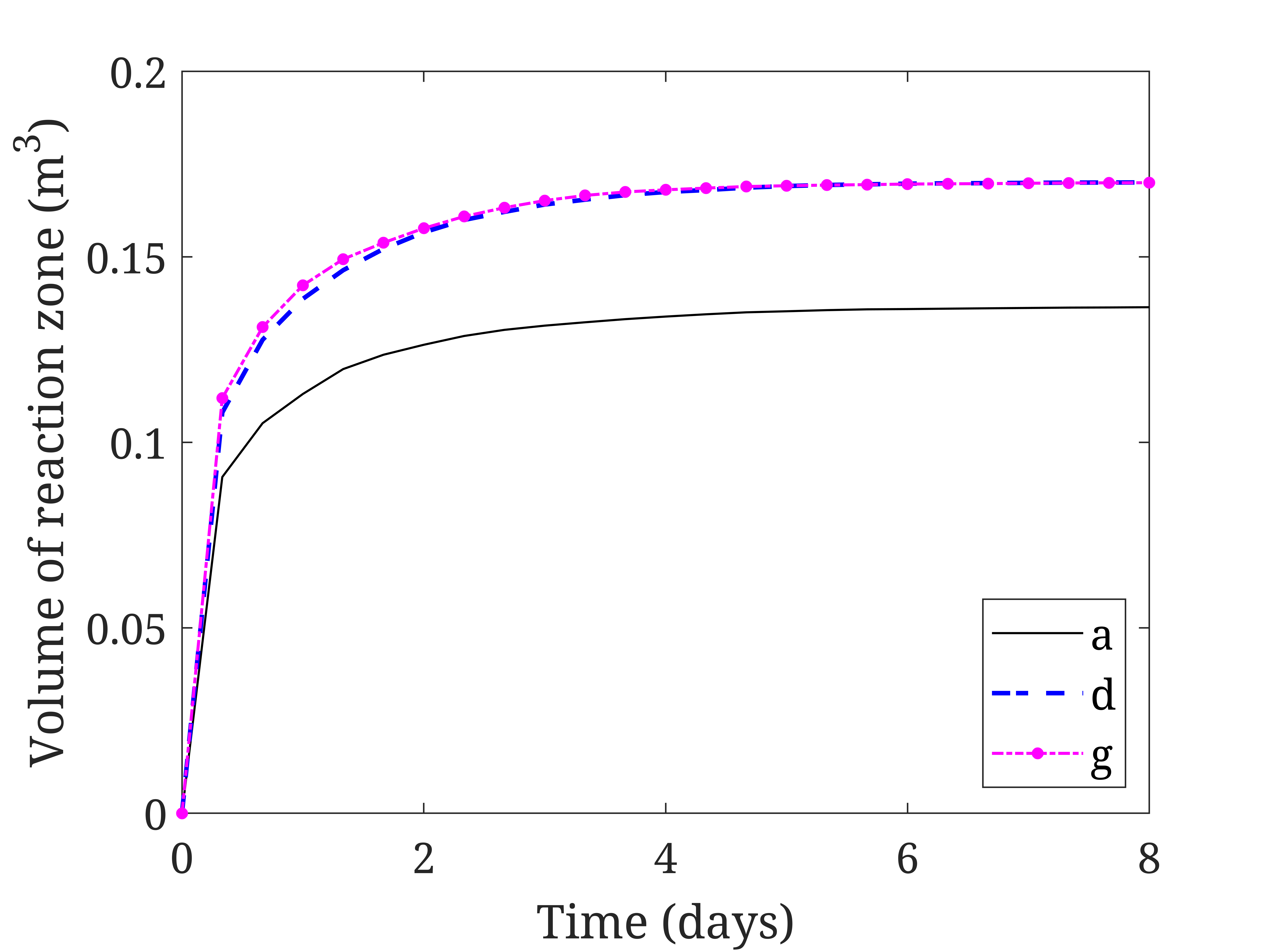}
    \caption{}
  \end{subfigure}
  \begin{subfigure}{0.48\linewidth}
    \includegraphics[trim=0cm 0cm 0cm 0cm,clip,width=\linewidth]{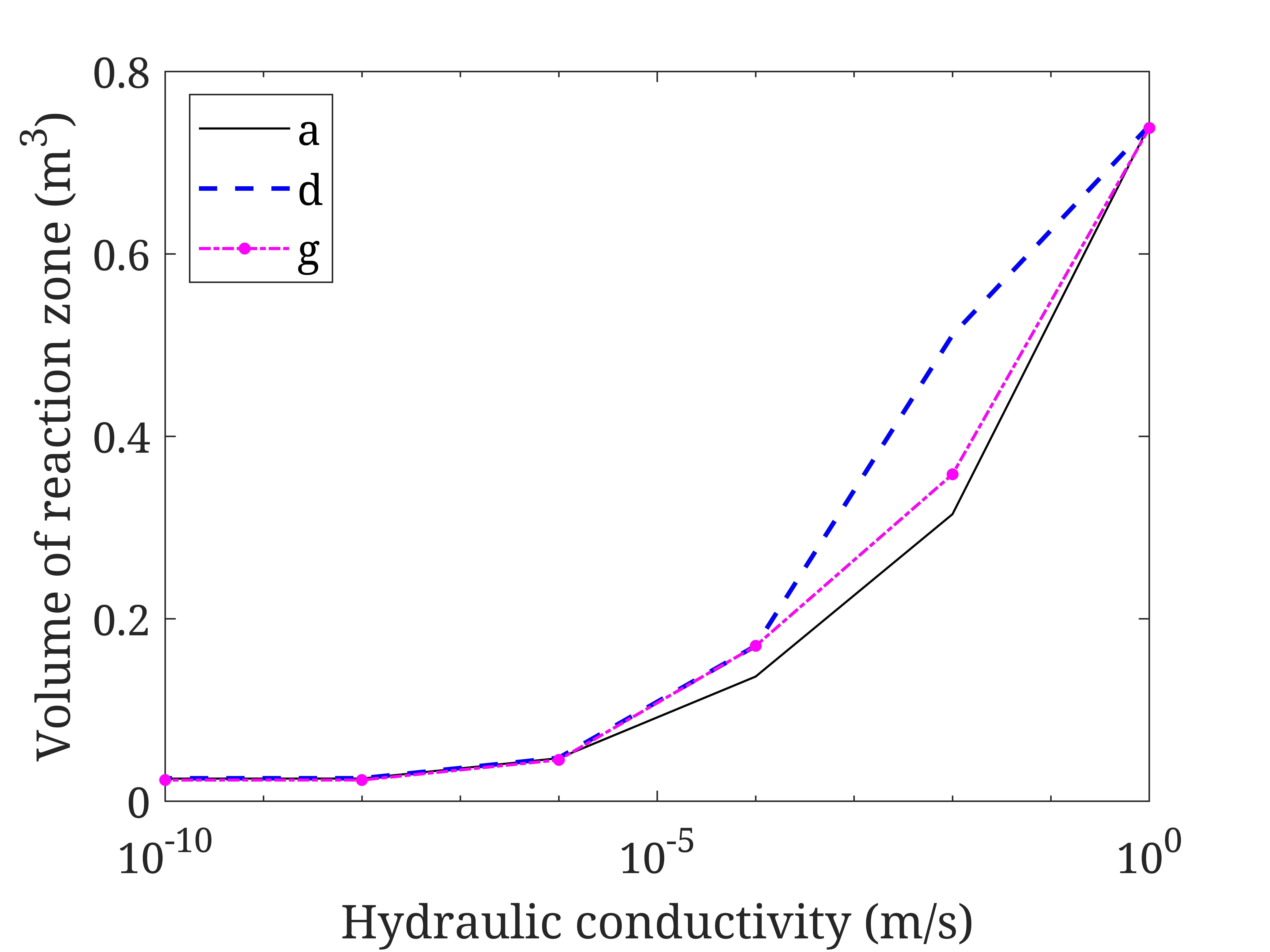}
    \caption{}
  \end{subfigure}
    \caption{Change in volume of reaction zone with (a) time, (b) bed form shapes, and (c) permeability.
    Since bed forms with equal \(\Delta \varphi_2\) show similar trends, one bed form is presented for each group.}
  \label{fig:reaction_zone_size_sensitivity}
\end{figure}

Hydraulic properties of sediments also influence solute transport and reaction processes. Hyporheic exchange volumes are lower in less conductive sediments, and the flux of solutes into the streambed is thus decreased. As a result, bed form morphology is negligible to \(V_{rz}\) because the magnitude of hyporheic exchange is below the threshold of influence. Conversely, bed form shape is a primary control on \(V_{rz}\) in highly conductive sediments (K $<$ \(10^{-6}\) m/s) because flow and solute exchange are not hindered by sediment permeability (Figure \ref{fig:reaction_zone_size_sensitivity}b). However, given the different flow regimes imposed by varying bed form geometries, this relationship may not be consistent in all environments. This inverse relationship also represents another layer of complexity that should be evaluated in future studies: under otherwise analogous environmental conditions (i.e., flow depth, surface water velocity), streambed sediments of varying permeability will form bed forms of different size and shape. As a result, the relative influence of bed form geometry and sediment permeability on hyporheic processes will shift across time and space as sediments are transported and the complex streambed topography evolves in response to episodic flow events. 

\begin{figure}[!ht]
  \centering
  \begin{subfigure}{0.48\linewidth}
  \centering
    \includegraphics[trim=0cm 0cm 0cm 0cm, clip, width=0.8\linewidth]{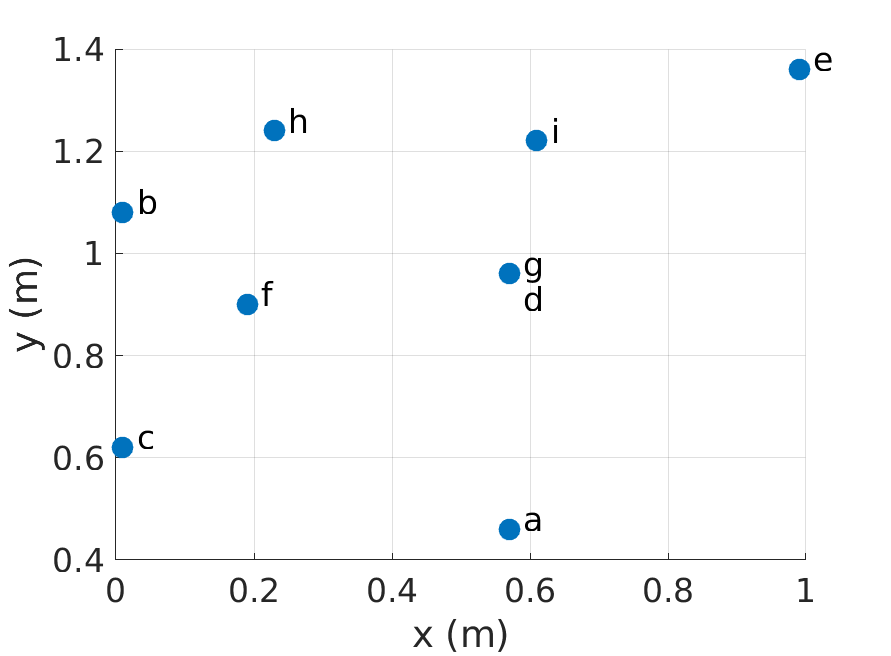}
    \caption{Maximum reaction rates spots}
  \end{subfigure}
  \begin{subfigure}{0.48\linewidth}
    \includegraphics[trim=2.8cm 2cm 5cm 2cm, clip, width=\linewidth]{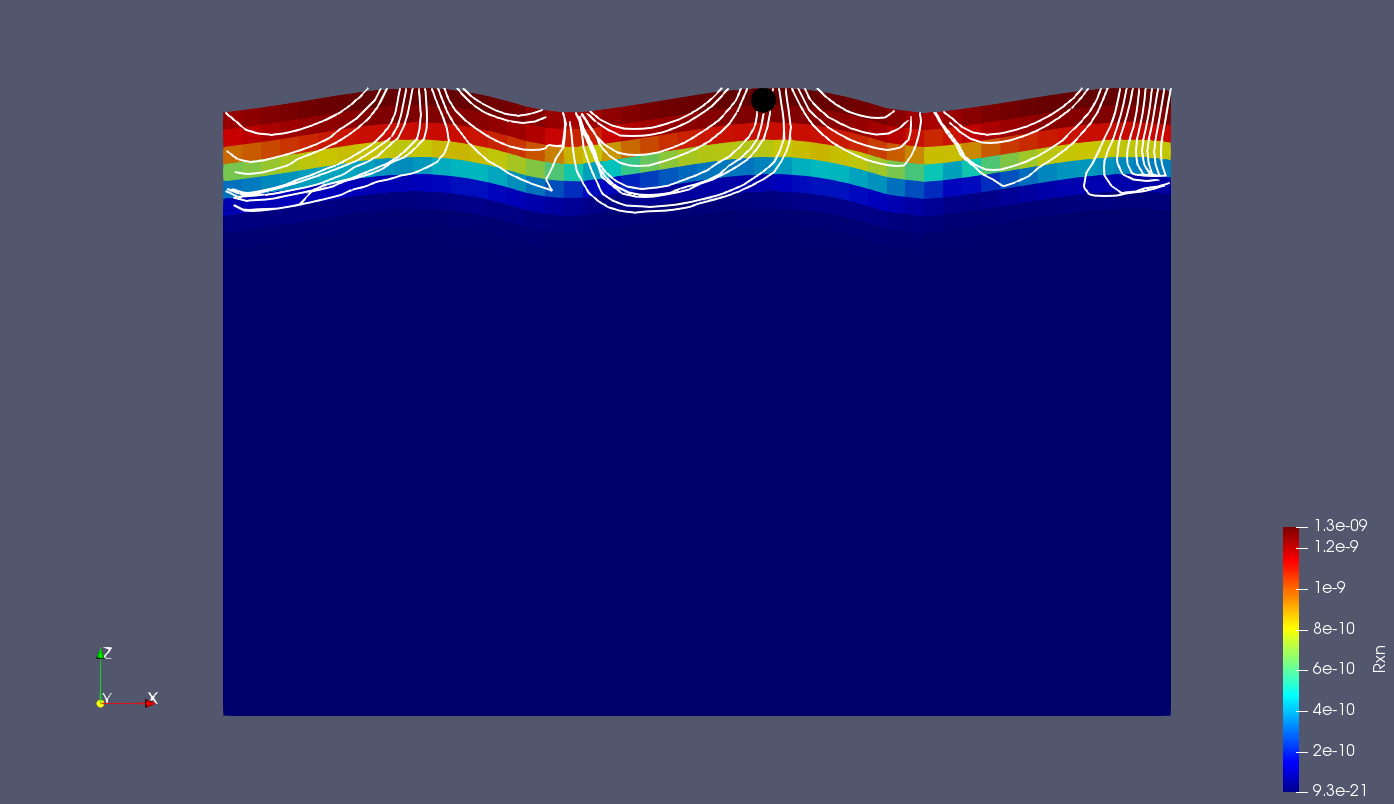}
    \caption{bed form (a) at y=0.46 (m)}
  \end{subfigure}
  \begin{subfigure}{0.48\linewidth}
    \includegraphics[trim=2.8cm 2cm 5cm 2cm, clip, width=\linewidth]{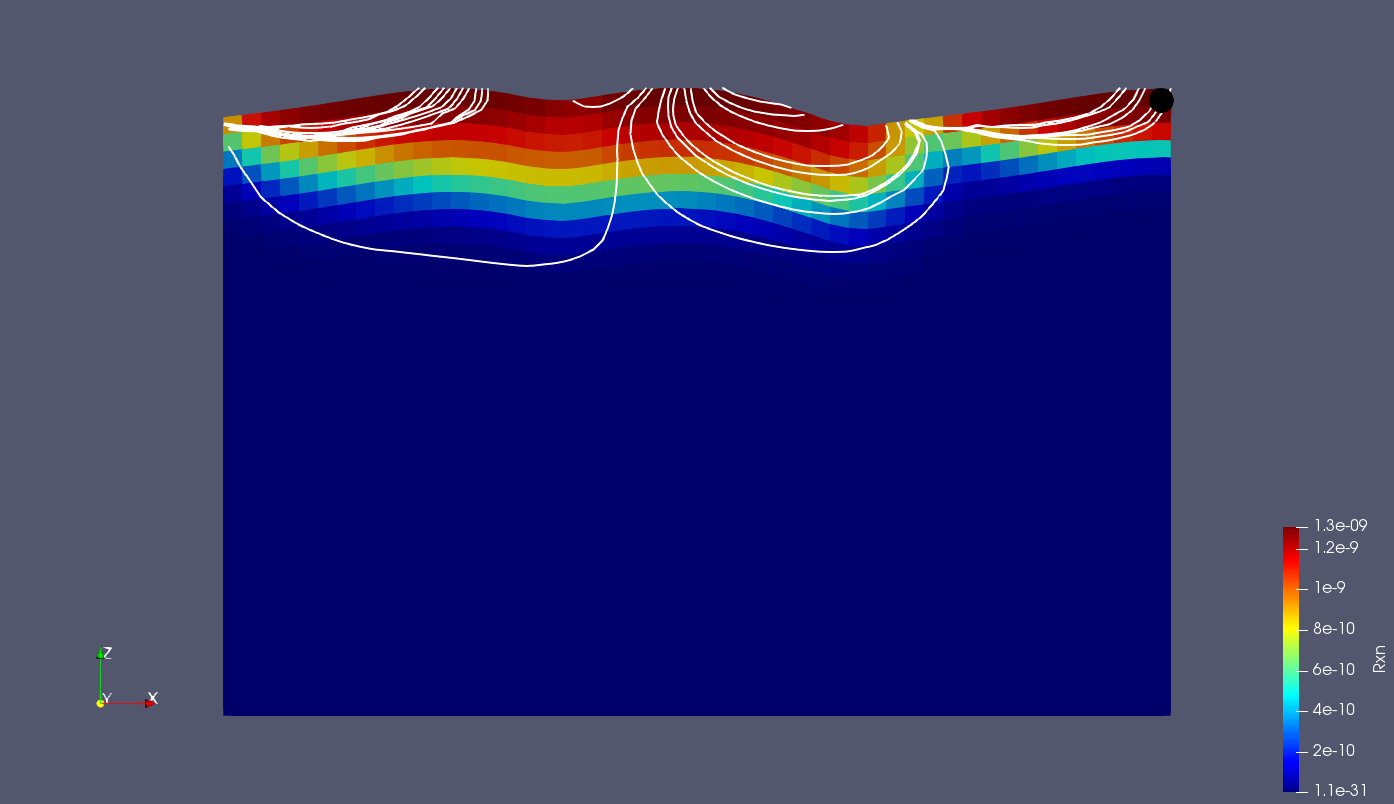}
    \caption{bed form (e) at y=0.96 (m)}
  \end{subfigure}
  \begin{subfigure}{0.48\linewidth}
    \includegraphics[trim=2.8cm 2cm 5cm 2cm, clip, width=\linewidth]{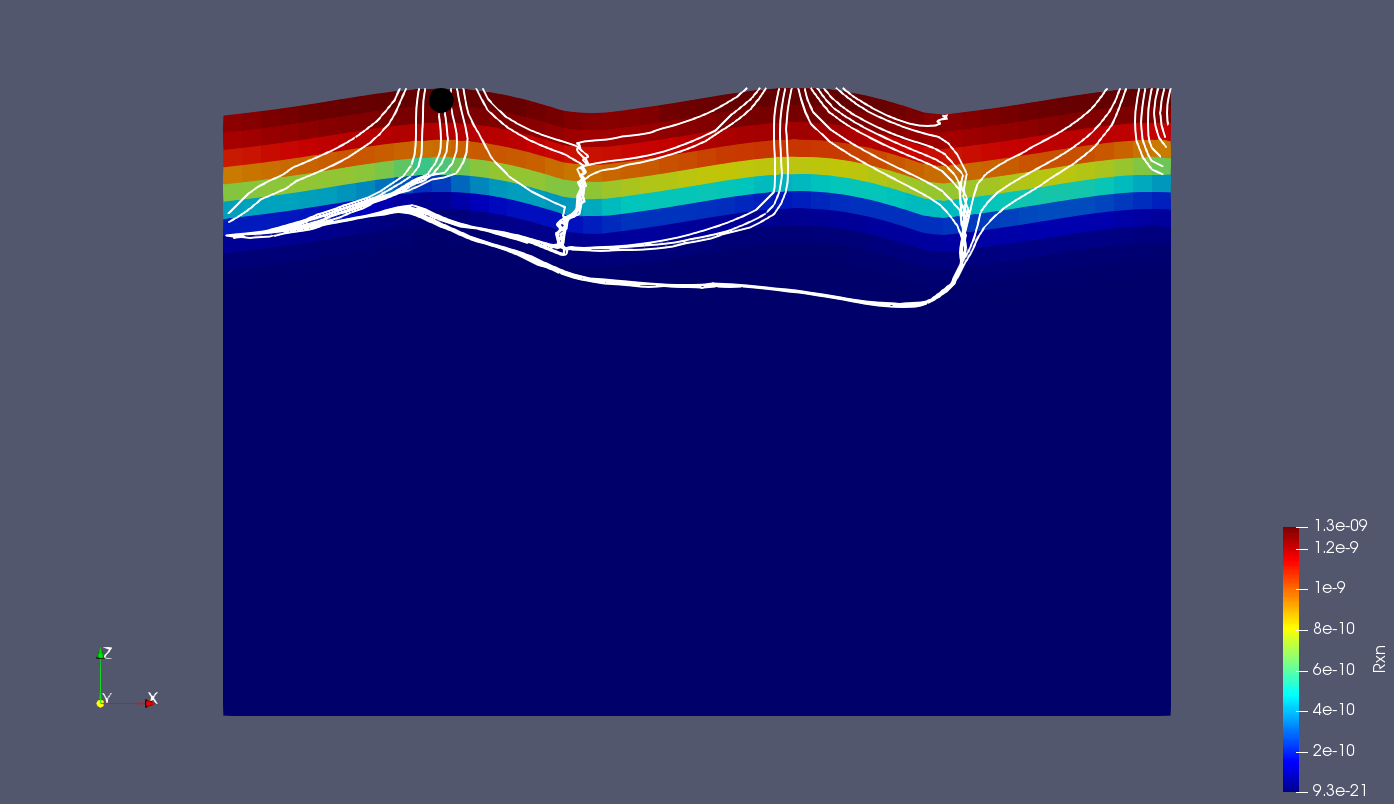}
    \caption{bed form (h) at y=1.24 (m)}
  \end{subfigure}
  \begin{subfigure}{0.6\linewidth}
    \includegraphics[width=\linewidth]{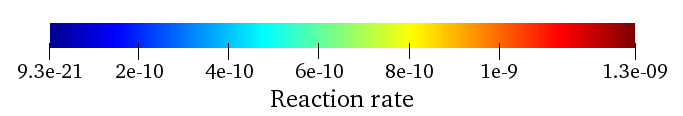}
  \end{subfigure}
    \caption{Hotspots; The contours display the reaction rate distribution, the black spots represent the location of the maximum reaction rates, and the white lines are the streamlines.}
  \label{fig:hotspots}
\end{figure}

Depending on the bed form geometry, the maximum reaction rate for identical biogeochemical processes may change in space. 
Figure \ref{fig:hotspots}a depicts the plan view of locations with maximum reaction rates in different bed forms. 
Figures \ref{fig:hotspots}b, \ref{fig:hotspots}c, and \ref{fig:hotspots}d display the locations of maximum reaction rate from a cross-sectional view. Here, bed forms (a), (e), and (h) are presented to show distinct locations of maximum reaction rate spots.


\section{Conclusions and Future Directions}

This work studied how bed form topographies control various hyporheic exchange processes. A framework to simulate surface and subsurface flow and reactive solute transport within the hyporheic zone is developed. The framework enables prediction of hotspots for different surface flow properties, bed form shapes, and sediment characteristics. While the framework is currently utilized for a simple reactions and the homogeneous sediment, it could also be used to represent heterogeneous sediments as well as biogeochemical processes such as nutrient transformation which involve multiple equilibrium and kinetically controlled reactions. 

Bed forms drive small-scale hyporheic exchange and host dynamic regions of biogeochemical activity. As a primary control on hyporheic residence times and nutrient availability, bed form geometry largely influences the rate and distribution of such microbial hotspots.
This work specifically shows that the reaction rates depend on exchange fluxes and thus on bed form-specific shape and morphologies. 
In particular, when bed form shape is represented by a series of sine curves, the phase shift of the second planform sinuosity (\(\Delta \varphi_2 \)) drives complex pressure distributions across the streambed that increase hyporheic exchange and the depth of solute penetration, often resulting in larger reaction zones and overall higher reaction rates. 
The results in this research suggest that out-of-phase bed forms generate more complex groundwater flow patterns with variable residence times, however, which reduce the efficiency of solute transformations. 

Future studies will need to address how heterogeneity in various bed forms affects the biogeochemical process and reaction rates. 
Bed form migration and erosional and depositional processes were outside of the scope of this study, but evolving bed topography would cause removal efficiency to vary temporally as the geometry shifted between in- and out-of-phase. 
Future research should also focus on how temporal variability in bed form geometry alters hyporheic flow patterns and shifts biogeochemical reaction rates, especially in dynamic environments with regular stage changes (i.e., tides, dam release) where erosional and depositional processes would likely be accelerated. 
A clear understanding of the role of hyporheic bed forms on hyporheic exchange processes is crucial to the protection of water quality, especially as anthropogenic contaminant input to the biosphere increases.

\section*{Acknowledgements}
The first author (FB) was funded by University of Cincinnati through faculty startup fund to the last author (MRS). The second author (CDW) was funded by the National Science Foundation (EAR-PF 1855193).


\bibliography{bedforms}

\begin{thebibliography}{10}
\expandafter\ifx\csname url\endcsname\relax
  \def\url#1{\texttt{#1}}\fi
\expandafter\ifx\csname urlprefix\endcsname\relax\def\urlprefix{URL }\fi
\expandafter\ifx\csname href\endcsname\relax
  \def\href#1#2{#2} \def\path#1{#1}\fi

\bibitem{rl1982sedimentary}
A.~J. RL, Sedimentary structures, their character and physical basis Volume 1,
  Elsevier, 1982.

\bibitem{ElliottBrooks1997}
A.~H. Elliott, N.~H. Brooks, Transfer of nonsorbing solutes to a streambed with
  bed forms: Laboratory experiments, Water Resources Research 33~(1) (1997)
  137--151.
\newblock \href {http://dx.doi.org/10.1029/96WR02783}
  {\path{doi:10.1029/96WR02783}}.

\bibitem{Gomez-Velez2015}
J.~D. Gomez-Velez, J.~W. Harvey, M.~B. Cardenas, B.~Kiel, Denitrification in
  the mississippi river network controlled by flow through river bedforms,
  Nature Geoscience 8~(12) (2015) 941--945.
\newblock \href {http://dx.doi.org/10.1038/ngeo2567}
  {\path{doi:10.1038/ngeo2567}}.

\bibitem{BOANO20101367}
F.~Boano, R.~Revelli, L.~Ridolfi, Effect of streamflow stochasticity on
  bedform-driven hyporheic exchange, Advances in Water Resources 33~(11) (2010)
  1367--1374, special Issue on ground water-surface water interactions.
\newblock \href
  {http://dx.doi.org/https://doi.org/10.1016/j.advwatres.2010.03.005}
  {\path{doi:https://doi.org/10.1016/j.advwatres.2010.03.005}}.

\bibitem{Buffington_Tonina2009}
J.~M. Buffington, D.~Tonina, Hyporheic exchange in mountain rivers ii: Effects
  of channel morphology on mechanics, scales, and rates of exchange, Geography
  Compass 3~(3) (2009) 1038--1062.
\newblock \href {http://dx.doi.org/10.1111/j.1749-8198.2009.00225.x}
  {\path{doi:10.1111/j.1749-8198.2009.00225.x}}.

\bibitem{Harvey2012}
J.~W. Harvey, J.~D. Drummond, R.~L. Martin, L.~E. McPhillips, A.~I. Packman,
  D.~J. Jerolmack, S.~H. Stonedahl, A.~F. Aubeneau, A.~H. Sawyer, L.~G. Larsen,
  C.~R. Tobias, Hydrogeomorphology of the hyporheic zone: Stream solute and
  fine particle interactions with a dynamic streambed, Journal of Geophysical
  Research: Biogeosciences 117~(G4) (2012) G00N11.
\newblock \href {http://dx.doi.org/10.1029/2012JG002043}
  {\path{doi:10.1029/2012JG002043}}.

\bibitem{Chen_Cardenas2018}
X.~Chen, M.~B. Cardenas, L.~Chen, Hyporheic exchange driven by
  three-dimensional sandy bed forms: Sensitivity to and prediction from bed
  form geometry, Water Resources Research 54~(6) (2018) 4131--4149.
\newblock \href {http://dx.doi.org/10.1029/2018WR022663}
  {\path{doi:10.1029/2018WR022663}}.

\bibitem{BAYANICARDENAS20081382}
M.~{Bayani Cardenas}, J.~L. Wilson, R.~Haggerty, Residence time of
  bedform-driven hyporheic exchange, Advances in Water Resources 31~(10) (2008)
  1382--1386.
\newblock \href
  {http://dx.doi.org/https://doi.org/10.1016/j.advwatres.2008.07.006}
  {\path{doi:https://doi.org/10.1016/j.advwatres.2008.07.006}}.

\bibitem{McClain2003}
M.~E. McClain, E.~W. Boyer, C.~L. Dent, S.~E. Gergel, N.~B. Grimm, P.~M.
  Groffman, S.~C. Hart, J.~W. Harvey, C.~A. Johnston, E.~Mayorga, W.~H.
  McDowell, G.~Pinay, Biogeochemical hot spots and hot moments at the interface
  of terrestrial and aquatic ecosystems, Ecosystems 6~(4) (2003) 301--312.
\newblock \href {http://dx.doi.org/10.1007/s10021-003-0161-9}
  {\path{doi:10.1007/s10021-003-0161-9}}.

\bibitem{Lautz2008}
L.~Lautz, R.~Fanelli, Seasonal biogeochemical hotspots in the streambed around
  restoration structures, Biogeochemistry 91 (2008) 85--104.
\newblock \href {http://dx.doi.org/10.1007/s10533-008-9235-2}
  {\path{doi:10.1007/s10533-008-9235-2}}.

\bibitem{Li2017}
L.~Li, K.~Maher, A.~Navarre-Sitchler, J.~Druhan, C.~Meile, C.~Lawrence,
  J.~Moore, J.~Perdrial, P.~Sullivan, A.~Thompson, L.~Jin, E.~W. Bolton, S.~L.
  Brantley, W.~E. Dietrich, K.~U. Mayer, C.~I. Steefel, A.~Valocchi,
  J.~Zachara, B.~Kocar, J.~Mcintosh, B.~M. Tutolo, M.~Kumar, E.~Sonnenthal,
  C.~Bao, J.~Beisman, Expanding the role of reactive transport models in
  critical zone processes, Earth-Science Reviews 165 (2017) 280--301.
\newblock \href {http://dx.doi.org/10.1016/j.earscirev.2016.09.001}
  {\path{doi:10.1016/j.earscirev.2016.09.001}}.

\bibitem{Thibodeaux1987}
L.~J. Thibodeaux, J.~D. Boyle, Bedform-generated convective transport in bottom
  sediment, Nature 325 (1987) 341--343.

\bibitem{Cardenas_Wilson2007a}
M.~B. Cardenas, J.~L. Wilson, Dunes, turbulent eddies, and interfacial exchange
  with permeable sediments, Water Resources Research 43~(8) (2007) W08412.
\newblock \href {http://dx.doi.org/10.1029/2006WR005787}
  {\path{doi:10.1029/2006WR005787}}.

\bibitem{CARDENAS2007301}
M.~B. Cardenas, J.~Wilson, Hydrodynamics of coupled flow above and below a
  sediment–water interface with triangular bedforms, Advances in Water
  Resources 30~(3) (2007) 301 -- 313.
\newblock \href {http://dx.doi.org/10.1016/j.advwatres.2006.06.009}
  {\path{doi:10.1016/j.advwatres.2006.06.009}}.

\bibitem{Marion2002}
A.~Marion, M.~Bellinello, I.~Guymer, A.~Packman, Effect of bed form geometry on
  the penetration of nonreactive solutes into a streambed, Water Resources
  Research 38~(10) (2002) 1209.
\newblock \href {http://dx.doi.org/10.1029/2001WR000264}
  {\path{doi:10.1029/2001WR000264}}.

\bibitem{BOANO2007148}
F.~Boano, R.~Revelli, L.~Ridolfi, Bedform-induced hyporheic exchange with
  unsteady flows, Advances in Water Resources 30~(1) (2007) 148--156.
\newblock \href
  {http://dx.doi.org/https://doi.org/10.1016/j.advwatres.2006.03.004}
  {\path{doi:https://doi.org/10.1016/j.advwatres.2006.03.004}}.

\bibitem{Sedimentary1982}
J.~R. Allen, Sedimentary Structures Their Character and Physical Basis,
  Vol.~30, Elsevier scientific, 1982.
\newblock \href {http://dx.doi.org/10.1016/S0070-4571(08)71008-2}
  {\path{doi:10.1016/S0070-4571(08)71008-2}}.

\bibitem{FLECKENSTEIN20101291}
J.~H. Fleckenstein, S.~Krause, D.~M. Hannah, F.~Boano, Groundwater-surface
  water interactions: New methods and models to improve understanding of
  processes and dynamics, Advances in Water Resources 33~(11) (2010)
  1291--1295, special Issue on ground water-surface water interactions.
\newblock \href
  {http://dx.doi.org/https://doi.org/10.1016/j.advwatres.2010.09.011}
  {\path{doi:https://doi.org/10.1016/j.advwatres.2010.09.011}}.

\bibitem{Packman2000}
A.~I. Packman, N.~H. Brooks, J.~J. Morgan, A physicochemical model for colloid
  exchange between a stream and a sand streambed with bed forms, Water
  Resources Research 36~(8) (2000) 2351--2361.
\newblock \href {http://dx.doi.org/10.1029/2000WR900059}
  {\path{doi:10.1029/2000WR900059}}.

\bibitem{Worman2002}
A.~W\"{o}rman, A.~I. Packman, H.~Johansson, K.~Jonsson, Effect of flow-induced
  exchange in hyporheic zones on longitudinal transport of solutes in streams
  and rivers, Water Resources Research 38~(1) (2002) 2--1--2--15.
\newblock \href {http://dx.doi.org/10.1029/2001WR000769}
  {\path{doi:10.1029/2001WR000769}}.

\bibitem{KAZEZYILMAZALHAN200626}
C.~M. Kazezyılmaz-Alhan, M.~A. Medina, Stream solute transport incorporating
  hyporheic zone processes, Journal of Hydrology 329~(1) (2006) 26 -- 38.
\newblock \href {http://dx.doi.org/10.1016/j.jhydrol.2006.02.003}
  {\path{doi:10.1016/j.jhydrol.2006.02.003}}.

\bibitem{Trauth2014}
N.~Trauth, C.~Schmidt, M.~Vieweg, U.~Maier, J.~H. Fleckenstein, Hyporheic
  transport and biogeochemical reactions in pool-riffle systems under varying
  ambient groundwater flow conditions, Journal of Geophysical Research:
  Biogeosciences 119~(5) (2014) 910--928.
\newblock \href {http://dx.doi.org/10.1002/2013JG002586}
  {\path{doi:10.1002/2013JG002586}}.

\bibitem{Trauth2015}
N.~Trauth, C.~Schmidt, M.~Vieweg, S.~E. Oswald, J.~H. Fleckenstein, Hydraulic
  controls of in-stream gravel bar hyporheic exchange and reactions, Water
  Resources Research 51~(4) (2015) 2243--2263.
\newblock \href {http://dx.doi.org/10.1002/2014WR015857}
  {\path{doi:10.1002/2014WR015857}}.

\bibitem{Li2020}
B.~Li, X.~Liu, M.~H. Kaufman, A.~Turetcaia, X.~Chen, M.~B. Cardenas, Flexible
  and modular simultaneous modeling of flow and reactive transport in rivers
  and hyporheic zones, Water Resources Research 56~(2) (2020) e2019WR026528.
\newblock \href {http://dx.doi.org/10.1029/2019WR026528}
  {\path{doi:10.1029/2019WR026528}}.

\bibitem{Yabusaki2017}
S.~B. Yabusaki, M.~J. Wilkins, Y.~Fang, K.~H. Williams, B.~Arora, J.~Bargar,
  H.~R. Beller, N.~J. Bouskill, E.~L. Brodie, J.~N. Christensen, M.~E. Conrad,
  R.~E. Danczak, E.~King, M.~R. Soltanian, N.~F. Spycher, C.~I. Steefel, T.~K.
  Tokunaga, R.~Versteeg, S.~R. Waichler, H.~M. Wainwright, Water table dynamics
  and biogeochemical cycling in a shallow, variably-saturated floodplain,
  Environmental Science \& Technology 51~(6) (2017) 3307--3317, pMID: 28218533.
\newblock \href {http://dx.doi.org/10.1021/acs.est.6b04873}
  {\path{doi:10.1021/acs.est.6b04873}}.

\bibitem{Boano2015}
F.~Boano, J.~W. Harvey, A.~Marion, A.~I. Packman, R.~Revelli, L.~Ridolfi,
  A.~Wörman, Hyporheic flow and transport processes: Mechanisms, models, and
  biogeochemical implications, Reviews of Geophysics 52~(4) (2014) 603--679.
\newblock \href {http://dx.doi.org/10.1002/2012RG000417}
  {\path{doi:10.1002/2012RG000417}}.

\bibitem{Janssen2012}
F.~Janssen, M.~B. Cardenas, A.~H. Sawyer, T.~Dammrich, J.~Krietsch, D.~de~Beer,
  A comparative experimental and multiphysics computational fluid dynamics
  study of coupled surface–subsurface flow in bed forms, Water Resources
  Research 48~(8) (2012) W08514.
\newblock \href {http://dx.doi.org/10.1029/2012WR011982}
  {\path{doi:10.1029/2012WR011982}}.

\bibitem{bardini2013small}
L.~Bardini, F.~Boano, M.~Cardenas, A.~Sawyer, R.~Revelli, L.~Ridolfi,
  Small-scale permeability heterogeneity has negligible effects on nutrient
  cycling in streambeds, Geophysical Research Letters 40~(6) (2013) 1118--1122.

\bibitem{sawyer2009hyporheic}
A.~H. Sawyer, M.~B. Cardenas, Hyporheic flow and residence time distributions
  in heterogeneous cross-bedded sediment, Water Resources Research 45~(8).

\bibitem{Tonina_Buffington2007}
D.~Tonina, J.~M. Buffington, Hyporheic exchange in gravel bed rivers with
  pool-riffle morphology: Laboratory experiments and three-dimensional
  modeling, Water Resources Research 43~(1) (2007) W01421.
\newblock \href {http://dx.doi.org/10.1029/2005WR004328}
  {\path{doi:10.1029/2005WR004328}}.

\bibitem{Chen2015}
X.~Chen, M.~B. Cardenas, L.~Chen, Three-dimensional versus two-dimensional bed
  form-induced hyporheic exchange, Water Resources Research 51~(4) (2015)
  2923--2936.
\newblock \href {http://dx.doi.org/10.1002/2014WR016848}
  {\path{doi:10.1002/2014WR016848}}.

\bibitem{Zhou2014}
Y.~Zhou, R.~W. Ritzi~Jr., M.~R. Soltanian, D.~F. Dominic, The influence of
  streambed heterogeneity on hyporheic flow in gravelly rivers, Groundwater
  52~(2) (2014) 206--216.
\newblock \href {http://dx.doi.org/10.1111/gwat.12048}
  {\path{doi:10.1111/gwat.12048}}.

\bibitem{Zheng2019}
L.~Zheng, M.~B. Cardenas, L.~Wang, D.~Mohrig, Ripple effects: Bed form
  morphodynamics cascading into hyporheic zone biogeochemistry, Water Resources
  Research 55~(8) (2019) 7320--7342.
\newblock \href {http://dx.doi.org/10.1029/2018WR023517}
  {\path{doi:10.1029/2018WR023517}}.

\bibitem{worman2006}
A.~W\"{o}rman, A.~I. Packman, L.~Marklund, J.~W. Harvey, S.~H. Stone, Exact
  three-dimensional spectral solution to surface-groundwater interactions with
  arbitrary surface topography, Geophysical Research Letters 33~(7) (2006)
  L07402.
\newblock \href {http://dx.doi.org/10.1029/2006GL025747}
  {\path{doi:10.1029/2006GL025747}}.

\bibitem{Stonedahl2010}
S.~H. Stonedahl, J.~W. Harvey, A.~Wörman, M.~Salehin, A.~I. Packman, A
  multiscale model for integrating hyporheic exchange from ripples to meanders,
  Water Resources Research 46~(12).
\newblock \href {http://dx.doi.org/10.1029/2009WR008865}
  {\path{doi:10.1029/2009WR008865}}.

\bibitem{Stonedahl2012}
S.~H. Stonedahl, J.~W. Harvey, J.~Detty, A.~Aubeneau, A.~I. Packman, Physical
  controls and predictability of stream hyporheic flow evaluated with a
  multiscale model, Water Resources Research 48~(10).
\newblock \href {http://dx.doi.org/10.1029/2011WR011582}
  {\path{doi:10.1029/2011WR011582}}.

\bibitem{Stonedahl2013}
S.~H. Stonedahl, J.~W. Harvey, A.~I. Packman, Interactions between hyporheic
  flow produced by stream meanders, bars, and dunes, Water Resources Research
  49~(9) (2013) 5450--5461.
\newblock \href {http://dx.doi.org/10.1002/wrcr.20400}
  {\path{doi:10.1002/wrcr.20400}}.

\bibitem{McLean1994}
S.~R. McLean, J.~M. Nelson, S.~R. Wolfe, Turbulence structure over
  two-dimensional bed forms: Implications for sediment transport, Journal of
  Geophysical Research: Oceans 99~(C6) (1994) 12729--12747.
\newblock \href {http://dx.doi.org/10.1029/94JC00571}
  {\path{doi:10.1029/94JC00571}}.

\bibitem{Rubin1987}
D.~M. Rubin,
  \href{https://books.google.com/books?id=p56tD8oajEgC}{Cross-bedding,
  Bedforms, and Paleocurrents}, Concepts in sedimentology and paleontology,
  Society of Economic Paleontologists and Mineralogists, 1987.
\newline\urlprefix\url{https://books.google.com/books?id=p56tD8oajEgC}

\bibitem{Rubin_Carter}
D.~M. Rubin, C.~L. Carter, Bedforms 4.0: Matlab code for simulating bedforms
  and cross-bedding, Tech. Rep. 2005-1272, U.S. Geological Survey Open-File
  Report, 13 p (2005).

\bibitem{ripple_dune}
Flows, sediments and bedforms.,
  \url{http://geologylearn.blogspot.com/2015/08/flows-sediments-and-bedforms.html}.

\bibitem{lunate}
Bedforms and cross-stratification produced by currents.,
  \url{https://sites.ualberta.ca/~jwaldron/gallerypages/bedforms.html}.

\bibitem{OpenFoam}
{OpenFOAM v7}, {The OpenFOAM Foundation Ltd}, https://openfoam.org (2019).

\bibitem{pflotran-paper}
G.~E. Hammond, P.~C. Lichtner, R.~T. Mills, Evaluating the performance of
  parallel subsurface simulators: An illustrative example with pflotran, Water
  Resources Research 50 (2014) 208--228.
\newblock \href {http://dx.doi.org/10.1002/2012WR013483}
  {\path{doi:10.1002/2012WR013483}}.

\bibitem{pflotran-web-page}
P.~C. Lichtner, G.~E. Hammond, C.~Lu, S.~Karra, G.~Bisht, B.~Andre, R.~T.
  Mills, J.~Kumar, J.~M. Frederick, {PFLOTRAN} {W}eb page,
  http://www.pflotran.org (2019).

\bibitem{pflotran-user-ref}
P.~C. Lichtner, G.~E. Hammond, C.~Lu, S.~Karra, G.~Bisht, B.~Andre, R.~T.
  Mills, J.~Kumar, J.~M. Frederick, {PFLOTRAN} user manual, Tech. rep.,
  http://documentation.pflotran.org (2019).

\bibitem{LaGriT}
{Los Alamos Grid Toolbox, LaGriT}, {Los Alamos National Laboratory},
  http://lagrit.lanl.gov (2020).

\bibitem{LAUNDER1974269}
B.~Launder, D.~Spalding, The numerical computation of turbulent flows, Computer
  Methods in Applied Mechanics and Engineering 3~(2) (1974) 269 -- 289.
\newblock \href {http://dx.doi.org/10.1016/0045-7825(74)90029-2}
  {\path{doi:10.1016/0045-7825(74)90029-2}}.

\bibitem{Menter2003}
F.~R. Menter, M.~Kuntz, R.~Langtry, Ten years of industrial experience with the
  {SST} turbulence model, The fourth international symposium on turbulence,
  heat and mass transfer, Antalya, Turkey, 2003.

\bibitem{dwivedi2018hot}
D.~Dwivedi, B.~Arora, C.~I. Steefel, B.~Dafflon, R.~Versteeg, Hot spots and hot
  moments of nitrogen in a riparian corridor, Water Resources Research 54~(1)
  (2018) 205--222.

\bibitem{dwivedi2018geochemical}
D.~Dwivedi, C.~I. Steefel, B.~Arora, M.~Newcomer, J.~D. Moulton, B.~Dafflon,
  B.~Faybishenko, P.~Fox, P.~Nico, N.~Spycher, et~al., Geochemical exports to
  river from the intrameander hyporheic zone under transient hydrologic
  conditions: East river mountainous watershed, colorado, Water Resources
  Research 54~(10) (2018) 8456--8477.

\bibitem{Bear1972}
J.~Bear, Dynamics of Fluids in Porous Media, Elsevier, New York, 1972.

\bibitem{Freeze1979}
R.~A. Freeze, J.~A. Cherry, Groundwater, Prentice-Hall, Englewood Cliffs, N.J.,
  1979.

\bibitem{Gu2007}
C.~Gu, G.~M. Hornberger, A.~L. Mills, J.~S. Herman, S.~A. Flewelling, Nitrate
  reduction in streambed sediments: Effects of flow and biogeochemical
  kinetics, Water Resources Research 43 (2007) 1--10.
\newblock \href {http://dx.doi.org/10.1029/2007WR006027}
  {\path{doi:10.1029/2007WR006027}}.

\bibitem{gomez2014hydrogeomorphic}
J.~D. Gomez-Velez, J.~W. Harvey, A hydrogeomorphic river network model predicts
  where and why hyporheic exchange is important in large basins, Geophysical
  Research Letters 41~(18) (2014) 6403--6412.

\bibitem{gomez2015denitrification}
J.~D. Gomez-Velez, J.~W. Harvey, M.~B. Cardenas, B.~Kiel, Denitrification in
  the mississippi river network controlled by flow through river bedforms,
  Nature Geoscience 8~(12) (2015) 941--945.

\bibitem{cardenas2008surface}
M.~B. Cardenas, Surface water-groundwater interface geomorphology leads to
  scaling of residence times, Geophysical Research Letters 35~(8).

\bibitem{cardenas2007potential}
M.~B. Cardenas, Potential contribution of topography-driven regional
  groundwater flow to fractal stream chemistry: Residence time distribution
  analysis of t{\'o}th flow, Geophysical Research Letters 34~(5).

\bibitem{kollet2008demonstrating}
S.~J. Kollet, R.~M. Maxwell, Demonstrating fractal scaling of baseflow
  residence time distributions using a fully-coupled groundwater and land
  surface model, Geophysical Research Letters 35~(7).

\bibitem{worman2007fractal}
A.~W{\"o}rman, A.~I. Packman, L.~Marklund, J.~W. Harvey, S.~H. Stone, Fractal
  topography and subsurface water flows from fluvial bedforms to the
  continental shield, Geophysical Research Letters 34~(7).

\bibitem{gardner2015high}
W.~P. Gardner, G.~Hammond, P.~Lichtner, High performance simulation of
  environmental tracers in heterogeneous domains, Groundwater 53~(S1) (2015)
  71--80.

\end{thebibliography}

\end{document}